\documentclass[acmsmall, nonacm]{acmart}

\author{Jos\'e Fragoso Santos}
\affiliation{
  \institution{Imperial College London}           
  \city{London}
  \country{UK}                   
}
\affiliation{
  \institution{INESC-ID/IST, Univ.~of Lisbon} 
  \state{Lisbon}
  \country{Portugal}                   
}

\author{Petar Maksimovi\'c}
\affiliation{
  \institution{Imperial College London}           
  \city{London}
  \country{UK}                   
}
\affiliation{
  \institution{Mathematical Institute SASA} 
  \city{Belgrade}
  \country{Serbia}                   
}

\author{Sacha-\'Elie Ayoun}
\affiliation{
  \institution{Imperial College London} 
  \city{London}
  \country{UK}              
}

\author{Philippa Gardner}
\affiliation{
  \institution{Imperial College London} 
  \city{London}
  \country{UK}              
}

\makeatletter
\def\mdseries@tt{m}
\makeatother
\usepackage[frozencache]{minted}
\usepackage{xcolor}
\usepackage{xparse}
\usepackage{xspace}
\usepackage{amssymb}
\usepackage{graphicx}
\usepackage{tabularx}
\usepackage{wasysym}
\usepackage{mathpartir}
\usepackage{url}
\usepackage{booktabs}
\usepackage{tikz}
\usepackage{pf2}
\usetikzlibrary{matrix,positioning,decorations.pathmorphing}
\usepackage{enumitem}
\usepackage{wrapfig}
\usepackage{caption}
\usepackage{makecell}
\usepackage{multirow}
\usepackage{makecell}

\usepackage{macros}
\usepackage{while_macros}

\newif\ifComments
\Commentstrue

\title[Gillian: Compositional Symbolic Execution for All]{Gillian: Compositional Symbolic Execution for All}

\begin{abstract}
We present \gillian, a language-independent framework for the development of compositional symbolic analysis tools. 
\gillian supports three flavours of analysis: whole-program symbolic testing, full verification, and bi-abduction. 
It comes with fully parametric meta-theoretical results and a modular implementation,  designed to minimise the instantiation effort required of the user.  
We evaluate \gillian by instantiating it to JavaScript and~C, and perform its analyses on a set of data-structure libraries, obtaining results that indicate that \gillian is robust enough to reason about real-world programming languages.

\end{abstract}

\begin{document}
\maketitle 

\section{Introduction}

Developing symbolic execution analyses for modern programming languages is a challenging and time-consuming task. 
The complexity of the underlying meta-theory and the associated tool development is substantial. 
However, symbolic execution tools share a number of features independent  of the target language: 
for instance, interaction with first-order solvers~\cite{z3} and the management of the variable store. 
Even so, the amount of effort required to transfer an analysis from one programming language to another is often prohibitive. 
While there has been work on streamlining this process by embedding the target language into a host language with support for symbolic analysis~\cite{bucur:asplos:2014,torlak:onward:2013}, these approaches do not scale well to fully-fledged real-world~languages. 


We present \gillian, a general framework for swift development of compositional symbolic analysis tools     
for real-world programming languages. 
\gillian is underpinned by \gil, a simple goto language
parametric on the memory model of the target language: that is, on a set of actions 
describing the fundamental ways in which the programs of the target language interact with their 
respective memories.   
%
%
\gillian comes with a fully parametric meta-theory that unifies the treatment of symbolic execution, compositional program reasoning based on separation logic (SL), and bi-abduction.
In particular, it supports three main styles of analysis: 
\begin{itemize}[leftmargin=2em]
\item \emph{whole-program symbolic execution}, where end-users write unit tests with symbolic inputs and 
outputs and use simple first-order assertions to describe the properties that the outputs must satisfy,
while \gillian tries to generate symbolic traces that invalidate those assertions; 
\item \emph{verification}, where end-users annotate the functions of their programs with 
SL-style specifications and use \gillian to verify that the functions meet their specifications; and
 \item \emph{bi-abduction}, where end-users simply provide a  program with no specifications or unit tests and 
 are given back a set of SL-style specifications describing the behaviour of every function 
 in the program up to a pre-established bound; if, in this process, a possible bug is found, users are given back the symbolic 
 trace that leads to the~bug. 
\end{itemize}

To obtain an instantiation of each type of analysis for their target languages, users need to provide a compiler from the target language to \gil, and then simply 
instantiate \gillian with a memory model that provides an action-level implementation of the corresponding analysis:
%
%
%
whole-program symbolic execution requires a symbolic implementation of the memory model of the 
target language; verification additionally requires that memory model to expose a set of \emph{core predicates} 
describing the memory's atomic constituents (essentially, SL assertions specific to that memory model); 
and bi-abduction requires the memory model to provide, on top of that, a mechanism for 
inferring missing resource whenever an action cannot be executed due to missing information. 
The implementation of a memory model with support for these three types of analyses (in OCaml) is substantially 
simpler than the implementation of each type of analysis from scratch.  

Importantly, the meta-theory of \gillian is fully parametric on the memory model of the target language. 
Hence, the soundness theorems of whole-program symbolic execution, verification, and bi-abduction are proven in 
general once and for all, and can be instantiated to each target language provided that the user 
proves the necessary memory-specific lemmas. 
Similarly to the implementation, proving the memory-specific lemmas is far 
easier than creating a new soundness proof for each instantiation and analysis. 
%

%


We evaluate \gillian by instantiating it to JavaScript and C. 
Using the obtained tools, we run the three analyses of \gillian on a series of data-structure libraries, purposefully written using the programming idioms specific to the two languages. 
For instance, our JavaScript examples rely on extensible objects, prototype-based inheritance, and closures, while our C examples use dynamic memory allocation, structures, and pointer arithmetic.  
The evaluation results indicate that \gillian is robust enough to reason about real-world programming~languages.

\myparagraph{Outline of the Paper} In \S\ref{sec:sem}, we present the parametric symbolic execution of \gillian. 
In \S\ref{sec:soundness}, we give a thorough account of our parametric soundness result.
The verification and bi-abduction analyses are then presented in detail in \S\ref{sec:verification} and \S\ref{sec:bi}, respectively.
In \S\ref{sec:rw}, we instantiate \gillian to obtain analysis tools for JavaScript and C.
We discuss the related work in \S\ref{sec:relwork} and conclude in \S\ref{sec:concls}.

Throughout the paper, to give the reader a better intuition, we illustrate how to instantiate \gillian to each 
type of analysis using as an example a simple \while language with static objects, reminiscent of that of~\cite{berdine:aplas:2005}. 
We also comment on our implementation choices and demonstrate how the implementation closely follows the theory.

\section{Parametric Symbolic Execution}
\label{sec:sem}

At the core of \gillian is a parametric interpreter for \gil, the simple intermediate goto language of \gillian.
We present the syntax of \gil and give its semantics in terms of state models~(\S\ref{subsec:gil:ssem}). A state model can be viewed as an interface through which a programming language interacts with its state, and is parametric on the underlying memory model. We introduce concrete and symbolic state and memory models, and demonstrate how to automatically lift memory models to appropriate state models (\S\ref{sec:concverssym}).
As the running example, we introduce a simple \while language: we give its syntax, actions, and compiler to \gil in \S\ref{subsec:while:compiler}, and present its concrete and symbolic memory models in \S\ref{subsec:while:csm}.

\subsection{\gil Syntax and Semantics}
\label{subsec:gil:ssem}

\gil is a simple goto language with top-level procedures. It is parametric on a set of actions, $\actions$, which provide a general mechanism for interacting with \gil states. This parametricity allows us to maintain \gil states opaque throughout the meta-theoretical development and to provide parametric soundness results, minimising the burden of proof for users of \gillian.

%
\smallskip
\begin{display}{The Syntax of \gil}
\hspace*{-0.15cm}
\begin{tabular}{l@{\quad\hspace*{0.95cm}}l}
   $\vl \in \vals \defeq \num \in \nums \mid \str \in \strs \mid \bool \in \bools \mid \loc, \varsigma \in \locs \mid \type \in \types \mid \fid \in \fids \mid \lst{\vl}$
   & 
   $\e \in \exprs \defeq \vl \mid \x \in \xs \mid \unop{\e} \mid \binop{\e_1}{\e_2}$
   \\[3pt]
    $\cm \in \cmds{\actions}$ $\defeq$ $\x := \e \mid \ifgoto{\e}{i} \mid \x := \e(\e') 
                                     \mid \x := \act(\e) \mid$ &    $\proc \in \procs{\actions}$ $\defeq$ $\procedure{f}{\x}{\lst{\cm}}$
 \vspace*{3pt}
    \\
    \hspace*{1.28cm}$\x := \symb{j} \mid \x := \fresh{j} \mid \return{\e} \mid  \fail{e} \mid \vanish $
    & 
    $\prog \in \progs{\actions}$ $:$ $\fids \pmap \procs{\actions}$
 \end{tabular} \\[0.2cm]
 
\noindent ~where $i, j \in \mathbb{N}$ and $\act \in \actions$.
\end{display}
\smallskip

\gil \emph{values}, $\vl \in \vals$, include numbers, strings, booleans, uninterpreted symbols, types, procedure identifiers, and lists of values. Uninterpreted symbols are mostly used to denote memory locations and instantiation-specific constants. Types are standard: they include, for example, the types of numbers, strings, booleans, and lists.
\gil~\emph{expressions}, $\e \in \exprs$, include values, program variables $\x$, and various unary and binary operators. 

\gil \emph{commands} include, first of all, the standard variable assignment, conditional goto, and dynamic procedure call.\footnote{The procedure call is dynamic in that the  identifier is obtained by evaluating the caller expression.} 
Next, we have three \gil-specific commands: action execution, 
$\x := \act(\e)$, executes the action $\act \in \actions$ with the argument obtained by evaluating $\e$; and two allocations, $\x := \symb{j}$ and $\x := \fresh{j}$, which generate fresh uninterpreted and interpreted symbols, respectively. These allocations are annotated with a jointly unique identifier $j \in \mathbb{N}$, addressed in \S\ref{subsec:alloc} and \S\ref{sec:concverssym}. Finally, we have three additional control-flow related commands: $\kwT{return}$ terminates the execution of 
the current procedure; $\kwT{fail}$ terminates the execution of the entire program with an error; 
and $\vanish$ silently terminates the execution of the entire program without generating a result. 


A \gil procedure, $\proc \in \procs{\actions}$ is of the form $\procedure{f}{\x}{\lst{\cm}}$, where $\f$ is its identifier, $\x$ is its formal parameter, 
and its body $\lst{\cm}$ is a sequence of \gil commands. 
%
A \gil program, $\prog \in \progs{\actions}$, is a finite partial function, mapping procedure identifiers 
to their corresponding procedures. 

\myparagraph{Semantics} 
The semantics of \gil is parameterised by a state model, $\gstate \in \gstates$, defined as follows. 


\begin{definition}[State Model]\label{def:state:module}
A state model $\gstate \in \gstates$ is a triple $\tup{\sset{\gstate}, \gval, \actions}$, consisting of: 
\etag{1} a set of states on which \gil programs operate, $\sset{\gstate} \ni \gstate$, \etag{2} a set of values stored in those states, $\gval \ni \gv$, and \etag{3} a set of 
actions that can be performed on those states, $\actions \ni \act$. All \gil states contain a variable store, $\sto : \xs \pmap \gval$, mapping program variables to values. 

A state model defines the following functions for acting on states ($\equiv_{pp}$ denotes pretty-printing for readability): 

\begin{itemize}
   \item $\kwT{setVar} : \sset{\gstate} \tmap \xs \tmap \gval \tmap \sset{\gstate}$ 
            \hfill ($\kwT{setVar}(\st, \x, \gv) \equiv_{pp} \st.\kwT{setVar}(\x, \gv)$) 
   \item $\kwT{setStore} : \sset{\gstate} \tmap (\xs \pmap \gval) \tmap \sset{\gstate}$ 
            \hfill ($\kwT{setStore}(\st, \sto) \equiv_{pp} \st.\kwT{setStore}(\sto)$) 
   \item $\kwT{getStore} : \sset{\gstate} \tmap (\xs \pmap \gval)$ 
            \hfill ($\kwT{getStore}(\st) \equiv_{pp} \st.\kwT{getStore}$) 
   \item $\kwT{ee} : \sset{\gstate} \tmap \exprs \pmap \gval$ 
            \hfill ($\kwT{ee}(\st, \e) \equiv_{pp} \st.\kwT{ee}(\e)$) 
   \item $\kwT{ea} : \actions \tmap \sset{\gstate} \tmap \gval \pmap \power{\sset{\gstate} \times \gval}$ 
            \hfill ($(\st', \gv') \in \kwT{ea}(\act, \st, \gv) \equiv_{pp} \action{\st}{\act}{\gv}{(\st', \gv')}$) 
\end{itemize}
\end{definition}

The intuition behind the state functions is as follows: \etag{1} $\kwT{setVar}(\st, \x, \gv)$  sets $\x$ to $\gv$ in the store of $\st$;
\etag{2} $\kwT{setStore}(\st, \x, \sto)$ replaces the store of $\st$ with $\sto$; 
\etag{3} $\kwT{getStore}(\st)$ obtains the store of $\st$; and 
\etag{4} $\kwT{ee}(\st, \e)$ evaluates the expression $\e$ in the store of $\st$; and
\etag{5} $\kwT{ea}(\act, \st, \gv)$ executes the action $\act$ on the state $\st$
with argument~$\gv$. 
Note that, since actions result in sets of state-value pairs, the \gil semantics may be  non-deterministic. 

 
A state model $\gstate =  \tup{\sset{\gstate}, \gval, \actions}$ is said to be \emph{proper} 
if and only if it defines the following three distinguished actions: $\kwT{assume}$, for extending the state with new information; and $\kwT{symb}$ and $\kwT{fresh}$, for generating new uninterpreted and interpreted symbols, respectively. Onward, we assume to work with proper state models. 


\smallskip
\begin{display}{\gil Semantic Domains for $\gstate =  \tup{\sset{\gstate}, \gval, \actions}$}
\begin{tabular}{@{~\quad}lr@{\ \ }c@{\ \ }l@{\qquad\quad}l}
   Call stacks: & $\cs \in \css{\gstate} $ & $\defeq$ & $ \tup{\f} \mid \tup{\f, \x, \sto, i} \cons \cs$ & where $\f \in \fids$, $\x \in \xs$, $\sto : \xs \pmap \gval$, $i \in \mathbb{N}$\\
   Configurations: & $\cf \in \cfs{\gstate}$    & $\defeq$ & $\tup{\prog, \st, \cs, i}$     & where $\prog \in \progs{\actions}$, $\st \in \sset{\gstate}$, $\cs \in \css{\gstate}$, $i \in \mathbb{N}$ \\
   Outcomes: & $\outcome \in \outcomes$  & $\defeq$ &  $\cont \mid \onormal{\gv} \mid \ofail{\gv}$ & where $\gv \in \gval$
 \end{tabular}
\end{display}
\smallskip


The \gil semantics is defined in Figure~\ref{fig:sem}. On each procedure call, it keeps track of the execution context of the caller, so that control can correctly be returned once the execution of the callee finishes. We achieve this by using \emph{call stacks}, $\cs \in \css{\gstate}$, which are \emph{non-empty} lists of stack frames. 
A top-level stack frame, $\tup{\f}$, only contains the identifier of the procedure that started the execution. 
An inner stack frame, $\tup{\f, \x, \sto, i}$, contains: 
\etag{1}~the identifier $\f$ of the procedure being executed;
\etag{2}~the variable $\x$ to which the return value of $\f$ will be assigned; 
\etag{3}~the store~$\sto$ of the caller of $\f$; and 
\etag{4}~the index $i$ to which control is transferred when the execution of $\f$ terminates. Additionally, we denote the argument of a function $\f$ by $\f.\kwT{arg}$.
 
\begin{figure}[!h]
\vspace*{-0.3cm}
{\footnotesize
\begin{mathpar}
\inferrule[\textsc{Assignment}]
  {
    \cmd(\prog, \cs, i) = \x := \e
    \and
    \ee{\st}{\e} = \gv
  }{\semtrans{\st, \cs, i}{\st.\kwT{setVar}(\x, \gv), \cs, i{+}1}{}{\prog}{}}  	 
 \and
\inferrule[\textsc{Action}]
  {
    \cmd(\prog, \cs, i) = \x := \act(\e)
    \quad
    \ee{\st}{\e} = \gv
    \quad
     \action{\st}{\act}{\gv}{(\st', \gv')}
  }{\semtrans{\st, \cs, i}{\st'.\kwT{setVar}(\x, \gv'), \cs, i{+}1}{}{\prog}{}}  	 
\\
\inferrule[\textsc{IfGoto - True}]
  {
    \cmd(\prog, \cs, i) = \ifgoto{\e}{j}
    \\\\
    \ee{\st}{\e} = \gv
    \\\\
    \actionT{\st}{assume}{\gv}{(\st', -)}
  }{\semtrans{\st, \cs, i}{\st', \cs, j}{}{\prog}{}}	 
\and
\inferrule[\textsc{IfGoto - False}]
  {
    \cmd(\prog, \cs, i) = \ifgoto{\e}{j}
    \\\\
    \ee{\st}{\lnot\e} = \gv
    \\\\
    \actionT{\st}{assume}{\gv}{(\st', -)}
  }{\semtrans{\st, \cs, i}{\st', \cs, i+1}{}{\prog}{}}  	 
\and
\vspace*{-0.1cm}
\inferrule[\textsc{Symb}]
  {
      \cmd(\prog, \cs, i) = \x := \symb{j}
      \\\\ 
      \actionT{\st}{symb}{j}{(\st', \gv)}
  }{
    \semtrans{\st, \cs, i}{\st'.\kwT{setVar}(\x, \gv), \cs, i{+}1}{}{\prog}{}
  }
 \and 
 \inferrule[\textsc{Fresh}]
  {
     \cmd(\prog, \cs, i) = \x := \fresh{j}
      \\\\ 
    \actionT{\st}{fresh}{j}{(\st', \gv)}
  }{
    \semtrans{\st, \cs, i}{\st'.\kwT{setVar}(\x, \gv), \cs, i{+}1}{}{\prog}{}
  }
  \and
\inferrule[\textsc{Call}]
  {
    \cmd(\prog, \cs, i) = \e(\e')
    \and
    \ee{\st}{\e} = \f
    \and
    \ee{\st}{\e'} = \gv
    \\\\
    \cs' = \tup{\f, \x, \st.\kwT{getStore}, i+1} \cons \cs
  }{\semtrans{\st, \cs, i}{\st.\kwT{setStore}(\btup{\f.\kwT{arg} \mapsto \gv}), \cs', 0}{}{\prog}{}}  	
\and
\inferrule[\textsc{Return}]
  {
     \cmd(\prog, \cs, i) = \return{\e}
      \and
    \ee{\st}{\e} = \gv   
     \\\\
    \cs = \tup{-, \x, \sto, j} \cons \cs'
    \and 
    \st' = \st.\kwT{setStore}(\sto)
  }{
    \semtrans{\st, \cs, i}{\st'.\kwT{setVar}(\x, \gv), \cs', j}{}{\prog}{}
  }
  \qquad
 \inferrule[\textsc{Top Return}]
  {
     \cmd(\prog, \cs, i) = \return{\e}
      \\\\
    \ee{\st}{e} = \gv
  }{
    \semtrans{\st, \tup{\fid}, i}{\st, \tup{\fid}, i}{}{\prog}{\onormal{\gv}}
  } 
  \qquad
 \inferrule[\textsc{Fail}]
  {
     \cmd(\prog, \cs, i) = \fail{e}
      \\\\
    \ee{\st}{\e} = \gv
  }{
    \semtrans{\st, \cs, i}{\st, \cs, i}{}{\prog}{\ofail{\gv}}
  } 
  %
\end{mathpar}}
%
%
\vspace*{-0.35cm}
\caption{Semantics of \gil: $\fullsemtrans{\st, \cs, i}{\st', \cs', j}{}{\prog}{\outcome'}{\cont}$}
\label{fig:sem}
\vspace{-0.3cm}
\end{figure}

To capture the flow of the execution, we use outcomes, $\outcome \in \outcomes$. \gil has three possible outcomes: 
\etag{1} continuation, $\cont$, signifying that the execution should 
proceed; 
\etag{2} return, $\onormal{\gv}$, signifying that there was a top-level return with value $\gvl$; 
and 
\etag{3} error, $\ofail{\gv}$, signifying that the execution \emph{failed} with value $\gvl$. In the rules, we elide the continuation outcome whenever it is clear from the context. 

Semantic transitions for \gil commands are of the form $\fullsemtrans{\st, \cs, i}{\st', \cs', j}{\gstate}{\prog}{\outcome}{\cont}$, 
meaning that, given a program $\prog$, the evaluation of the $i$-th command of the top procedure of the call stack $\cs$ in the state $\st$ generates the state~$\st'$, call stack~$\cs'$, and outcome~$\outcome$, and the next command to be evaluated is the $j$-th command of the top procedure of~$\cs'$. 

\begin{wrapfigure}{R}{0.40\textwidth}
\vspace*{-0.45cm}
\begin{minted}[fontsize=\scriptsize]{ocaml}
module type State = sig
  type t  (** Type of Gil states  *)
  type vt (** Type of Gil values  *)
  type a  (** Type of Gil actions *)

  val init : t
  val setVar : t -> Var.t -> vt -> t
  val setStore : t -> (Var.t, vt) Map.t -> t
  val store : t -> (Var.t, vt) Map.t
  val ee : t -> Expr.t -> vt
  val ea : a -> t -> vt -> (t * vt) list
  val assume : t -> vt -> (t * vt) list
  val fresh : t -> vt -> (t * vt) list
  val symb : t -> vt -> (t * vt) list
  ...
end

module type Allocator = sig
  type t   (** Type of allocation records   *)
  type vts (** Type of unintepreted symbols *)
  type vtf (** Type of interpeted symbols   *)

  val alloc_s : t -> int -> t * vts
  val alloc_f : t -> int -> t * vtf
end
\end{minted}
\vspace*{-0.3cm}
\caption{OCaml State/Allocator Signature}\label{fig:ocaml:state:sig}
\vspace{-0.3cm}
\end{wrapfigure}

\myparagraph{Implementation}
In the implementation, procedure calls and actions may have multiple parameters. Further, we have several additional commands: the \emph{unconditional goto}, $\goto{i}$; \emph{procedure application}, $\x := \apply(\e, \e)$, for modelling functions which take a variable number of arguments; the \emph{external procedure call}, $\x := \extern \e(\lst\e)$, for modelling language features that step out of the program, such as the \texttt{eval} command of JavaScript or system calls in C;\emph{argument collection}, $\arguments$, which returns a \gil list containing the arguments with which the current procedure was called; and the \emph{phi-node} command, $\x := \phinode{\lst{x : \lst{\x}}}$, which allows \gil programs to be written in Single-Static-Assignment (SSA) style~\cite{SSA}. These commands can all be compiled to the \gil of the paper, with the exception of the external procedure call. We do not provide meta-theoretical guarantees for \gil programs that use external procedure calls.

The general \gil interpreter follows the \gil semantics given in Figure~\ref{fig:sem} and is implemented as an OCaml functor parameterised by an OCaml module with type \texttt{State}.
The \texttt{State} module type, whose signature is partially given in Figure~\ref{fig:ocaml:state:sig}, follows our formal definition of state 
models as per Definition~\ref{def:state:module}. In addition, it contains functions related to, for example, state and value simplification, as well as interaction with the first-order solver.

\subsubsection{\gil Allocation}
\label{subsec:alloc}
Fresh value generation is a common source of technical clutter often 
omitted or hand-waved in the formal presentation of program analyses.
\gillian relieves the user of the framework 
from needing to reason about this issue by
having built-in fresh-value \emph{allocators}, inspired by the work of~\citet{banerjee:csf:2002}.

\begin{definition}[Allocator]
\label{def:am}
An allocator $\allocator \in \allocators$ is a pair, $\tup{\sset{\allocator}, \gval}$, consisting of: 
\etag{1} a set $\sset{\allocator} \ni \arec$ of allocation records\footnote{Intuitively, an allocation record maintains information about already allocated values; this approach is complementary to the free set approach of~\citet{Raza2009Footprints}, where information is maintained about values that can still be allocated.}; and
\etag{2} the set $\gval$ of values that may be allocated.
It exposes the function $\allocf : \sset{\allocator} \tmap \mathbb{N} \tmap \power{\gval} \pmap \sset{\allocator} \times \gval$,
which satisfies the well-formedness constraint
$(\arec', \gvl) = \allocf(\arec, j, Y) \implies \gv \in Y$,
and is pretty printed as $\alloc{\arec}{j}{\arec', \gvl}{Y}$.
\end{definition}

Informally, $\allocf(\arec, j, Y)$ generates a fresh value~$\gv$ taken from 
$Y \subseteq \gval$, associates it with the \emph{allocation site}\footnote{An allocation site $j$ is the program point that is associated with either the $\symb{j}$ or the $\fresh{j}$ command.} uniquely identified by the natural number $j$, and returns it together with a updated allocation record. We discuss allocators and their properties in more detail in \S\ref{subsec:symb:soundness}, in the context of our soundness results.

\myparagraph{Implementation} We implement allocators as shown in Figure~\ref{fig:ocaml:state:sig}. As we do not have the expressive power to pass arbitrary subsets in OCaml, we instead require two separate types, $\mathtt{vts}$/$\mathtt{vtf}$ used generating fresh uninterpreted/interpreted symbols, each generated 
by a dedicated allocation function, \texttt{alloc\_s}/\texttt{alloc\_f}. 
We show how allocators can be used in practice in \S\ref{sec:concverssym}.

\subsection{\while: Syntax, Actions, and Compilation to \gil}\label{subsec:while:compiler}
We demonstrate how to instantiate \gillian using a simple \while language with static objects.
Its syntax includes: the variable assignment; the skip command; sequencing; the if-then-else conditional; the while loop; the static function call; the return statement; assume and assert statements for driving the symbolic analysis; 
and statements for operating on static objects: object creation, 
property lookup, property mutation, and object disposal. For simplicity, we assume that the semantics of expressions and the variable store are the same for \while and \gil.

\medskip
\begin{display}{The Syntax of \while}
\begin{tabular}{l}
   $\cw \in \cws \, \defeq \,  \x := \e \mid \wskip \mid \cw_1; \cw_2 \mid \wif{\e}{\cw_1}{\cw_2} \mid \wwhile{\e}{\cw} \mid \x := \fid(\e) \mid \wreturn{\e} \mid $  \\ 
   $\hspace*{2.13cm}\wassume{\e} \mid \wassert{\e} \mid \x := \wobj{\prop_i: \e_i\mid_{i=1}^n} \mid \wlookup{\x}{\e}{\prop} \mid \wmutate{\e}{\prop}{\e'} \mid \wdispose{\e}$ 
 \end{tabular}
\end{display}
\medskip

In order to compile \while to \gil, we first have to pick a set of actions $\actionswhile$ for acting on \while states. As we have four operations on objects---allocation, lookup, mutation, and disposal---assigning an action to each of those would be a reasonable first attempt. However, since \gillian has a built-in allocator for 
generating fresh locations, we do not need a separate action for object allocation, arriving at the set of actions $\actionswhile = \lbrace \lookupAction, \mutateAction, \disposeAction \rbrace$. 

\begin{figure}[!t]
{\footnotesize
\begin{mathpar}
\inferrule[\textsc{Assignment}]
  {}{
     {\begin{array}[t]{l}
      \compWhile(\x := \e, \wpc) \semeq \\
      {\begin{array}{l}
        \wpc:  \x := \e  \\ 
         \nextpc{\wpc + 1}
       \end{array}} 
      \end{array}}
  }
 \quad 
%
 %
 %
\quad 
\inferrule[\textsc{Assume}]
  {}{
     {\begin{array}[t]{l}
      \compWhile(\wassume{\e}, \wpc) \semeq \\
      {\begin{array}{l}
         \wpc: \ifgoto{\e}{(\wpc + 2)} \\ 
          \wpc+1: \vanish \\
         \nextpc{\wpc + 2}
       \end{array}} 
      \end{array}}
  }
\quad 
\inferrule[\textsc{Assert}]
  {}{
     {\begin{array}[t]{l}
      \compWhile(\wassert{\e}, \wpc) \semeq \\
      {\begin{array}{l}
         \wpc: \ifgoto{\e}{(\wpc + 2)} \\ 
          \wpc+1: \fail{\e} \\
         \nextpc{\wpc + 2}
       \end{array}} 
      \end{array}}
  }
 \and 
 \inferrule[\textsc{Lookup}]
  {}{
     {\begin{array}[t]{l}
      \compWhile(\wlookup{\x}{\e}{\prop}, \wpc) \semeq \\
      {\begin{array}{l}
          \wpc: \x := \lookupAction(\litlst{\e, \prop}) \\
         \nextpc{\wpc + 1}
       \end{array}} 
      \end{array}}
  }
   \and 
 %
%
  %
 \end{mathpar}}
 \vspace*{-0.35cm}
\caption{\while-to-GIL Compiler: $\compWhile : \cws \tmap \mathbb{N} \tmap \listtype{\cmds{\actionswhile}} \times \mathbb{N}$ (excerpt)}
 \vspace*{-0.3cm}
\label{while:compiler}
\end{figure}

A part of the \while-to-\gil compiler is given in Figure~\ref{while:compiler} (cf.~Appendix~\ref{app:s2}). It is modelled as a function 
$\compWhile : \cws \tmap \mathbb{N} \tmap \listtype{\cmds{\actionswhile}} \times \mathbb{N}$, mapping a 
\while statement $\cw \in \cws$ and a natural number $\wpc$ (read: \emph{program counter}) 
to a sequence of \gil commands and the next available program counter, $\npc$ (denoted in Figure~\ref{while:compiler} using the notation $\nextpc{\npc}$). 
For instance, if $\compWhile(\cw, \wpc) = (\lst{\cm}, \npc)$, then the while 
statement $\cw$ compiles to the sequence of \gil commands given by $\lst{\cm}$ and 
that the commands in $\lst{\cm}$ are labelled with indexes $\wpc$ to $\npc-1$. 

The compilation rules are straightforward; we explain the ones given in Figure~\ref{while:compiler}.
We compile the \while assignment to a \gil assignment, shallowly embedding \while variables to \gil variables.
The assume statement, $\wassume{\e}$, compiles to a goto statement, 
$\ifgoto{\e}{(\wpc + 2)}$, which branches on the value of the expression to be assumed, followed by a silent cutting of the branch in which it does not hold by using the \gil command $\vanish$. 
The assert statement, $\wassert{\e}$, is compiled to the same goto statement that branches on $\e$, but this time, if the branch in which $\e$ does not hold is reached, the execution will terminate with error by using the \gil command $\kwT{fail}$. 
Finally, the lookup of \while is compiled as a call to the corresponding action, $\lookupAction$, whose parameter is a \gil list containing the expression denoting the address of the object, $\e$, and the looked-up property, $\prop$.

\subsection{Concrete and Symbolic States}
\label{sec:concverssym} 

Reasoning about programs can, at a high level, be separated into reasoning about the \emph{variable store} and about the \emph{memory model} of the programming language in question. \gillian simplifies this process by providing built-in reasoning about the variable store, leaving to the user only to take care of the memory model. In particular, it is possible to lift a given memory to a \gil state by coupling said memory with an appropriate variable store and allocator. In this section, we illustrate this lifting for \emph{concrete} and \emph{symbolic} memories, obtaining concrete and symbolic states. 

Concrete memories store concrete values, $\vl \in \vals$. Symbolic memories store logical expressions, $\sexp \in \sexps$, generated by the grammar $\sexp \in \sexps \ \defeq \ \vl \mid \lx \in \lxs \mid \unop{\sexp} \mid \binop{\sexp_1}{\sexp_2}$, 
where $\lx$ ranges over a set of \emph{logical variables}, $\lxs$. 
The formal definitions of these two memory models are as follows.
 


\begin{definition}[Concrete Memory Model]\label{def:cmem}
A concrete memory model $\cmemory \in \cmemories$ 
is a pair $\tup{\sset{\cmemory}, \actions}$,
consisting of a set of concrete memories, $\sset{\cmemory} \ni \cmem$, and a set of actions $\actions \ni \act$. 
A concrete memory model additionally defines a function $\cea$, for concrete action execution  
on memories: 
$$\cea : \actions \tmap \sset{\cmemory} \tmap \vals \pmap \sset{\cmemory} \times \vals
\qquad \qquad \qquad ({\cea}(\act, \cmem, \vl) \equiv_{pp} \cmem.\act(\vl))$$
\end{definition}

\begin{definition}[Symbolic Memory Model]\label{def:smem}
A symbolic memory model $\smemory \in \smemories$ 
is a pair $\tup{\sset{\smemory}, \actions}$,
consisting of a set of symbolic memories, $\sset{\smemory} \ni \smem$, and a set of actions $\actions \ni \act$. 
A symbolic memory model additionally defines a function $\sea$ for symbolic action execution
on memories: 
 $$\sea : \actions\hspace*{0.02cm}{\tmap}\hspace*{0.02cm}\sset{\smemory}\hspace*{0.02cm}{\tmap}\hspace*{0.02cm}\sexps\hspace*{0.02cm}{\tmap}\hspace*{0.02cm}\pcs\hspace*{0.02cm}{\pmap}\hspace*{0.02cm}\power{\sset{\smemory} \times \sexps \times \pcs} 
  \quad ((\smem', \sexp', \pc') \in {\sea}(\act, \smem, \sexp, \pc) \equiv_{pp} \action{\smem}{\act}{\sexp, \pc}{(\smem', \sexp', \pc')})$$
where $\pc \in \pcs \subset \sexps$ denotes a boolean logical expression.
\end{definition}

%
%

%
%
%

From a concrete memory, $\cmem \in \sset{\cmemory}$, we construct a concrete state $\st$ by coupling
$\cmem$ with a concrete store, $\sto : \xs \pmap \vals$, and a concrete allocation record, $\arec \in \sset{\allocator}$. 
Analogously, from a symbolic memory, $\smem \in \sset{\smemory}$, we construct a symbolic state $\sst$ by 
coupling $\smem$ with a symbolic store, $\ssto : \xs \pmap \sexps$, and a symbolic allocation record, 
$\sarec \in \sset{\hat\allocator}$. Symbolic states also include a boolean logical expression $\pc \in \pcs$, referred to as the \emph{path condition} of $\sst$. 
Path conditions~\cite{symb:exec:survey} bookeep the constraints on the symbolic variables that led the execution to the 
current symbolic state. 
We formally describe the liftings from concrete and symbolic memories to the appropriate states below, with $\actionsZ = \set{\kwT{assume}, \kwT{fresh}, \kwT{symb}}$.

\begin{definition}[Concrete State Constructor (\cstateConstr)]\label{def:concrete:lifting}
Given an allocator $\allocator = \tup{\sset{\allocator}, \vals}$, the 
concrete state constructor $\cstateConstr : \cmemories \tmap \gstates$ is defined as  
$\cstateConstr(\tup{\sset{\cmemory}, \actions}) \semeq \tup{\sset{\gstate}, \vals, \actions  \dunion \actionsZ}$,
where: 

{\small
\begin{tabular}{lll}
   \multicolumn{3}{l}{\mybullet $\sset{\gstate} = \sset{\cmemory} \times \pfunset{\xs}{\vals} \times \sset{\allocator}$} \\ 
  \mybullet $\kwT{setVar}(\tup{\cmem, \sto, \arec}, \x, \vl)$ 
       & $\semeq$ & 
       $\tup{\cmem, \sto[\x \mapsto \vl], \arec}$ \\ 
  \mybullet $\kwT{setStore}(\tup{\cmem, \_, \arec}, \sto)$
     & $\semeq$ & 
     $\tup{\cmem, \sto, \arec}$ \\ 
  \mybullet $\kwT{getStore}(\tup{\_, \sto, \_})$
    & $\semeq$ & 
    $\sto$ \\
  \mybullet $\kwT{ee}(\tup{\_, \sto, \_}, \e)$
    & $\semeq$ & 
    $\ceval{\e}{\sto}$ \\
 \mybullet $\kwT{ea}(\act, \tup{\cmem, \sto, \arec}, \vl)$
    & $\semeq$ & 
    $\lbrace (\tup{\cmem', \sto, \arec}, \vl') \mid (\cmem', \vl') = \cea(\act, \cmem, \vl) \rbrace$ \\
   \mybullet $\kwT{assume}(\st, \vl)$
    & $\semeq$ & 
    $\lbrace (\st, \vl) \mid \vl = \true \rbrace$ \\
    \mybullet $\kwT{symb}(\tup{\cmem, \sto, \arec}, j)$
    & $\semeq$ &
     $\lbrace (\tup{\cmem, \sto, \arec'}, \varsigma) \mid \alloc{\arec}{j}{\arec', \varsigma}{\locs} \rbrace $ \\
    \mybullet $\kwT{fresh}(\tup{\cmem, \sto, \arec}, j)$
    & $\semeq$ &
     $\lbrace (\tup{\smem, \ssto, \arec'}, \vl) \mid \alloc{\arec}{j}{\arec', \vl}{\vals} \rbrace $ \\
    
\end{tabular}}
\end{definition}

\begin{definition}[Symbolic State Constructor ($\sstateConstr$)]\label{def:symbolic:lifting}
Given an allocator $\hat\allocator = \tup{\sset{\hat\allocator}, \sexps}$, the 
symbolic state constructor $\sstateConstr : \smemories \tmap \gstates$ is defined as 
$\sstateConstr(\tup{\sset{\smemory}, \actions}) \semeq \tup{\gsstate, \sexps, \actions  \dunion \actionsZ}$,
where: 

{\small
\begin{tabular}{lll}
   \multicolumn{3}{l}{\mybullet $\sset\gsstate = \sset{\smemory} \times \pfunset{\xs}{\sexps} \times \sset{\hat\allocator} \times \pcs$} \\ 
  \mybullet $\kwT{setVar}(\tup{\smem, \ssto, \sarec, \pc}, \x, \sexp)$ 
       & $\semeq$ & 
       $\tup{\smem, \ssto[\x \mapsto \sexp], \sarec, \pc}$ \\ 
  \mybullet $\kwT{setStore}(\tup{\smem, \_, \sarec, \pc}, \ssto)$
     & $\semeq$ & 
     $\tup{\smem, \ssto, \sarec, \pc}$ \\ 
  \mybullet $\kwT{getStore}(\tup{\_, \ssto, \_, \_})$
    & $\semeq$ & 
    $\ssto$ \\
  \mybullet $\kwT{ee}(\tup{\_, \ssto, \_, \_}, \e)$
    & $\semeq$ & 
    $\ssto(\e)$ \\
 \mybullet $\kwT{ea}(\act, \tup{\smem, \ssto, \sarec, \pc}, \sexp)$
    & $\semeq$ & 
    $\lbrace (\tup{\smem', \ssto, \sarec, \pc \, \wedge \pc'}, \sexp') \mid (\smem', \sexp', \pc') \in \sea(\act, \smem, \sexp) \rbrace$ \\
   \mybullet $\kwT{assume}(\tup{\smem, \ssto, \sarec, \pc}, \pc')$
    & $\semeq$ & 
    $\lbrace (\tup{\smem, \ssto, \sarec, \pc \gand \pc'}, \true) \mid \pc \gand \pc' \, \text{\sat} \rbrace$ \\
  \mybullet $\kwT{symb}(\tup{\smem, \ssto, \sarec, \pc}, j)$
    & $\semeq$ &
    $\lbrace (\tup{\cmem, \sto, \sarec', \pc}, \varsigma) \mid \alloc{\sarec}{j}{\sarec', \varsigma}{\locs} \rbrace $ \\
  \mybullet $\kwT{fresh}(\tup{\smem, \ssto, \sarec, \pc}, j)$
    & $\semeq$ &
    $\lbrace (\tup{\cmem, \sto, \sarec', \pc}, \lx) \mid \alloc{\sarec}{j}{\sarec', \lx}{\lxs} \rbrace $ \\
  
  %
\end{tabular}}
\end{definition}

The concrete/symbolic lifting constructs all of the functions that a state model exposes, with the help of the action execution function of the parameter concrete/symbolic memories and the $\kwT{alloc}$ function of the parameter concrete/symbolic allocator. 
In both cases, the construction of $\kwT{setVar}$, $\kwT{setStore}$, and 
 $\kwT{getStore}$ is straightforward. The remaining cases are described below.
 
\myparagraph {\prooflab{EvalExpr}} 
In the concrete case, expression evaluation is performed concretely (we use $\ceval{\e}{\sto}$ to denote the
standard expression evaluation of $\e$ with respect to $\sto$). In the symbolic case, it amounts to substituting all the program variables in $\e$ with their associated logical 
expressions given by the store (we denote this substitution by $\ssto(\e)$). In the implementation, \gillian's first-order solver applies a number of algebraic identities to simplify the expression resulting from $\ssto(\e)$. It is also in charge of discharging satisfiability/entailment questions (e.g.~$\pc \vdash \sexp = \sexp'$), by encoding them into the Z3 solver of~\citet{z3}.

\myparagraph{\prooflab{Action}} Action execution on states amounts to calling action execution on the parameter memory.
As symbolic actions, unlike concrete actions, may branch, they additionally 
generate a logical expression, $\pc'$, describing the conditions under which the chosen 
branch is taken. Hence, the path condition of the obtained state is a conjunction of $\pc'$ with the path 
condition $\pc$ of the original state. 

\myparagraph{\prooflab{Assume}} The function $\kwT{assume}(\st, \gv)$ extends the state $\st$ with information that the value $\gv$ \emph{holds}. 
In the concrete case, $\kwT{assume}(\st, \bool)$ returns the singleton set containing the original state when $\bool = \true$ and the empty set otherwise.
In the symbolic case, $\kwT{assume}(\sst, \pc')$ returns $\sst$ with its path condition strengthened with $\pc'$ if this new path condition is satisfiable, and the empty set otherwise. 

\myparagraph{\prooflab{Symb/Fresh}} The functions $\kwT{symb}$ and $\kwT{fresh}$ generate symbols 
using the parameter allocator. We use the arrow parameter of the allocator to indicate the set from which to pick 
the freshly generated value: $\kwT{symb}$ picks an uninterpreted and $\kwT{fresh}$ picks an interpreted symbol. As symbolic values are logical expressions, it makes sense to pick a fresh logical variable when generating a fresh interpreted symbol and then later impose constraints on it via the $\kwT{assume}$ function.


\begin{wrapfigure}{r}{0.45\textwidth}
\begin{minted}[fontsize=\scriptsize]{ocaml}
module type SMemory = sig
  type t            (** Type of symbolic memories *)
  type a            (** Type of actions           *)
  type vt = LExpr.t (** Type of symbolic values   *)

  val ea : a -> t -> vt -> vt -> (t * vt * vt) list
end
\end{minted}
\medskip
\begin{minted}[fontsize=\scriptsize]{ocaml}
module SState
  (SMem : SMemory)
  (Alloc : Allocator with type vtus = Loc.t
                      and type vtis = LVar.t) :
  (State with type a = SMem.a) = struct
    type vt = LExpr.t
    type mt = SMem.t
    type at = Alloc.t
    type t = mt * vt Store.t * at * LExpr.t
    type a = SMem.a
    ...
end
\end{minted}
\vspace*{-0.25cm}
\caption{OCaml Signatures: Symbolic Memories and States}\label{fig:make:state}
\vspace*{-0.5cm}
\end{wrapfigure}

\myparagraph{Implementation}
The concrete and symbolic state constructors are implemented as OCaml functors,
respectively parameterised by OCaml modules with types \texttt{CMemory} and  \texttt{SMemory} (cf.~Figure~\ref{fig:make:state}). 
Both functors are additionally parameterised by a module with type \texttt{Allocator}
for the generation of fresh values. 
The \texttt{CMemory} and \texttt{SMemory} module types precisely follow our formal characterisation of 
concrete and symbolic state models as per Definitions~\ref{def:cmem} and~\ref{def:smem}. 
Furthermore, the functors \texttt{CState} and \texttt{SState}  also follow the concrete and symbolic 
liftings described in Definitions~\ref{def:concrete:lifting} and~\ref{def:symbolic:lifting}.

\subsection{\while: Concrete and Symbolic Memories}
\label{subsec:while:csm}

We show how to instantiate \gillian to obtain a concrete/symbolic interpreter
for \while by combining the \while-to-\gil compiler of \S\ref{subsec:while:compiler} with the concrete/symbolic \while memory model. 

The first step towards defining a memory model is to pick the carrier set of the model, 
in this case: the set of \while memories. 
We define a concrete \while memory as a partial mapping from symbols (corresponding to object locations) and strings 
(corresponding to property names) to values; formally: $\cmem \in \wcmems: \locs \times \strs \pmap \vals$. 
Analogously, we define a symbolic memory to be a partial mapping from 
logical expressions and strings to logical expressions; formally: $\smem \in \wsmems: \sexps \times \strs \pmap \sexps$. 
Property names are not lifted to logical expressions in the symbolic case, since the \while statements
that act on object properties do not allow the property name to be resolved dynamically. 

\begin{figure}[!b]
\vspace*{-0.1cm}
\begin{minipage}{0.40\textwidth}
{\small
\begin{mathpar}
\inferrule[C-Lookup]
  {\cmem = \_ \dunion \wcell{\loc}{\prop}{\vl}}
  {\action{\cmem}{\lookupAction}{\litlst{\loc, \prop}}{(\cmem, \vl)}} 
 \\
\inferrule[C-Mutate-Present]
  {
    \cmem = \cmem' \dunion \wcell{\loc}{\prop}{\_}
    \quad
    \cmem'' = \cmem' \dunion \wcell{\loc}{\prop}{\vl}
  }{\action{\cmem}{\mutateAction}{\litlst{\loc, \prop, \vl}}{(\cmem'', \vl)}} 
 \\
\inferrule[C-Mutate-Absent]
  {
    (\loc, \prop) \notin \mathrm{dom}(\cmem)
    \and
    \cmem' = \cmem \dunion \wcell{\loc}{\prop}{\vl}
  }{\action{\cmem}{\mutateAction}{\litlst{\loc, \prop, \vl}}{(\cmem', \vl)}} 
  \\
\inferrule[C-Dispose]
  {
    \memproj{\cmem}{\loc} = (\_, \cmem')
  }{\action{\cmem}{\disposeAction}{\loc}{(\cmem', \true)}} 
\end{mathpar}}
\end{minipage}
\begin{minipage}{0.59\textwidth}
{\small
\begin{mathpar}
\inferrule[S-Lookup]
  {
     \pc \vdash \sexp = \sexp' 
     \and
     \smem = \_ \dunion \wcell{\sexp'}{\prop}{\sexp_v}
  }{\action{\smem}{\lookupAction}{\litlst{\sexp, \prop}, \pc}{\litset{(\smem, \sexp_v, \true)}}} 
 \\
\inferrule[S-Mutate-Present]
  {
  	\pc \vdash \sexp = \sexp'' 
	\quad
    \smem = \smem' \dunion \wcell{\sexp''}{\prop}{\_}
    \quad
    \smem'' = \smem' \dunion \wcell{\sexp''}{\prop}{\sexp'}
  }{\action{\smem}{\mutateAction}{\litlst{\sexp, \prop, \sexp'}, \pc}{\litset{(\smem'', \true, \true)}}} 
 \\
\inferrule[S-Mutate-Absent]
  {
  	\memproj{\smem}{\sexp, \prop, \pc} = \emptyset
	\quad
    \smem' = \smem \dunion \wcell{\sexp}{\prop}{\sexp'}
  }{\action{\smem}{\mutateAction}{\litlst{\sexp, \prop, \sexp'}, \pc}{\litset{(\smem', \true, \true)}}} 
  \\
\inferrule[S-Dispose]
  {
    \memproj{\smem}{\sexp, \pc} = (\_, \smem')
  }{\action{\smem}{\disposeAction}{\sexp, \pc}{\litset{(\smem', \true, \true)}}}
\end{mathpar}}
\end{minipage}
\vspace{-0.3cm}
\caption{\while: Actions in Concrete and Symbolic Memories}\label{fig:while:memories}\vspace{-0.3cm}
\end{figure}

Having chosen the concrete and symbolic carrier sets, we can now define the functions $\cea$ and~$\sea$, for acting 
on concrete and symbolic memories, respectively. 
In particular, these functions must support the actions 
$\actionswhile = \lbrace \lookupAction, \mutateAction, \disposeAction \rbrace$ 
introduced in \S\ref{subsec:while:compiler}.
The definition of the \while concrete/symbolic actions is given in Figure~\ref{fig:while:memories}
(concrete on the left and symbolic on the right), and is straightforward.
In the rules, we use several projection operators: $\memproj{\cmem}{\loc}$ splits the concrete memory $\cmem$ into a pair $(\cmem', \cmem'')$, where $\cmem'$ contains the memory cells at location $\loc$ and $\cmem''$ contains the others; $\memproj{\smem}{\sexp, \pc}$ splits the symbolic memory $\smem$ into a pair $(\smem', \smem'')$, where $\smem'$ contains the memory cells at location corresponding to $\sexp$ under $\pc$  and $\smem''$ contains the others; and $\memproj{\smem}{\sexp, \prop, \pc}$ returns a set of possible locations in $\smem$ corresponding to $\sexp$ under $\pc$ that have property $\prop$.
We describe the rules \prooflab{C-Dispose} and \prooflab{S-Dispose} in detail.

\myparagraph {\prooflab{C-Dispose}} To dispose of the object at location $\loc$, we split the memory $\cmem$ using $\memproj{\cmem}{\loc}$ and  return the part of the memory that does not contain the cells at location~$\loc$, and the value $\true$. 
The value $\true$ is returned to comply with the expected type of 
$\cea$; it is not used by the compiled code. 

\myparagraph {\prooflab{S-Dispose}} Analogously to \prooflab{C-Dispose}, we split the symbolic memory $\smem$ using $\memproj{\smem}{\sexp, \pc}$ and return the part of the memory that does not contain the cells corresponding to $\sexp$ under $\pc$, the value $\true$, and the path condition $\true$. 
The returned path condition indicates that the action does not branch
and that, therefore, the path condition of the corresponding state will effectively not be updated.

%

\section{Parametric Soundness}\label{sec:soundness}

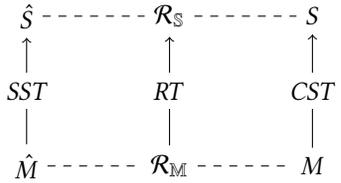
\begin{wrapfigure}{R}{0.35\textwidth}
\vspace*{-0.4cm}
\begin{tikzpicture}
\node (SS) {$\hat\gstate$};
\node (SM) [below= 1.4cm of SS] {$\smemory$};
\node (RM) [right= 1.25cm of SM] {$\memrelset$};
\node (CM) [right= 1.25cm of RM] {$\cmemory$};
\node (CS) at (CM |- SS) {$\gstate$};
\node (RS) at (RM |- SS) {$\strelset$};
\draw[dashed] (SS) to (RS);
\draw[dashed] (RS) to (CS);
\draw[dashed] (SM) to (RM);
\draw[dashed] (RM) to (CM);
\draw[->] (SM) to node [fill=white] {$\sstateConstr$} (SS);
\draw[->] (RM) to node [fill=white] {$RT$} (RS) ;
\draw[->] (CM) to node [fill=white] {$\cstateConstr$} (CS);
\end{tikzpicture}
\vspace*{-0.3cm}
\caption{Proof Infrastructure}\label{fig:proof:infrastructure}
\vspace*{-0.4cm}
\end{wrapfigure}

Proving the soundness of symbolic analyses is a time-consuming task 
that often requires a considerable number of auxiliary lemmas and definitions.
The complexity of such proofs becomes unwieldy
as we move towards real-world programming languages with multiple program constructs and intricate runtime environments, which tends to detract from mathematical rigour in favour of less time consuming, but also less trustworthy, informal arguments. 
\gillian streamlines the development of soundness 
proofs for the instantiations of the framework, by focussing the user's proof effort \emph{only} on the target language memory and the actions that it exposes. 

We propose a proof infrastructure, illustrated in Figure~\ref{fig:proof:infrastructure}, consisting of: 
\etag{1} a class of soundness relations between state models, $\strelset$, which are preserved by the semantics of \gil;  
\etag{2} a class of soundness relations between memory models, $\memrelset$, which, when lifted 
             to states, yield relations in~$\strelset$; and 
\etag{3} the lifting mechanism, $RT$.
With this infrastructure in place, proving the soundness of a given symbolic semantics in terms of a given 
concrete semantics amounts to proving that the soundness relation between the two corresponding 
memory models is in $\memrelset$.

The section is structured as follows: \S\ref{subsec:parametric:soundness} describes a class of soundness 
relations that are preserved by the semantics of \gil; \S\ref{subsec:symb:soundness} describes the mechanism 
for lifting a soundness relation between memories to a soundness relation between states; and \S\ref{subsec:while:soundness}
shows how to leverage the proposed proof infrastructure to prove the soundness of the \while 
symbolic analysis. 

%

\subsection{Parametric Soundness}\label{subsec:parametric:soundness}

The standard approach for defining soundness~\cite{cousot:popl:1977} can be coarsely described as in 
the diagram of Figure~\ref{fig:soundness:properties} (left), where: 
\etag{1} $\sst_1$ is an abstract state over-approximating a concrete state, $\st_1$;  
\etag{2} $\sst_2$ is the abstract state obtained by abstractly executing a given command 
on $\sst_1$; 
and \etag{3} $\st_2$ is the state obtained by concretely executing the same 
command on $\st_1$. 
In this setting, the abstract semantics is \emph{sound} with respect to 
the concrete one if $\sst_2$ also over-approximates~$\st_2$. 

In the context of abstract analyses that may branch,  
this elegant characterisation of soundness 
cannot describe what it means for a single abstract trace 
to be sound. 
For instance, Figure~\ref{fig:soundness:properties} (mid) shows a scenario with
two possible abstract transitions from the original abstract 
state. The only way to ensure that the final abstract state over-approximates 
the final concrete state is to \emph{merge} the two final abstract 
states,\footnote{The union of abstract states as in the standard collecting semantics~\cite{cousot:ai:survey}
is a form of merging.} forcing us to reason about all possible
abstract traces at the same time. 

We propose a characterisation of soundness that allows us to describe  
both what it means for a single abstract trace to be sound independently 
of all the others and also to recover the standard notion of 
soundness when given the set of all possible abstract traces. 
This property is inspired by work on symbolic execution~\cite{survey:ecse,survey:acm}, 
where one can talk about the soundness of single symbolic execution trace 
by strengthening the path condition of the initial symbolic state with the path condition of the final state.
This effectively filters out all of the initial concrete states for which 
the concrete execution diverges from the path of the given symbolic trace. 

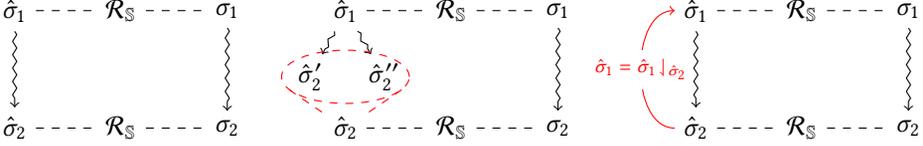
\begin{figure}[!h]
\vspace*{-0.5cm}
\begin{tikzpicture}
\node (SS1) {$\sst_1$};
\node (SS2) [below= 1cm of SS1] {$\sst_2$};
\node (RS2) [right= 0.8cm of SS2] {$\strelset$};
\node (CS2) [right= 0.8cm of RS2] {$\st_2$};
\node (CS1) at (CS2 |- SS1) {$\st_1$};
\node (RS1) at (RS2 |- SS1) {$\strelset$};
\draw[dashed] (SS1) to (RS1);
\draw[dashed] (RS1) to (CS1);
\draw[dashed] (SS2) to (RS2);
\draw[dashed] (RS2) to (CS2);
\draw[->, line join=round, decorate, decoration={zigzag, segment length=5, amplitude=.8,post=lineto, post length=2pt}] (SS1) to (SS2);
\draw[->, line join=round, decorate, decoration={zigzag, segment length=5, amplitude=.8,post=lineto, post length=2pt}] (CS1) to (CS2);
\end{tikzpicture}
\quad
\begin{tikzpicture}
\node (SS1) {$\sst_1$};
\node (SS2) [below= 1cm of SS1] {$\sst_2$};
\node (RS2) [right= 0.8cm of SS2] {$\strelset$};
\node (CS2) [right= 0.8cm of RS2] {$\st_2$};
\node (CS1) at (CS2 |- SS1) {$\st_1$};
\node (RS1) at (RS2 |- SS1) {$\strelset$};
\node (SS21) [below left= 0.3cm and -0.1cm of SS1] {$\sst_2'$};
\node (SS22) [below right= 0.3cm and -0.1cm of SS1] {$\sst_2''$};
\draw[dashed] (SS1) to (RS1);
\draw[dashed] (RS1) to (CS1);
\draw[dashed] (SS2) to (RS2);
\draw[dashed] (RS2) to (CS2);
\draw[->, line join=round, decorate, decoration={zigzag, segment length=5, amplitude=.8,post=lineto, post length=2pt}] (SS1) to (SS21);
\draw[->, line join=round, decorate, decoration={zigzag, segment length=5, amplitude=.8,post=lineto, post length=2pt}] (SS1) to (SS22);
\draw[->, line join=round, decorate, decoration={zigzag, segment length=5, amplitude=.8,post=lineto, post length=2pt}] (CS1) to (CS2);
\draw [dashed, color=red] (0,-0.87cm) ellipse (0.85cm and 0.35cm);
\draw[dashed, color=red] (-0.75,-1.04) to (SS2);
\draw[dashed, color=red] (0.75,-1.04) to (SS2);
\end{tikzpicture}
\begin{tikzpicture}
\node (SS1) {$\sst_1$};
\node (SS2) [below= 1cm of SS1] {$\sst_2$};
\node (RS2) [right= 0.8cm of SS2] {$\strelset$};
\node (CS2) [right= 0.8cm of RS2] {$\st_2$};
\node (CS1) at (CS2 |- SS1) {$\st_1$};
\node (RS1) at (RS2 |- SS1) {$\strelset$};
\draw[dashed] (SS1) to (RS1);
\draw[dashed] (RS1) to (CS1);
\draw[dashed] (SS2) to (RS2);
\draw[dashed] (RS2) to (CS2);

\path[every node/.style={font=\sffamily\scriptsize}]
	(SS2) edge[color=red, ->, bend left=90] node [midway,fill=white] {$\sst_1 = \rfun{\sst_1}{\sst_2}$} (SS1);
	
\draw[->, line join=round, decorate, decoration={zigzag, segment length=5, amplitude=.8,post=lineto, post length=2pt}] (SS1) to (SS2);
\draw[->, line join=round, decorate, decoration={zigzag, segment length=5, amplitude=.8,post=lineto, post length=2pt}] (CS1) to (CS2);
\end{tikzpicture}
 \vspace*{-0.35cm}
 \caption{Soundness Properties}\label{fig:soundness:properties}
 \vspace*{-0.4cm}
\end{figure}

\myparagraph{Restriction Operators}
As \gillian states are general, we cannot use path conditions as a device
to pinpoint the set of concrete traces that need to be modelled by 
a given abstract trace. 
Instead, we introduce \emph{restriction operators} on abstract states.
Informally, the restriction of an abstract state~$\sst_1$ with another abstract state $\sst_2$, written  $\rfun{\sst_1}{\sst_2}$, denotes the state $\sst_1$ strengthened with some information coming from $\sst_2$. In Figure~\ref{fig:soundness:properties} (right), we give further intuition on how we will use restriction: for example, if $\sst_1$ over-approximates $\st_1$ and it also holds that $\sst_1 = \rfun{\sst_1}{\sst_2}$, meaning that~$\sst_1$ does not learn anything from~$\sst_2$, then $\sst_2$ will also over-approximate $\st_2$.
%

We define restriction operators algebraically. 
A \emph{restriction} operator $\frestriction : X \tmap X \pmap X$  on a set $X$, 
written $\rfun{x_1}{x_2}$ for $\frestriction\!(x_1, x_2)$, is a binary associative function 
satisfying the following properties: 

\begin{minipage}{\textwidth}
\begin{mathpar}
  \inferrule[Idempotence]{}{\rfun{x}{x} = x}
\and
  \inferrule[Right Commutativity]{}{\rfun{(\rfun{x_1}{x_2})}{x_3} = \rfun{(\rfun{x_1}{x_3})}{x_2}}
\and
  \inferrule[Weakening]{\rfun{x_1}{\rfun{x_2}{x_3}} = x_1}{\rfun{x_1}{x_2} = x_1 \, \wedge \, \rfun{x_1}{\x_3} = x_1}
\end{mathpar}
\end{minipage}

\smallskip
\noindent meaning that self-restriction does not gain information, the order of applied restrictions does not influence the accumulated information gain, and that if $x_1$ cannot gain information from the combined knowledge of $x_2$ and $x_3$, then it cannot gain information from either $x_2$ or $x_3$. It is easily verifiable that every restriction operator  $\frestriction : X \tmap X \pmap X$ induces a pre-order $(X, \rleq)$,
given by $x_1 \rleq x_2 \iff \rfun{x_1}{x_2} = x_1$. 

A restriction operator on states $\frestriction : \sset{\gstate} \tmap \sset{\gstate} \pmap \sset{\gstate}$ is said to be 
\emph{preserved} by a state model $\gstate = \tup{\sset{\gstate}, \gval, \actions}$ 
if all of the state-generating functions exposed by the state model are monotonic with respect to the pre-order induced by $\frestriction$; put formally: 

\noindent
\begin{minipage}{\textwidth}
{\small
\vspace*{0.1cm}
\begin{mathpar}
     \inferrule[RMono-SetVar]{}{\st.\kwT{setVar}(\x, \vl) = \st' \implies \st' \rleq \st}
      \qquad
     \inferrule[RMono-SetStore]{}{\st.\kwT{setStore}(\sto) = \st' \implies \st' \rleq \st}
     \qquad
     \inferrule[RMono-Action]{}{\action{\st}{\act}{v}{(\st', -)} \implies \st' \rleq \st}
   \end{mathpar}}
\end{minipage}

\vspace*{-0.2cm}
We say that $\frestriction$ is a restriction operator on a state model
$\gstate = \tup{\sset{\gstate}, \gval, \actions}$, if $\frestriction$ is a restriction 
operator on the carrier set $\sset{\gstate}$ and $\frestriction$ is preserved by $\gstate$. 
Restriction operators are extended from states to configurations straightforwardly: 
$\rfun{\tup{\prog, \st, \cs, i}}{\tup{\prog, \st', -, -}} \semeq \tup{\prog, \rfun{\st}{\st'}, \cs, i}$.

\myparagraph{Compatibility}
In the following, we assume a pre-order $\leq$ on abstract states, writing $\sst_2 \leq \sst_1$ to mean that 
the models of $\sst_2$ are contained in the models of $\sst_1$ (we say that $\sst_2$ is more precise than~$\sst_1$). This pre-order may differ from that induced by the chosen restriction operator 
on abstract states. 
Consider the symbolic execution setting: the fact that the path condition of a state~$\sst_2$ implies 
the path condition of a state $\sst_1$ does not necessarily mean that all the models of $\sst_2$ 
are contained in the models of $\sst_1$, as these two states may describe different memories. 
However, the chosen restriction operator must be compatible with the pre-order on states. 
Formally, we say that a pre-order $(X, \leq)$ is \emph{compatible} with a restriction operator $\frestriction$ 
on $X$ \emph{iff} the following properties hold: 

\noindent
\begin{minipage}{\textwidth}
\vspace*{-0.1cm}
\begin{mathpar}
\inferrule[$\frestriction$-$\leq$ Compatibility]{}{\rfun{x_1}{x_2} \, \leq \, x_1}
\and
\inferrule[$\leq$-$\frestriction$ Compatibility]{x_1 \leq x_2}{x_1 \rleq x_2}
\and
\inferrule[Strengthening]{x_1 \leq x_1' \quad x_2 \rleq x_2' }{\rfun{x_1}{x_2} \, \leq \, \rfun{x_1'}{x_2'}}
\end{mathpar}
\end{minipage}

\vspace*{0.1cm}
These properties essentially describe that restriction increases $\leq$-precision, that $\leq$-precision implies $\rleq$-precision, and how $\leq$-precision and $\frestriction$ combine under $\rleq$-precision.
Unsurprisingly, the pre-order $\rleq$ induced by $\frestriction$ is indeed compatible with $\frestriction$.


\myparagraph{Soundness Relations}
We are now in the position to describe the class of soundness relations that are preserved by the semantics of \gil. 
This class is formally given in Definition~\ref{def:soundness:relations}, which makes use of the notion of induced pre-order.
Formally, given a relation $\rel \, \in X \times Y$ between two sets $X$ and $Y$, the pre-order 
on $X$ induced by $\rel$, written $\leq_{\rel}$, is defined as follows: 
$x_1 \leq_{\rel} x_2$ if and only if $\lbrace y \mid x_1 \, \rel \, y \rbrace \subseteq \lbrace y \mid x_2 \, \rel \, y \rbrace$. 
We elide the $\rel$ in $\leq_{\rel}$ when it is clear from the context.

\begin{definition}[Soundness Relation - States]\label{def:soundness:relations}
Given two state models, $\hat\gstate = \tup{\sset{\hat\gstate}, \hat\gval, \actions}$ and $\gstate = \tup{\sset{\gstate}, \gval, \actions}$, 
a soundness relation $\soundRel$ for $\hat\gstate$ with respect to $\gstate$ is a triple
$\tup{\frestriction, \stRel, \vlRel}$, consisting of: 
\etag{1}~a restriction operator $\frestriction$ on $\hat\gstate$;
\etag{2}~a binary relation $\stRel \, \subseteq \sset{\hat\gstate} \times \sset{\gstate}$; and 
\etag{3}~a ternary relation $\vlRel \subseteq \sset{\hat\gstate} \times \hat\gval \times \gval$, 
such that $\frestriction$ is compatible with the pre-order induced by $\stRel$ (denoted by $\leq$) and the following constraints hold: 

\noindent
\begin{minipage}{\textwidth}
{\small
\begin{mathpar}
     \inferrule[Store]{}{\stR{\sst}{\st} \implies \vlR{\sst}{\sst.\kwT{store}}{\st.\kwT{store}}}
     \and 
     \inferrule[EvalExpr]{}{\leqStR{\sst}{\sst'}  \gand \stR{\sst}{\st} 
            \implies \vlR{\rfun{\sst'}{\sst}}{\ee{\sst'}{\e}}{\ee{\st}{\e}}}
\and
      \inferrule[SetVar]{}{
        {\begin{array}{l}
             \leqStR{\sst}{\sst'} \gand \stR{\sst}{\st}  \gand \vlR{\sst}{\hat\vl}{\vl} \\
              \quad \implies \stR{\rfun{\sst'.\kwT{setVar}(\x, \hat\vl)}{\sst}}{\st.\kwT{setVar}(\x, \vl)}
         \end{array}}}
     \and 
     \inferrule[SetStore]{}{
       {\begin{array}{l}
         \leqStR{\sst}{\sst'} \gand \stR{\sst}{\st}  \gand \vlR{\sst}{\hat\sto}{\sto} \\
            \quad  \implies  \stR{\rfun{\sst'.\kwT{setStore}(\hat\sto)}{\sst}}{\st.\kwT{setStore}(\sto)}
        \end{array}}}
\and
     \inferrule[Action]{}{\action{\sst'}{\act}{\hat\vl}{(\sst'', \hat\vl')} \gand \leqStR{\sst}{\rfun{\sst'}{\sst''}} \gand \ \stR{\sst}{\st} \gand  \vlR{\sst}{\hat\vl}{\vl} \\
            \qquad \implies \exists \, \st', \vl' \, . \, \action{\st}{\act}{\vl}{(\st', \vl')} \gand \stR{\rfun{\sst'}{\sst}}{\st'} \gand  \vlR{\rfun{\sst'}{\sst}}{\hat\vl'}{\vl'}}
     \and
     \inferrule[Weakening]{}{
       {\begin{array}{l}
        \sst \rleq \sst' \gand \vlR{\sst}{\hat\vl}{\vl}  \\ 
        \qquad \implies \vlR{\sst'}{\hat\vl}{\vl}
        \end{array}}
     }
  \end{mathpar}}
\end{minipage}

\noindent where $\vlR{\sst}{\hat\sto}{\sto}$ is shorthand for: $\domain(\hat\sto) = \domain(\sto) = X$ and $\forall \x \in X \, . \, \vlR{\sst}{\hat\sto(\x)}{\sto(\x)}$.
\end{definition}

In a nutshell, if $\soundRel = \tup{\frestriction, \stRel, \vlRel}$ is a soundness relation for $\hat\gstate = \tup{\sset{\hat\gstate}, \hat\gval, \actions}$ 
in terms of $\gstate = \tup{\sset{\gstate}, \gval, \actions}$, then: 
\etag{1} $\stR{\sst}{\st}$ means that $\sst$ is an over-approximation of $\st$, and
\etag{2} $\vlR{\sst}{\vl_1}{\vl_2}$ means that, considering the information in $\sst$, $\vl_1$ is an over-approximation of $\vl_2$.  
The constraints imposed on the state functions guarantee that $\stRel$ and $\vlRel$ are preserved by the 
\gil interpreter. 
Below, we discuss the \textsc{Action} and \textsc{Weakening} constraints; the other ones can be understood analogously.

\myparagraph{\prooflab{Action}} This constraint states that if we have an abstract action execution 
$\action{\sst'}{\act}{\hat\vl}{(\sst'', \hat\vl')}$, then, for any abstract state $\sst$,  
concrete state $\st$, and concrete value $\vl$, such that: 
\etag{1} $\leqStR{\sst}{\rfun{\sst'}{\sst''}}$ (meaning that $\sst$ is more precise than the original state $\sst'$ restricted to the final state $\sst''$), 
\etag{2} $\stR{\sst}{\st}$ (meaning that $\sst$ is an over-approximation of $\st$), and 
\etag{3} $\vlR{\sst}{\hat\vl}{\vl}$ (meaning that, according to $\sst$, $\hat\vl$ is an over-approximation of $\vl$); 
then, there must exist a concrete state $\st'$ and a concrete value $\vl'$, such that: 
\etag{i} $\action{\st}{\act}{\vl}{(\st', \vl')}$, 
\etag{ii} $\stR{\rfun{\sst'}{\sst}}{\st'}$ (meaning that the restriction of the final state, $\sst'$ to $\sst$ is an over-approximation of the final concrete state, $\st'$), and 
\etag{iii} $\vlR{\rfun{\sst'}{\sst}}{\hat\vl'}{\vl'}$ (meaning that~$\hat\vl'$ is an over-approximation of $\vl'$ according to $\rfun{\sst'}{\sst}$). 

\myparagraph{\prooflab{Weakening}}
This constraint states that, if $\sst$ is more $\rleq$-precise than $\sst'$
and, according to $\sst$, $\hat\vl$ is an over-approximation of $\vl$, 
then $\hat\vl$ is also an over-approximation of $\vl$ according to $\sst'$. 

Next, Theorem~\ref{theo:soundness} states that the semantics of \gil preserves soundness relations. 
It uses the standard liftings of $\vlRel$, $\stRel$, and $\leq_{\stRel}$ to call stacks and configurations (cf. Appendix~\ref{app:s3}).

\begin{theorem}[Soundness]\label{theo:soundness}
Let $\soundRel = \tup{\frestriction, \stRel, \vlRel}$ be a soundness relation
 for $\hat\gstate = \tup{\sset{\hat\gstate}, \hat\gval, \actions}$ in terms of $\gstate = \tup{\sset{\gstate}, \gval, \actions}$
 and $\leq$ the pre-order induced by $\stRel$. It holds that: 
 $$ 
 \hat\cf' \ssemarrow \hat\cf'' \gand \hat\cf \leq \rfun{\hat\cf'}{\hat\cf''} \gand \stR{\hat\cf}{\cf}
    \implies \exists \, \cf' \, . \, \cf \ssemarrow \cf' \gand \stR{\rfun{\hat\cf''}{\hat\cf}}{\cf'}
 $$
\end{theorem}

From there, by choosing $\hat\cf \equiv \rfun{\hat\cf'}{\hat\cf''}$, we obtain the desired soundness result.

\begin{corollary}[Soundness]\label{cor:soundness}
Let $\soundRel = \tup{\frestriction, \stRel, \vlRel}$ be a soundness relation
 for $\hat\gstate = \tup{\sset{\hat\gstate}, \hat\gval, \actions}$ in terms of $\gstate = \tup{\sset{\gstate}, \gval, \actions}$
 and $\leq$ the pre-order induced by $\stRel$. It holds that: 
 $$ 
 \hat\cf \ssemarrow \hat\cf' \gand \stR{(\rfun{\hat\cf}{\hat\cf'})}{\cf}
    \implies \exists \, \cf' \, . \, \cf \ssemarrow \cf' \gand \stR{\hat\cf'}{\cf'}
 $$
\end{corollary}

This corollary states that, if we have an abstract \gil trace and the 
initial concrete configuration~$\cf$ is over-approximated by the initial abstract configuration strengthened with the information of the final abstract configuration, then there exists a concrete \gil trace starting from $\cf$ such that the final abstract configuration is an over-approximation of the final concrete configuration. 

\subsection{Concrete-Symbolic Soundness}\label{subsec:symb:soundness}

We identify a class of relations between symbolic memory models and concrete memory models 
from which one can construct soundness relations between the corresponding symbolic
state models and concrete state models.  

\myparagraph{Symbolic Memory Interpretation}
Intuitively, a symbolic memory model  $\smemory = \tup{\sset{\smemory}, \actions, \sea}$ 
is related to a concrete memory model $\cmemory = \tup{\sset{\cmemory}, \actions, \cea}$,
if there is an interpretation function $\interp$ mapping the memories in 
$\sset{\smemory}$ to memories in $\sset{\cmemory}$. 
Memory interpretation must preserve actions, meaning that: if an action can be executed 
in a given symbolic memory $\smem$ under a path condition $\pc$, then it can also 
it can also be executed in any interpretation of $\smem$ that satisfies 
both the original path condition, $\pc$, and the path condition generated by 
the action, $\pc'$. Furthermore, the output concrete memory must be an interpretation 
of the output symbolic memory. The notion of interpretation is made precise in 
Definition~\ref{def:interpretation}. 

\begin{definition}[Symbolic Memory Interpretation]\label{def:interpretation}
Given a symbolic memory model $\smemory = \tup{\sset{\smemory}, \actions, \sea}$ 
and a concrete memory model $\cmemory = \tup{\sset{\cmemory}, \actions, \cea}$, 
an interpretation of $\smemory$ with respect to $\cmemory$ is a function 
$\interp : \sset{\smemory} \tmap (\lxs \pmap \vals) \pmap \sset{\cmemory}$ 
such that:
\begin{equation}
\begin{array}{l}
\action{\smem}{\act}{\sexp, \pc}{(\smem', \sexp', \pc')} 
  \gand  \cmem = \interp(\smem, \lenv)
  \gand \seval{\pc \, \wedge \, \pc'}{\lenv} = \true \\
   \qquad \qquad \implies \exists \, \cmem' \, . \, 
     \cmem' = \interp(\smem', \lenv) 
     \gand 
     \caction{\cmem}{\act}{\seval{\sexp}{\lenv}}{(\cmem', \seval{\sexp'}{\lenv})}
\end{array}\label{eq:interp:restriction}
\end{equation}
\end{definition}

Note that the interpretation function is parametric on a \emph{logical environment}, $\lenv$, which maps symbolic variables to concrete values. Onward, we write $\seval{\sexp}{\lenv}$ to denote interpretation of logical expressions under $\lenv$, defined in the standard way.

\myparagraph{Allocator Interpretation and Restriction} Before we proceed to the lifting of memories to states, we introduce the notions of interpretation and restriction for allocators. The interpretation is straightforwardly defined as follows.

\begin{definition}[Symbolic Allocator Interpretation]
Given a symbolic allocator $\hat\allocator = \tup{\sset{\hat\allocator}, \hat\gval}$ 
and a concrete allocator $\allocator = \tup{\sset{\allocator}, \gval}$, 
an interpretation of $\hat\allocator$ with respect to $\allocator$ is a function 
$\allocint : \sset{\hat\allocator} \tmap (\hat\gval \pmap \gval) \pmap \sset{\allocator}$ 
such that:
\[
\begin{array}{l}
\alloc{\hat\arec}{j}{\hat\arec', \hat\gvl}{Y}
  \gand \arec = \allocint(\hat\arec, \lenv) 
 \implies  
 	\exists \, \arec'. \, 
	 \arec' = \allocint(\hat\arec', \lenv) \gand
     \alloc{\arec}{j}{\arec', \lenv(\hat\gvl)}{\lenv(Y)}
\end{array}\]
\end{definition}

On the other hand, restriction is more involved. A restriction operator on allocator records, $\frestriction : \sset{\allocator} \tmap \sset{\allocator} \pmap \sset{\allocator}$, is defined algebraically, in the same way as in \S\ref{subsec:parametric:soundness}. The intuition behind it, however, is different. In this setting, it makes sense to talk about $\rfun{\arec}{\arec'}$ only if $\arec'$ has allocated the same values as $\arec$, and possibly more. Then, $\rfun{\arec}{\arec'}$ describes an allocator record that necessarily follows the allocation of~$\arec'$ on values already allocated by $\arec'$, but not yet by $\arec$.  

A restriction operator is said to be
\emph{preserved} by an allocator $\allocator = \tup{\sset{\allocator}, \gval}$ 
if it satisfies the following two properties: 
\begin{mathpar}
     \inferrule[RMono-Alloc]{}{\alloc{\arec}{j}{\arec', \gvl}{Y} \implies \arec' \rleq \arec}
     \qquad
\inferrule[FutureToPastAlloc]{}{
     \alloc{\arec}{j}{\arec', \gvl}{Y} \gand  \arec'' \rleq \rfun{\arec}{\arec'}  \\\\
     \implies \alloc{\arec''}{j}{\rfun{\arec'}{\arec''}, \gvl}{Y}}
   \end{mathpar}
We say that $\frestriction$ is a restriction operator on an allocator 
$\allocator = \tup{\sset{\allocator}, \gval}$, if $\frestriction$ is a restriction 
operator on the carrier set $\sset{\allocator}$ and $\frestriction$ is preserved by $\allocator$. 

\begin{wrapfigure}{R}{0.3\textwidth}
\centering
\vspace*{-0.5cm}
\begin{tikzpicture}
\node (AL1) {$\arec$};
\node (AL2) [below= 0.7cm of SS1] {$(\arec', \gvl)$};
\node (RAL2) [right= 1cm of AL2] {$(\arec', \gvl)$};
\node (RAL1) at (RAL2 |- AL1) {$\rfun{\arec}{\arec'}$};
\draw[->] (AL1) to node [left] {$j$} (AL2);
\draw[->] (RAL1) to node [right] {$j$} (RAL2);
\node (X) [below right= 0.12cm and 0.38cm of AL1] {$\Longrightarrow$};
\end{tikzpicture}

\vspace*{-0.3cm}
\caption{\prooflab{FutureToPastAlloc}}
\label{fig:futuresimple}
\vspace*{-0.3cm}
\end{wrapfigure}

We turn our attention now to the \prooflab{FutureToPastAlloc} property, which merits further discussion. In~\S\ref{subsec:parametric:soundness}, we showed how to use restriction to transfer state-level information from the final abstract state to the initial abstract state in order to filter out initial concrete states for which the concrete execution diverges from the given abstract trace. Here, similarly, restriction allows us to use future allocation information to direct present allocation to the desired traces. This is essential, as both the allocation of locations and symbolic variables can be non-deterministic. 
We discuss a simplified version of \prooflab{FutureToPastAlloc} given diagrammatically in Figure~\ref{fig:futuresimple}; the full version is a straightforward generalisation.
This property states that, if we allocate using $\arec$ at site $j$ and obtain $\arec'$ and a value~$\gvl$, then allocation using $\arec$ restricted to $\arec'$ at the same site must yield the same allocator and the same value. Combined with the definition of the lifting from memories to states, given shortly, this property effectively directs concrete allocation to the appropriate traces.

\myparagraph{Lifting Interpretations}
Given an interpretation $\interp : \sset{\smemory} \tmap (\lxs \pmap \vals) \pmap \sset{\cmemory}$  
of a symbolic memory model $\smemory = \tup{\sset{\smemory}, \actions, \sea}$  in terms of 
a concrete memory model $\cmemory = \tup{\sset{\cmemory}, \actions, \cea}$, 
the candidate soundness relation $\interpLift(\interp) = \tup{\frestriction, \stRel, \vlRel}$ for 
$\sstateConstr(\smemory)$ in terms of $\cstateConstr(\cmemory)$ is defined as~follows:  
$$
\begin{array}{lll}
\rfun{\tup{\smem, \ssto, \sarec, \pc}}{\tup{\_, \_, \sarec', \pc'}} & \semeq  & \tup{\smem, \ssto, \rfun{\sarec}{\sarec'}, \pc \gand \pc'}  \\ 
\stR{\sst}{\st} & \semeq & \exists \, \lenv \, . \, (\st, \lenv) \in \modls(\sst) \\
\vlR{\tup{\_, \_, \_, \pc}}{\sexp}{\vl} & \semeq & \exists \, \lenv \, . \, \seval{\pc}{\lenv} = \true \gand \seval{\sexp}{\lenv} = \vl
\end{array}
$$
where: 
$$
   \modls(\tup{\smem, \ssto, \pc, \sarec})  \semeq 
           \left\{ (\tup{\cmem, \sto, \arec}, \lenv) \mid
                   \seval{\pc}{\lenv} = \true \gand
                   \cmem = \interp(\smem, \lenv) \gand 
                   \sto = \seval{\ssto}{\lenv} \gand  
                   \arec = \allocint(\sarec, \seval{.}{\lenv})  
            \right\} 
$$

Theorem~\ref{theo:lifting} states that, given a memory interpretation $\interp$, 
$\interpLift(\interp)$ is a soundness relation
as per Definition~\ref{def:soundness:relations}. 
Hence, we conclude, appealing to Theorem~\ref{theo:soundness}, that 
$\interpLift(\interp)$ is preserved by the 
semantics of \gil. 
This means that, in order to prove the soundness of their analyses, the users
of \gillian only have to provide an interpretation function linking their symbolic 
memories to their concrete memories and prove that that interpretation 
preserves the actions exposed by the memories, meaning that 
they must satisfy Equation~\ref{eq:interp:restriction}.

\begin{theorem}[Soundness Relation - Lifting]\label{theo:lifting}
Let $\interp$ be an interpretation of a symbolic memory model  $\smemory$  in terms of 
a concrete memory model $\cmemory$; then, $\interpLift(\interp) = \tup{\frestriction, \stRel, \vlRel}$ 
is a soundness relation for $\sstateConstr(\smemory)$ in terms of $\cstateConstr(\cmemory)$.
\end{theorem}

\subsection{\while: Sound Symbolic Analysis}\label{subsec:while:soundness}

The interpretation of \while symbolic memories in terms of \while concrete memories, $\winterp$, 
is inductively defined as follows: 
\begin{mathpar}
\inferrule[Empty]{}{\winterp(\emptymem, \lenv) \semeq \emptymem} 
\and 
\inferrule[Cell]
   {
     \loc = \seval{\sexp}{\lenv} 
     \and
     \vl = \seval{\sexp'}{\lenv}
   }{\winterp(\wcell{\sexp}{\prop}{\sexp'}, \lenv) \semeq \wcell{\loc}{\prop}{\vl}} 
 \and 
\inferrule[Union]
   {
     \cmem_1 = \winterp(\smem_1, \lenv) 
     \and
     \cmem_2 = \winterp(\smem_2, \lenv)
   }{\winterp(\smem_1 \dunion \smem_2, \lenv) \semeq  \cmem_1 \dunion \cmem_2}  
\end{mathpar}
This interpretation is standard: \while memories are interpreted piecemeal 
and the interpretations are composed together using the disjoint union operator.  

Lemma~\ref{lemma:while:soundness} states that $\winterp$ preserves the actions of \while, 
$\actionswhile = \lbrace \lookupAction, \mutateAction, \disposeAction \rbrace$. 
Its proof is straightforward, requiring only a case analysis 
on the rules given in Figure~\ref{fig:while:memories}.
This is much simpler than the customary inductive proofs on semantic derivations, which underpin standalone soundness proofs. 

\begin{lemma}[\while: Memory Interpretation]\label{lemma:while:soundness}
$\winterp$ is an interpretation of $\wsmems$ with respect to~$\wcmems$. 
\end{lemma}

We conclude with the instantiation of Theorem~\ref{theo:soundness} to 
\while. 

\begin{theorem}[\while: Soundness]\label{theo:while:soundness}
Given $\tup{\frestriction, \stRel, \vlRel} = \interpLift(\winterp)$, it holds that:
 $$ 
 \hat\cf \ssemarrow_{\wsub} \hat{\cf'} \gand  \stR{(\rfun{\hat\cf}{\hat{\cf'}})}{\cf}
    \implies \exists \, \cf'. \, \cf \semarrow_\wsub \cf' \gand \stR{\hat{\cf'}}{\cf'}
 $$
\end{theorem}

\section{Parametric Verification}
\label{sec:verification}

We extend the semantics of \gil with support for separation logic (SL) specifications. 
This allows the general semantics to: \etag{i} use SL specifications to jump over procedure calls 
instead of re-executing a procedure at each call site; and \etag{ii} use symbolic execution 
to verify SL specifications efficiently instead of re-implementing an SL proof system from scratch. 
To achieve this, we introduce a parametric assertion language, whose semantics 
is defined in terms of an underlying state model. 
In this way, users of \gillian gain access 
to  an out-of-the-box verification tool for their 
target languages, while only having to provide a minimal 
description of their own specific assertions  (onward, \emph{core predicates})
in terms of the actions of the state model. 

The section is structured as follows: \S\ref{subsec:naive:verification} discusses the type 
of verification guarantee that can be obtained from the analysis presented in \S\ref{sec:sem}
and its main shortcomings; 
\S\ref{subsec:assertions} presents our parametric assertion language together with 
its semantics; \S\ref{subsec:verification} describes the extension of the semantics of 
\gil to account for the use of SL-specifications and predicates; and, finally, \S\ref{subsec:while:verification}
shows how to leverage the proposed infrastructure to obtain a verification tool 
for \while.

\subsection{Na\"ive Verification}
\label{subsec:naive:verification}

The symbolic analysis presented in \S\ref{sec:sem} is \emph{sound} in the following sense: if 
we pick a concrete execution that follows the same path as that of the given symbolic execution 
and if the initial symbolic state is an over-approximation of the initial concrete state; then, we 
are guaranteed to terminate in a concrete state that is also over-approximated by the final symbolic 
state. The key point here is that we are only referring to concrete executions that follow \emph{the same path} as that of the given symbolic execution; nothing is said about the others. 

Note, however, that if the symbolic execution does not branch, then no restriction needs to be imposed on the corresponding concrete executions.
This intuition is captured by Theorem~\ref{theo:verification}, below, using the notion 
of restriction (cf.~\S\ref{subsec:parametric:soundness}): essentially, a symbolic 
execution is guaranteed not to branch if the initial symbolic state is as $\rleq$-precise 
as the final symbolic state.

\begin{theorem}[Na\"ive Verification]\label{theo:verification}
Let $\soundRel = \tup{\frestriction, \stRel, \vlRel}$ be a soundness relation
 for $\gsstate = \tup{\sset{\gsstate}, \gsval, \actions}$ in terms of $\gstate = \tup{\sset{\gstate}, \gval, \actions}$
 and $\leq$ the pre-order induced by $\stRel$. It holds that: 
 $$
 \hat\cf \ssemarrow \hat\cf' \gand \hat\cf \rleq \hat\cf' \gand \stR{\hat\cf}{\cf}
    \implies \exists \, \cf' \, . \, \cf \ssemarrow \cf' \gand \stR{\hat\cf'}{\cf'}
 $$
\end{theorem}

Theorem~\ref{theo:verification} gives us a standard verification guarantee for programs that do not 
branch. Most programs, however, do branch and are, therefore, not in the conditions of the theorem. 
To account for branching, we need a mechanism for merging symbolic execution traces. 
With such a mechanism, one can re-interpret Theorem~\ref{theo:verification} as follows: 
if a concrete execution follows the same path as that of \underline{\emph{one of}} the symbolic 
execution traces that are merged together in the given symbolic trace, then the conclusions
of the theorem hold.  

In general, merging symbolic traces is not easy, as different traces may describe 
different structures in memory. In the symbolic execution literature, many techniques 
address this problem, such as \emph{predicate abstraction}~\cite{pasreanu:lmcs:2007,jhala:tacas:2006} 
or the combination of guarded unions with carefully crafted merging algorithms~\cite{torlak:pldi:2014}. 
Here, we use SL predicates for describing inductive data structures in memory, 
allowing us to \emph{fold} symbolic traces that operate on different \emph{unfoldings} 
of the same data structure.

\subsection{Parametric Assertion Language}
\label{subsec:assertions}

\begin{wrapfigure}{R}{0.43\textwidth}
\begin{minipage}{0.43\textwidth}
\smallskip
\vspace*{-0.45cm}
\begin{display}{\gil Parametric Assertions}
\begin{tabular}{l}
   $\Pass, \Qass \in \Passes{\corePreds}$ $\defeq$ $\emp \mid \pc \mid \cPred{\e} \mid \aPred{\pn}{\e} \mid \astar{\Pass}{\Qass}$
   \\[3pt]
   $\pred \in \predicates$ $\defeq$ $\fullpredicate{\pn}{\x}{\Pass_0}{\Pass_n}$
   \\[3pt]
   $\pval \in \pvals{\gval}$ $\defeq$ $\pn(\gvl)$, where $\gvl \in \gval$
 \end{tabular}
\end{display}
\end{minipage}
\vspace*{-0.3cm}
\end{wrapfigure}
 
Given a set of core predicates $\corePreds$ to be provided by the user of the
framework, \gil assertions, $\Pass, \Qass \in \Passes{\corePreds}$, include:
the empty memory assertion $\emp$, 
boolean logical expressions, $\pc \in \pcs$, core predicate assertions, $\cPred{\e}$
with $\corePred \in \corePreds$, user-defined predicate assertions, $\pn(\e)$
(where $\pn \in \pns \subset \strs$, the set of predicate names),
and the standard separating conjunction, $\astar{\Pass}{\Qass}$. 
User-predicate definitions are of the form
$\fullpredicate{\pn}{\x}{\Pass_0}{\Pass_n}$ where: $\pn$ is the predicate name, $\x$ its argument, and $\Pass_0, ..., \Pass_n$ are $n$ 
alternative predicate definitions, each a \gil assertion.  
This is a presentational simplification, as predicates in general can have more than one argument.
To model folded predicate assertions in a state with 
values $\gvl \in \gval$, we introduce the notion of value predicates. 
A value predicate $\pval \in \pvals{\gval}$, is a pair 
$(\pn, \gvl)$, written $\pn(\gvl)$ for legibility, consisting of a predicate name, 
$\pn$, and a value, $\gvl$. Onward, we refer to value predicates simply as predicates.


\myparagraph{Predicate States}
Unsurprisingly, in order to define the interpretation of assertions, we require that the 
carrier set of the parameter state model, $\sset{\gstate}$, forms a partial commutative 
monoid~\cite{calcagno:lics:2007}, under a given composition operator, $\stcomp$, and  
neutral element $\stzero$. Furthermore, we require all the state-generating functions
to be frame preserving. Put formally: \\
\noindent
\begin{minipage}{\textwidth}
{\small
\vspace*{0.1cm}
\begin{mathpar}
     \inferrule[Frame-SetVar]
            {\st.\kwT{setVar}(\x, \vl) = \st'}
            {(\st \stcomp \st_f).\kwT{setVar}(\x, \vl) = \st' \stcomp \st_f}
      \qquad
     \inferrule[Frame-SetStore]
            {\st.\kwT{setStore}(\sto) = \st'}
            {(\st \stcomp \st_f).\kwT{setStore}(\sto) = \st' \stcomp \st_f}
     \qquad
     \inferrule[RMono-Action]
            {\action{\st}{\act}{v}{(\st', -)} }
            {\action{(\st \stcomp \st_f)}{\act}{v}{((\st' \stcomp \st_f, -)}}
   \end{mathpar}}
\end{minipage}
Finally, we extend a given 
state model with support for predicates. We say that a state model supports 
predicates if it exposes an action $\kwT{getP}$ for recovering a given predicate 
from the state and an action $\kwT{setP}$ for adding a new predicate to the state. 
Given an arbitrary state model $\gstate = \tup{\sset{\gstate}, \gval, \actions}$, we construct 
a new state model with support for predicates by coupling 
$\sset{\gstate}$ with a list of predicates $\lst{\pval} \in \lst{\pvals{\gval}}$.
There, $\kwT{getP}$ simply removes the required predicate from $\lst{\pval}$, 
while $\kwT{setP}$ adds it to $\lst{\pval}$. 
This lifting is formally described~below. 

\begin{definition}[Predicate State Constructor (\vpstateConstr)]\label{def:pred:lifting}
The predicate state constructor $\vpstateConstr : \gstates \tmap \gstates$ is defined as  
$\vpstateConstr(\tup{\sset{\gstate}, \gval, \actions}) \semeq \tup{\sset{\gstate'}, \gval, \actions \dunion \lbrace \kwT{setP}, \kwT{getP}\rbrace}$,
where: 

{\small
\begin{tabular}{lll}
   \mybullet $\sset{\gstate'}$ &$\semeq$ & $\sset{\gstate} \times \lst{\pvals{\gval}}$ \\ 
  \mybullet $\kwT{setVar}_{\predsub}(\tup{\st, \lst{\pval}}, \x, \gvl)$ 
       & $\semeq$ & 
       $\tup{\kwT{setVar}(\st,  \x, \gvl), \lst{\pval}}$ \\ 
  \mybullet $\kwT{setStore}_{\predsub}(\tup{\st, \lst{\pval}}, \sto)$
     & $\semeq$ & 
     $\tup{\kwT{setStore}(\st, \sto), \lst{\pval}}$ \\ 
  \mybullet $\kwT{store}_{\predsub}(\tup{\st, -})$
    & $\semeq$ & 
    $\kwT{store}(\st)$ \\
  \mybullet $\kwT{ee}_{\predsub}(\tup{\st, -}, \e)$
    & $\semeq$ & 
    $\kwT{ee}(\st, \e)$ \\
 \mybullet $\kwT{ea}_{\predsub}(\act, \tup{\st, \lst{\pval}}, \gvl)$
    & $\semeq$ & 
    $\lbrace (\tup{\st', \lst{\pval}}, \gvl') \mid (\st', \gvl') \in \kwT{ea}(\act, \st, \gvl) \rbrace$, \textbf{if} $\act \not\in \lbrace \kwT{getP}, \kwT{setP} \rbrace$ \\
 \mybullet $\kwT{ea}_{\predsub}(\kwT{setP}, \tup{\st, \lst{\pval}}, \litlst{\pn, \gvl})$
    & $\semeq$ & 
    $(\tup{\st, (\pn(\gvl)\cons\lst{\pval})}, -)$ \\
    \mybullet $\kwT{ea}_{\predsub}(\kwT{getP}, \tup{\st, \lst{\pval}}, \litlst{\pn, \gvl})$
    & $\semeq$ & 
    $(\tup{\st, \lst{\pval}_1 \cat \lst{\pval}_2}, -)$, 
    where $\lst{\pval} = \lst{\pval}_1 \cat \litlst{\, \pn(\gvl) \,} \cat \lst{\pval}_2$
\end{tabular}}
\end{definition}

\myparagraph{Core Predicate Interpretation}
In order to define the semantics of their assertions, the users of the framework need to 
describe the meaning of their core predicates in terms of the actions exposed by the 
parameter state model. 
Informally, each core predicate $\corePred \in \corePreds$ is associated with two actions: 
\etag{i} the \emph{getter action}, $\fgetter{\corePred}$, for recovering its footprint from the state; 
and \etag{ii} the \emph{setter action}, $\fsetter{\corePred}$, for extending the state with the 
footprint of the core predicate.

\begin{definition}[Core Predicate Action Interpretation]
A core predicate action interpretation is a 4-tuple $\actInterp \in \actInterps = \tup{\corePreds, \actions, \fset, \fget}$
consisting of a set of core predicates $\corePreds$, a set of actions $\actions$, and two functions 
$\fset, \fget : \corePreds \tmap \actions$; we write $\fsetter{\delta}$ for $\fset(\delta)$ and $\fgetter{\delta}$ for $\fget(\delta)$. 
A core predicate action interpretation $\tup{\corePreds, \actions, \fset, \fget}$ is said to be \emph{well-formed} 
with respect to a state model 
 $\gstate = \tup{\sset{\gstate}, \gval, \actions}$ if and only if, 
 for all core predicates $\corePred \in \corePreds$, it holds that: 
\begin{equation}\label{eq:well:formedness}
   \action{\st}{\fgetter{\delta}}{\gvl}{\st'} \iff 
      \action{\st'}{\fsetter{\delta}}{\gvl}{\st}     
\end{equation}
\end{definition} 

This well-formedness constraint captures the intuition behind getters and setters and states that they have to be, essentially, each other's inverses.

Given a state model $\gstate = \tup{\sset{\gstate}, \gval, \actions}$ and a core action 
interpretation $\tup{\corePreds, \actions, \fset, \fget}$, the induced action interpretation of an 
assertion $\Pass \in \Passes{\corePreds}$ is a pair of functions consisting of the 
\emph{getter} and \emph{setter} of $\Pass$, respectively, $\fgetter{\Pass}$ and $\fsetter{\Pass}$. 
Formally, Figure~\ref{fig:assertions:interpretation}
defines two induced functions: 
\begin{itemize}
   \item $\setInterpS{\corePreds}{\gstate} : \Passes{\corePreds} \tmap \sset{\gstate} \tmap (\lxs \dunion \xs \pmap \gval) \pmap \sset{\gstate}$
      \hfill
      ($\st' = \setInterpS{\corePreds}{\gstate}(\Pass, \st, \subst) \equiv_{pp} \actionSetterS{\st}{\Pass}{\subst}{\st'}$)
    \item $\getInterpS{\corePreds}{\gstate} : \Passes{\corePreds} \tmap \sset{\gstate} \tmap (\lxs \dunion \xs \pmap \gval) \pmap \sset{\gstate}$  
         \hfill
         ($\st' = \getInterpS{\corePreds}{\gstate}(\Pass, \st, \subst) \equiv_{pp}  \actionGetter{\st}{\Pass}{\subst}{\st'}$)
 \end{itemize}
\medskip
mapping each assertion $\Pass \in \Passes{\corePreds}$ to its getter and setter, respectively. 
The definition makes use of a \emph{substitution},~$\subst$, essentially a function mapping 
variables and logical variables to values. 

\medskip
Lemma~\ref{lemma:asrt:interp} states that, if we are given a well-formed core predicate interpretation, 
the induced getter and setter of every assertion $\Pass$ satisfy the well-formedness constraint
stated in Equation~\ref{eq:well:formedness}.
Finally, Theorem~\ref{teo:asrt:soundness} establishes that assertion interpretation preserves soundness relations. 
Hence, if we execute the getter/setter of a given assertion $\Pass$ in a given abstract state, and if we are given an 
initial concrete state that is over-approximated by the initial abstract state strengthened with the information 
of the final abstract state, then the concrete execution of the getter/setter of $\Pass$ will yield a final state
that is over-approximated by the final abstract state.

\begin{lemma}[Assertion Interpretation]\label{lemma:asrt:interp}
Let $\tup{\corePreds, \actions, \fset, \fget}$ be a well-formed core predicate interpretation 
with respect to a predicate state model $\gstate = \tup{\sset{\gstate}, \gval, \actions}$; then, it holds that: 
$\actionSetterS{\st}{\Pass}{\subst}{\st'}$ if and only if 
$\actionGetter{\st'}{\Pass}{\subst}{\st}$. 
\end{lemma}

\begin{theorem}[Assertion Interpretation - Soundness]\label{teo:asrt:soundness}
Let $\soundRel = \tup{\frestriction, \stRel, \vlRel}$ be a soundness relation
 for $\hat\gstate = \tup{\sset{\hat\gstate}, \hat\gval, \actions}$ in terms of $\gstate = \tup{\sset{\gstate}, \gval, \actions}$
 and $\leq$ the pre-order induced by $\stRel$; and let $\tup{\corePreds, \actions, \fset, \fget}$ be a 
 well-formed core predicate action interpretation for $\hat\gstate$ and $\gstate$; 
 then, it holds that: 
 \begin{align}
\begin{array}{l}
 \actionSetterS{\sst}{\Pass}{\hat\subst}{\sst'} \gand  
   \sst'' \leq \rfun{\sst}{\sst'} \gand 
   \stR{\sst''}{\st} \gand
    \vlR{\sst''}{\hat\subst}{\subst}  \\ 
   \qquad \qquad \implies 
   \exists \, \st' . \ \
       \actionSetterS{\st}{\Pass}{\subst}{\st'} \gand 
       \stR{\rfun{\sst'}{\sst''}}{\st'} 
 \end{array}
  \\
 \begin{array}{l}
  \actionGetter{\sst}{\Pass}{\hat\subst}{\sst'} \gand 
     \sst'' \leq \rfun{\sst}{\sst'} \gand 
      \stR{\sst''}{\st} \gand 
      \vlR{\sst''}{\hat\subst}{\subst} \\
      \qquad \qquad \implies 
       \exists \, \st' . \ \ 
         \actionGetter{\st}{\Pass}{\subst}{\st'} \gand 
           \stR{\rfun{\sst'}{\sst''}}{\st'} 
 \end{array}
\end{align}
\end{theorem}

\begin{figure}[!t]
{\small
\begin{minipage}{0.45\textwidth}
\begin{mathpar}
\inferrule[set - star]{
   \actionSetterS{\st}{\Pass}{\subst}{\st'}
   \\\\
   \actionSetterS{\st'}{\Qass}{\subst}{\st''}
}{
   \actionSetterS{\st}{(\astar{\Pass}{\Qass})}{\subst}{\st''}
}
\quad 
\inferrule[get - star]{
   \actionGetter{\st}{\Pass}{\subst}{\st'}
   \\\\
   \actionGetter{\st'}{\Qass}{\subst}{\st''}
  }{
    \actionGetter{\st}{(\astar{\Pass}{\Qass})}{\subst}{\st''}
  } 
\end{mathpar}
\end{minipage}
\begin{minipage}{0.05\textwidth}
\end{minipage}
\begin{minipage}{0.45\textwidth}
\begin{mathpar}
\inferrule[set - boolean expr]
  {
    \ee{\st}{\subst(\e)} = \gvl
    \\\\
    \actionT{\st}{assume}{\gvl}{\st'}
  }{
    \actionSetterS{\st}{\pc}{\subst}{\st'}
  }
\quad 
\inferrule[get - boolean expr]
  {
      \ee{\st}{\lnot \, \subst(\e)} = \gvl 
      \\\\
      \st.\kwT{assume}(\gvl) = \emptyset 
  }{
    \actionGetter{\st}{\pc}{\subst}{\st}
 } 
\end{mathpar}
\end{minipage}

\medskip
\medskip
\begin{minipage}{0.45\textwidth}
\begin{mathpar}
\inferrule[set - pred]{
  \ee{\st}{\litlst{\pn, \subst(\e)}} = \gvl
  \\\\
  \action{\st}{setP}{\gvl}{\st'}
}{
   \actionSetterS{\st}{\aPred{\pn}{\e}}{\subst}{\st'}
}
\and 
\inferrule[get - pred]{
  \ee{\st}{\litlst{\pn, \subst(\e)}} = \gvl
  \\\\
  \action{\st}{getP}{\gvl}{\st'}
}{
    \actionGetter{\st}{\aPred{\pn}{\e}}{\subst}{\st'}
  } 
\end{mathpar}
\end{minipage}
\begin{minipage}{0.05\textwidth}
\end{minipage}
\begin{minipage}{0.45\textwidth}
\begin{mathpar}
\inferrule[set - core pred]{
   \ee{\st}{\subst(\e)} = \gvl
   \\\\
   \action{\st}{\fsetter{\delta}}{\gvl}{\st'} 
}{
   \actionSetterS{\st}{\cPred{\e}}{\subst}{\st'}
}
\and 
\inferrule[get - core pred]{
   \ee{\st}{\subst(\e)} = \gvl
   \\\\
   \action{\st}{\fgetter{\delta}}{\gvl}{\st'}
}{
   \actionGetter{\st}{\cPred{\e}}{\subst}{\st'}
} 
\end{mathpar}
\end{minipage}}
\vspace*{-0.2cm}
\caption{Assertion Interpretation: $\actionSetterS{\st}{\Pass}{\subst}{\st'}$ and $\actionGetter{\st}{\Pass}{\subst}{\st'}$}\label{fig:assertions:interpretation}
\vspace*{-0.4cm}
\end{figure}


\myparagraph{Implementation} 
The predicate state constructor is implemented as an OCaml functor parameterised 
by an OCaml module of type \texttt{State}. The \texttt{PState} functor precisely follows 
the lifting described in Definition~\ref{def:pred:lifting}: the action type of the returned state extends the action type of 
the parameter state with the actions \texttt{getP} and \texttt{setP}
of the formalism. 
We also define the module type \texttt{CorePred}, describing core predicate 
action interpretations. \texttt{CorePred} modules must define the \texttt{getter} and 
\texttt{setter} functions, mapping core predicates to their corresponding 
getter and setter~actions. 

\begin{figure}[!h]
\vspace*{-0.2cm}
\begin{minipage}{0.48\textwidth}
\begin{minted}[fontsize=\scriptsize]{ocaml}
type 'a pred_act = GetP | SetP | OtherAct of 'a

module PState (St : State) :
  (State with type a = St.a pred_act) = struct
    type vt = St.vt
    type t = St.t * (PName.t * vt) list
    type a = St.a pred_act
    ...
end
\end{minted}
\end{minipage}
\begin{minipage}{0.48\textwidth}
\begin{minted}[fontsize=\scriptsize]{ocaml}
module type CorePred = sig 
    type a (** Type of actions         *)
    type t (** type of core predicates *)

    val getter : t -> a 
    val setter : t -> a 
end 
\end{minted}
\end{minipage}
\vspace*{-0.5cm}
\caption{OCaml Predicate State Constructor}
\vspace*{-0.6cm}
\end{figure}

\subsection{Parametric Verification Semantics}
\label{subsec:verification}

We extend the syntax of \gil commands with logical commands for interacting with 
predicate assertions and specifications in state models with support for predicates. 
Besides the original commands $\cm \in \cmds{\actions}$, we include: 
a command $\vfold{\pn}{\e}{j; \left(\lx_i: \e_i\right)\!\mid_{i=0}^n}$ for folding 
the predicate denoted by $\pn(\e)$\footnote{(ignoring the additional symbolic paraphernalia for the moment)}; 
a command $\vunfold{\pn}{\e}$ for unfolding the predicate denoted by $\pn(\e)$; and, 
a command $\vcall{\x}{\e_1}{\e_2}{j; \left(\lx_i: \e_i\right)\!\mid_{i=0}^n}$ for executing the procedure denoted by 
$\e_1$ with the argument denoted by $\e_2$ \emph{using the procedure specification} instead of 
executing the body of the procedure. 
Observe that the standard procedure call command remains part of the syntax, as it is included in $\cm \in \cmds{\actions}$; 
to avoid confusion, we refer to the logical procedure call as spec call.  
Finally, we write $\vprog$ and $\vproc$ to respectively denote \gil programs and procedures 
that use logical commands. 

\smallskip
\begin{display}{Extended \gil Syntax}
\begin{tabular}{l@{\qquad\qquad\qquad\qquad}l}
  $\vcm \in \vcmds{\actions}$ $\defeq$ $\cm \in \cmds{\actions} \mid \vcall{\x}{\e_1}{\e_2}{j; \left(\lx_i: \e_i\right)\!\mid_{i=0}^n}\mid$ &   
                $\vproc \in \vprocs{\actions}$ $\defeq$ $\vprocedure{f}{\x}{\lst{\vcm}}$
  \\
    \hspace*{1.32cm}$\vfold{\pn}{\e}{j; \left(\lx_i: \e_i\right)\!\mid_{i=0}^n} \mid \vunfold{\pn}{\e}$
    & 
    $\vprog \in \vprogs{\actions}$ $:$ $\fids \pmap \vprocs{\actions}$
 \end{tabular}
\end{display}

\gil specifications have the form $\spec{\Pass}{\f}{\x}{\Qass}{\e}$, where $\Pass$ and $\Qass$ are the 
pre- and post-condition of the procedure with identifier $\f$ and parameter $\x$, and $\e$ denotes 
its return value. 
Procedures are allowed to have multiple specifications; we use $\getspec{\vprog}{\f}{j}$ to 
refer to the $j$-th specification of the procedure with identifier $\f$ in the program $\vprog$. 
Note that both specifications and predicate definitions may include logical variables. 
Accordingly, the fold and spec call commands allow the programmer 
to provide the appropriate bindings for these variables. 
At the implementation level, \gillian includes a search algorithm that automatically finds 
the appropriate bindings. A description of this algorithm, however, is out of the scope of this paper. 

An excerpt of the semantics of \gil logical commands is given in Figure~\ref{fig:verification:semantics} (cf.~Appendix~\ref{app:s4}).
The rules have the same structure as those of \gil commands (given in Figure~\ref{fig:sem}).
Below, we give a brief description of their behaviour. 

%

\vspace*{-0.3cm}
\begin{figure}[!h]
{\footnotesize
\begin{mathpar}
  \inferrule[\textsc{Fold}]
  {
     \cmd(\vprog, \cs, i) = \vfold{\pn}{\e_0}{j; \left(\lx_i: \e_i\right)\!\mid_{i=1}^n}
      \\\\
      \left(\gvl_i = \ee{\st}{\e_i}\right)\mid_{i=0}^n 
      \\\\
      \getppred{\vprog}{\pn}{} = \fullpredicate{\pn}{\x}{\Pass_0}{\Pass_n}
       \\\\
      \subst = [x \mapsto \gvl_0, \lx_1 \mapsto \gvl_1, ..., \lx_n \mapsto \gvl_n]
      \\\\
      \actionGetter{\st}{\Pass_j}{\subst}{\st'}
      \and 
      \action{\st'}{setP}{\litlst{\pn, \gvl}}{\st''}  
  }{
    \vsemtrans{\st, \cs, i}{\st'', \cs, i{+}1}{}{\vprog}{}
  }
  \and
\inferrule[\textsc{Spec Call}]
  {
     \cmd(\vprog, \cs, i) = \vcall{\x}{\e}{\e_0}{j; \left(\lx_i: \e_i\right)\!\mid_{i=1}^n}
      \\\\
      \f = \ee{\st}{\e}
      \and 
      \left(\gvl_i = \ee{\st}{\e_i}\right)\mid_{i=0}^n 
      \\\\
      \getspec{\vprog}{\f}{j} = \spec{\Pass}{\f}{\x_0}{\Qass}{\e'}
       \\\\
      \subst = [x_0 \mapsto \gvl_0, ..., \lx_n \mapsto \gvl_n]
      \\\\
      \ee{\st}{\subst(\e')} = \gvl
      \and
      \actionGetter{\st}{\Pass}{\subst}{\st'}
      \and 
      \actionSetterS{\st'}{\Qass}{\subst}{\st''}  
  }{
    \vsemtrans{\st, \cs, i}{\st''.\kwT{setVar}(\x, v), \cs, i{+}1}{}{\vprog}{}
  }
\end{mathpar}}
\vspace*{-0.4cm}
\caption{Verification Semantics of \gil: $\vfullsemtrans{\st, \cs, i}{\st', \cs', j}{}{\vprog}{\outcome'}{\outcome}$ (excerpt)}
\label{fig:verification:semantics}
\vspace{-0.4cm}
\end{figure}

\myparagraph{\prooflab{Fold}} The rule folds the predicate denoted by  $\pn(\e)$. First, 
it obtains the $j$-th definition of the predicate to be folded, $\Pass_j$, and uses its 
getter, $\fgetter{\Pass_j}$, to remove the corresponding footprint from the current state; then, 
it uses the action $\kwT{setP}$ to extend the current state with the folded predicate.

\myparagraph{\prooflab{Spec Call}} The rule executes the procedure with the identifier denoted by $\e$ with 
the argument denoted by $\e_0$ using its $j$-th specification. 
In order to ``execute'' the procedure, it uses the getter of the precondition, $\fgetter{\Pass}$, 
to remove the corresponding footprint from the current state, and the setter of the postcondition, 
$\fsetter{\Qass}$, to produce the corresponding footprint in the obtained state. 


\medskip
Finally, Theorem~\ref{theo:verification:preds} states that the execution of logical 
commands preserves correctness relations. 
At first glance, it coincides with the Na\"ive Verification Theorem 
(Theorem~\ref{theo:verification}). However, since logical commands offer us a mechanism 
to merge symbolic traces together, the theorem is applicable 
to programs which branch and loop.

\begin{theorem}[Verification]\label{theo:verification:preds}
Let $\soundRel = \tup{\frestriction, \stRel, \vlRel}$ be a soundness relation
 for $\gsstate = \tup{\sset{\gsstate}, \gsval, \actions}$ in terms of $\gstate = \tup{\sset{\gstate}, \gval, \actions}$
 and $\leq$ the pre-order induced by $\stRel$. It holds that: 
 $$ 
 \hat\cf~\vssemarrow~\hat\cf' \gand \hat\cf = \rfun{\hat\cf}{\hat\cf'} \gand \stR{\hat\cf}{\cf}
    \implies \exists \, \cf' \, . \, \cf~\vssemarrow~\cf' \gand \stR{\hat\cf'}{\cf'}
 $$
\end{theorem}

\subsection{\while: Verification}
\label{subsec:while:verification}

The only core predicate for \while is the $\wpcell([\loc, \prop, v])$ predicate, whose footprint is a single heap cell and which states that the property $\prop$ of the object at location $\loc$ has value $v$. We also introduce the setter and the getter actions for the $\wpcell$ core predicate, $\wscell$ and $\wgcell$. The rules for $\wscell$ are given below (cf.~Appendix~\ref{app:s4}). One can observe that those rules are similar to the rules of $\mutateAction$, except that they produce the resource of the $\wpcell$ core predicate. In fact, we could have chosen to have only $\wscell$ and $\wgcell$, but pay the price of complicating \while-to-\gil to pre-emptively remove the resource that $\wscell$ produces and re-produce the resource that $\wgcell$ consumes, cluttering the presentation.

\begin{minipage}{\textwidth}
{\small
\begin{mathpar}
\inferrule[C-SetCell]
  {
     (\loc, \prop) \not\in \domain(\cmem)
     \quad 
     \cmem' = \cmem \dunion \wcell{\loc}{\prop}{\vl}
  }{\action{\cmem}{\wscell}{\litlst{\loc, \prop, \vl}}{(\cmem', \true)}} 
  \quad
\inferrule[S-SetCell]
  {
   \memproj{\smem}{\sexp_l, \prop, \pc} = \emptyset
     \and 
      \smem' = \smem \dunion \wcell{\sexp_l}{\prop}{\sexp_v}
  }{\action{\smem}{\wscell}{\litlst{\sexp_l, \prop, \sexp_v}, \pc}{\litset{(\smem', \true, \true)}}} 
\end{mathpar}}
\end{minipage}

\smallskip
In summary, we have that $\corePreds_{\wsub} =  \lbrace \wpcell \rbrace$ and  $\actionswhile = \lbrace \lookupAction, \mutateAction, \disposeAction, \wscell, \wgcell \rbrace$ and proceed to state the property required for Theorem~\ref{theo:verification:preds} to hold for \while.

\begin{lemma}[\while: Core Predicate Interpretation]
The core predicate action interpretation $\tup{\corePreds_{\wsub}, \actionswhile, [\wpcell \mapsto \wscell], [\wpcell \mapsto \wgcell]}$ is well-formed 
with respect to $\cstateConstr(\wcmems)$ and $\sstateConstr(\wsmems)$.
\end{lemma}

\section{Parametric Bi-abductive Analysis}
\label{sec:bi}

Bi-abductive symbolic analyses are quickly becoming the gold-standard of automatic 
compositional testing~\cite{ohearn:lics:2018}. These analyses allow developers to find bugs in their code even when no 
specifications or (unit) tests are given. 
The foremost example of this is Infer~\cite{calcagno2015moving}, a fully automatic compositional tool aimed at 
lightweight bug-finding for static languages (C, C++, Java, Objective C), which is part of the 
code review pipeline at Facebook.

 Bi-abductive analyses symbolically analyse one procedure at a time, relying on a built-in 
 mechanism for automatic inference of resource when the footprint of the command at
 hand is not part of the current state. 
We extend \gillian with a parametric bi-abductive analysis that only requires  
the parameter state model to expose a minimal mechanism for inferring missing 
resource, while the bi-abductive analysis keeps track of all the generated 
\emph{fixes} and combines them to generate a set of specifications
describing the behaviour of the analysed procedure up to a bound.   

The section is structured as follows: \S\ref{subsec:biabduction}  presents our parametric 
bi-abductive analysis; and \S\ref{subsec:biabduction:while}
shows how to instantiate it for the \while language.

\subsection{Parametric Bi-abductive Analysis}
\label{subsec:biabduction}

Instead of redesigning a bi-abductive analysis from scratch, we maintain the semantics of \gil unaltered and, instead, instrument the parameter state model with a mechanism for on-the-fly 
correction of action errors during execution. 
This approach streamlines the formalism, avoiding redundancy, and leads to a modular 
implementation, which has no code duplication. 

\myparagraph{Action Fixes} 
The first step toward having automatic error correction is the appropriate modelling of 
action errors. To this end, we modify the signature of the action execution function, $\kwT{ea}$, 
exposed by every state model, in order to 
account for the cases in which the resource required by the parameter action, $\act$,
is not part of the given state. 
\begin{itemize}
   \item $\kwT{ea} : \actions \tmap \sset{\gstate} \tmap \gval \pmap \power{\sset{\gstate} \times \gval} \dunion \power{\Passes{\corePreds}}$ 
            \hfill ($(\st', \vl') \in \kwT{ea}(\act, \st, \vl) \equiv_{pp} \succAction{\st}{\act}{\gvl}{\st', v'}$) \\ 
            $\phantom{xx}$ \hfill  ($\Pass \in \kwT{ea}(\act, \st, \vl) \equiv_{pp} \failAction{\st}{\act}{v}{\Pass}$)
\end{itemize}
We write $\succAction{\st}{\act}{v}{\st', \gvl'}$ to mean that $\act$ terminates successfully 
when executed on $\st$, generating the state $\st'$ and the value $\gvl'$, and
$\failAction{\st}{\act}{v}{\Pass}$ to mean that the state model 
was not able to execute the action $\act$ due to not having the required resource. 
The assertion $\Pass$, which we refer to as the action \emph{fix}, describes a possible way 
of completing the original state so that the action can be performed successfully. 
Unsurprisingly, fixes are assumed to work\footnote{This is for the user of 
\gillian to prove.}; meaning that, if we extend the original 
state with the generated fix, it must be possible to execute the action successfully. 
This constraint is formally described by the equation below:
\begin{equation}
\label{eq:bi-abd}
\failAction{\st}{\act}{\gvl}{\Pass} \gand \actionSetterS{\st}{}{\Pass}{\st''}
  \implies \exists \, \st', \gvl'.  \succAction{\st''}{\act}{\gvl}{\st', \gvl'}
\end{equation}

We note that, since we model fixes using assertions, the underlying parameter state model must have support for predicates. Additionally, only the fixes that talk about logical variables that existed in the initial state are admissible.

\myparagraph{Bi-Abductive Analysis}
In order to define the bi-abductive analysis, we instrument the parameter state model
so that every time an action generates a fixable error, the corresponding fix is 
applied and the execution continues.
The bi-abductive execution also keeps track of all the applied 
fixes to allow for the generation of procedure specifications from from 
bi-abductive execution traces. 

Given a state model with support for predicates $\gstate = \tup{\sset{\gstate}, \gval, \actions}$, we construct 
a new state model with support for automatic error fixing by coupling 
$\sset{\gstate}$ with an assertion $\Pass \in \Passes{\corePreds}$, representing the 
fixes that were generated so far. Every time an action generates an error, we extend the current state 
$\st \in \sset{\gstate}$ with the footprint of the corresponding fix and continue with the execution. 
The lifting is formally described below. We write $\upcast{\vl}$ to denote the upcast 
of a value to a boolean expression.

\begin{definition}[BiState Constructor (\bistateConstr)]\label{def:bi:lifting}
The bi-abductive state constructor $\bistateConstr : \gstates \tmap \gstates$ is defined as  
$\bistateConstr(\tup{\sset{\gstate}, \gval, \actions}) \semeq \tup{\sset{\gstate'}, \gval, \actions}$,
where: 

\medskip
{\small
\begin{tabular}{lll}
   \mybullet $\sset{\gstate'}$ &$\semeq$ & $\sset{\gstate} \times \Passes{\corePreds}$ \\ 
  \mybullet $\kwT{setVar}_{\bisub}(\tup{\st, \Pass}, \x, \gvl)$ 
       & $\semeq$ & 
       $\tup{\kwT{setVar}(\st,  \x, \gvl), \Pass}$ \\ 
  \mybullet $\kwT{setStore}_{\bisub}(\tup{\st, \Pass}, \sto)$
     & $\semeq$ & 
     $\tup{\kwT{setStore}(\st, \sto), \Pass}$ \\ 
  \mybullet $\kwT{store}_{\bisub}(\tup{\st, -, -})$
    & $\semeq$ & 
    $\kwT{store}(\st)$ \\
  \mybullet $\kwT{ee}_{\bisub}(\tup{\st, -, -}, \e)$
    & $\semeq$ & 
    $\kwT{ee}(\st, \e)$ \\
 \mybullet $\kwT{ea}_{\bisub}(\act, \tup{\st, \Pass}, \gvl)$
    & $\semeq$ & 
    $\left\{ (\tup{\st', \Pass}, \gvl') \mid \succAction{\st}{\act}{\vl}{\st', \gvl'} \right\}$ if $\act \neq \kwT{assume}$\\ 
 \mybullet $\kwT{ea}_{\bisub}(\kwT{assume}, \tup{\st, \Pass}, \gvl)$
    & $\semeq$ & 
    $\left\{ (\tup{\st', \Pass * \pc}, \gvl') \mid \upcast{\vl} = \pc \gand \succActionT{\st}{assume}{\vl}{\st', \gvl'} \right\}$\\ 
 \mybullet $\kwT{ea}_{\bisub}(\act, \tup{\st, \Pass}, \gvl)$
    & $\semeq$ & 
         $\left\{{\begin{array}{l}
               \tup{\st'', \Pass * \Qass}, \gvl') \\
               \qquad \mid
               \failAction{\st}{\act}{\vl}{\Qass} \gand 
               \actionSetterS{\st}{\Qass}{\zsubst}{\st'} \gand
               \succAction{\st'}{\act}{\vl}{\st'', \gvl'}
          \end{array}}\right\}$ 
\end{tabular}}
\end{definition}

Finally, Theorem~\ref{theo:bi:lifting} connects successful
bi-abductive executions to successful parameter-state executions. 
Given a bi-abductive execution $\vsemtrans{\tup{\st, \litlst{~}}, \cs, i}{\tup{\st, \Pass}, \cs', j}[*]{\bistateConstr(\gstate)}{\vprog}{\outcome}$, 
we construct a parameter-state execution by extending its initial state $\st$ with the missing resource computed during the
bi-abductive execution, described by $\Pass$. 

\begin{theorem}[Bi-abduction]\label{theo:bi:lifting} 
$$
\begin{array}{l}
\vsemtrans{\tup{\st, \emp}, \cs, i}{\tup{\st', \Pass}, \cs', j}[*]{\bistateConstr(\gstate)}{\prog}{\outcome}
  \gand 
  \actionSetterS{\st}{}{\Pass}{\st''} 
   \implies 
  \vsemtrans{\st'', \cs, i}{\st', \cs', j}[*]{\gstate}{\prog}{\outcome}
\end{array}
$$
\end{theorem}


\begin{wrapfigure}{R}{0.34\textwidth}
\vspace*{-0.47cm}
\begin{minted}[fontsize=\scriptsize]{ocaml}
module BiState
  (St : State)
  (CP : CorePred with type a = St.a) :
    (State with type a = St.a) = struct
      type vt = St.vt
      type t = St.t * CP.t Asrt.t
      type a = St.a
      ...
 end
\end{minted}
\vspace*{-0.3cm}
\end{wrapfigure}

\myparagraph{Implementation}
The bi-abductive state constructor is implemented as an OCaml functor parametric on 
an OCaml module of type \texttt{State} together with a module of type \texttt{CorePred}. 
The \texttt{BiState} functor precisely follows 
the lifting described in Definition~\ref{def:bi:lifting}. 
It is straightforward to check that the carrier type of the bi-abductive state, \texttt{t}, 
matches the carrier sets of the \emph{lifted} state, $\sset{\gstate} \times \Passes{\corePreds}$. 
The \texttt{CorePred} module is required for extending the parameter state with the footprint 
of the generated fixes during the bi-abductive~analysis. 

\subsection{\while: Bi-abductive Analysis}
\label{subsec:biabduction:while}

Let us recall the actions of the \while: $\actionswhile = \lbrace \lookupAction, \mutateAction, \disposeAction, \wscell, \wgcell \rbrace$. Out of those, the two that may fail and can be fixed are $\lookupAction$ and $\wgcell$; $\mutateAction$ will always succeed, $\wscell$ may fail because the resource to be set could still be in the heap, but this cannot be corrected; and $\disposeAction$ may fail because the object might not be in the heap, but this also cannot be corrected. For $\lookupAction$ and $\wgcell$, the only cause of error is the absence of the looked-up property in the heap, and the fix is, simply, to add it. We give the fix for $\lookupAction$; the fix for $\wgcell$ is similar.
\begin{mathpar}
\inferrule[S-Lookup - Missing Prop]
  {
      \memproj{\smem}{\sexp, \pc} = (\smem', -)  
      \and 
      \smem' = \uplus_{i=0}^n (-.\prop_i \mapsto -)
      \and 
      \pc \vdash \prop \not\in \lbrace \prop_i\!\mid_{i=0}^n \rbrace
      \and 
      \lx \text{ fresh}
  }{\action{\smem}{\lookupAction}{\litlst{\sexp, \prop}, \pc}{\rfail{\wpcell\langle[\sexp, \prop, \lx]\rangle}}} 
\end{mathpar}

The following lemma is easily shown to hold by case analysis on the fixes.

\begin{lemma}[\while: Correctness of Fixes]
The fixes of \while satisfy Equation~(\ref{eq:bi-abd}).
\end{lemma}

%



\section{\gillian in the real world: JavaScript and C}
\label{sec:rw}

We instantiate the \gillian framework to obtain analysis tools for JavaScript and C, two real-world, widely-used programming languages. For each language, we present the concrete and symbolic memory models along with one of their actions, and discuss the   challenges encountered in the process. We highlight the trustworthiness of the obtained analyses: we compile JavaScript to \gil by linking to JS-2-JSIL~\cite{javert}, a thoroughly tested compiler for JavaScript; and we compile C to \gil by linking to CompCert~\cite{ccert1,ccert2}, the first verified C compiler. We evaluate the obtained tools by successfully performing whole-program symbolic testing, full verification, and automatic compositional testing on a series of data-structure libraries, including binary search trees, key-value maps, priority queues, and singly-and doubly-linked lists, purposefully written using the programming idioms specific to the two languages. This demonstrates that \gillian can be used to quickly obtain analysis tools sufficiently robust to reason about complex real-world languages.

\subsection{\gillian-JS: Trustworthy Analysis of ES5 Strict Programs}

\newcommand\sexprl{\sexp_\loc}
\newcommand\sexprp{\sexp_p}
\newcommand\heap[0]{h}
\newcommand\sheap[0]{\hat{h}}
\newcommand\dom{d}
\newcommand\sdom{\hat{d}}
\newcommand{\none}{\diameter}
\newcommand{\pointsto}{\mapsto}
\newcommand{\project}[2]{#1\!\triangleleft\!#2}

\myparagraph{The JavaScript Memory Model} We inherit the concrete and symbolic memory models of JavaScript from~\citet{javert2}, adapting them slightly to the setting of~\gillian. 

Concrete JavaScript memories, $\cmem \in \jscmems$, consist of a concrete heap and a concrete domain table. The concrete heap, $\heap : \locs \times \strs \pmap \vals$, maps object locations (modelled as symbols) and property names (modelled as strings) to \gil values. This is similar to the \while semantics except that, in the values, we designate a distinguished symbol $\none$ (read: none) to denote property absence: that is, if $\heap(\loc, \prop) = \none$, then the object at $\loc$ does not have property $\prop$. This ability to explicitly address the absence of properties is required because JavaScript has extensible objects.
The concrete domain table, $\dom : \locs \pmap \power{\strs}$, maps objects to the sets of properties that they may have: that is, if $\prop \notin \dom(\loc)$, then the object located at $\loc$ is guaranteed not to have property~$\prop$. This, together with the use of $\none$, ensures that we have a semantics of JavaScript that respects the frame property~\cite{reynolds:lics:2002}. 

Symbolic JavaScript memories, $\smem \in \jssmems$, consist of a symbolic heap and a symbolic domain table. The symbolic heap, $\sheap : \sexps \times \sexps \pmap \sexps$, maps pairs of logical expressions (property names are also modelled as logical expressions, as JavaScript has dynamic property access) to logical expressions. 
The signature of the symbolic domain table, analogously, is: $\sdom : \sexps \pmap \sexps$.

We note that heaps and domain tables are connected both in the concrete and in the symbolic semantics via the heap-domain invariant: for a given location, if its domain is defined, then all of the properties of that object that are in the heap must also be in its domain.

Below, we present a rule for one of the actions of the memory model, $\jsgetcell$, which is meant to receive an object location and a property and retrieve the value of the property. This rule illustrates a branching action, which passes the constraint $\sexprp = \sexp_i$ back into the state after understanding that we have full knowledge about the object in question (meaning that the properties of the object in the heap, collected by $\memproj{\sheap}{\sexprl}$, coincide with its domain) and that the looked-up property~$\sexprp$ may be equal to one of its properties. 

\begin{minipage}{\textwidth}
{\footnotesize
\vspace*{0.1cm}
\begin{mathpar} 
   %
      \inferrule[\textsc{SGetProp - Branch - Found}]
   { 
       \smem = (\sheap, \sdom) 
		\quad
		\left(\pc \vdash \sexprl = \sexprl' \gand \sdom(\sexprl') = \memproj{\sheap}{\sexprl'}\right)
       \quad
       \pc \gand (\sexprp = \sexp_i) \text{ \sat}       
       \quad
       \sheap = \_ \dunion (\wcell{\sexprl'}{\sexp_i}{\sexp}) 
   }{  \action{\smem}{\jsgetcell}{\litlst{\sexprl, \sexprp}, \pc}{(\smem, \sexp, \sexprp = \sexp_i)}}
   %
 \end{mathpar}
 }
 \end{minipage}

\myparagraph{Implementation} We take the implementations of the concrete and symbolic memory models of JavaScript (ES5 Strict) from JaVerT 2.0 and plug them directly into \gillian, with minor modifications in the symbolic memory model related to differences in design in the fixes of the bi-abduction. Additionally, we compile the runtime environment of JaVerT 2.0, which consists of JSIL implementations of the internal and built-in functions of ES5 Strict, to \gil using a JSIL-to-GIL compiler. 

\myparagraph{Trustworthiness} 
As JS-2-JSIL is a well-tested, trusted compiler from ES5 Strict to JSIL, the trustworthiness of the analysis of \gillian-JS rests on the correctness of JSIL-to-GIL. 
However, JSIL-to-GIL is very simple, as the two languages have similar constructs, with the only noteworthy difference being that JSIL has commands that directly work with objects, whereas \gil has the actions. 
To gain confidence in the correctness of JSIL-to-GIL, we have replayed and passed the 8797 concrete tests of the Test262 official JavaScript test suite~\cite{test262} also passed by JS-2-JSIL.

\subsection{\gillian-C: Towards Certified Verification of C Programs}

\newcommand{\Cmem}{\mathtt{mem}}
\newcommand{\Cmems}{\ensuremath{M_{\mathbb C}}}
\newcommand{\Csmems}{\ensuremath{\widehat{M}_{\mathbb{C}}}}
\newcommand{\Cperm}{\ensuremath{\mathcal{P}\mkern-1.5mu\mathit{erm}}}
\newcommand{\Csper}{\ensuremath{\widehat\Cperm}}
\newcommand{\Cblock}{\ensuremath{\mathcal{B}\mathit{lock}}}
\newcommand{\Cbyte}{\ensuremath{\mathcal{B}\mkern-1.5mu\mathit{yte}}}
\newcommand{\Csbyt}{\ensuremath{\widehat\Cbyte}}
\newcommand{\Cchunk}{\ensuremath{\mathtt{memory\_chunk}}}
\newcommand{\Cval}{\ensuremath{\mathtt{val}}}
\newcommand{\Cmval}{\ensuremath{\mathit{mv}}}
\newcommand{\Csv}{\ensuremath{\mathit{sv}}}
\newcommand{\Cmvals}{\ensuremath{\vals}}
\newcommand{\Smval}{\ensuremath{\hat{\mathit{mv}}}}
\newcommand{\Smvals}{\ensuremath{\sexps}}
\newcommand{\ptable}{\ensuremath{\mathit{pt}}}
\newcommand{\sptable}{\ensuremath{\widehat{\mathit{pt}}}}
\newcommand{\offset}{\mathit{off}}

\newcommand{\cload}{{\sf load}}
\newcommand{\mchunk}{\mathit{mch}}

\newcommand{\csize}{\mathit{size}}
\newcommand{\calignment}{\mathit{alignment}}
\newcommand{\ctype}{\mathit{type}}
\newcommand{\cdecsymb}{{\sf decodeSymb}}

\myparagraph{The C Memory Model} We inherit the concrete memory model of C from CompCert~\cite{compcertmem}, with minor adaptations related to encoding C values in \gil. We define a symbolic memory model, deriving it from the concrete model and using ideas from CompCertS~\cite{CompCertS}, a memory model for CompCert that uses symbolic values to improve precision. For clarity, we elide the details from the CompCert memory model related to concurrency.

Concrete C memories, $\cmem \in \Cmems$, consist of a concrete heap and a concrete permission table. The concrete heap, $\heap : \locs \times \ints \pmap \Cmvals$, maps memory locations (modelled as symbols) and offsets (modelled as integers) to concrete C memory values. A C memory value, $\Cmval \in \Cmvals$, is either a byte (integer in the range [0, 255]) or a three-element \gil list $[l, \offset, k]$, denoting the $k^{\mathrm{th}}$ byte of the pointer to location $l$ with offset $\offset$.
  The concrete permission table, $\dom : \locs \times \ints \pmap \ints$, maps memory locations and offsets to their permissions, which indicate the allowed operations on the associated cell. We model permissions as integers from 1 to 4, representing, respectively, $\mathtt{Nonempty}$, $\mathtt{Readable}$, $\mathtt{Writable}$, and $\mathtt{Freeable}$, where $\mathtt{Nonempty}$ indicates that the cell only admits pointer comparison.


  Symbolic C memories, $\smem \in \Csmems$, consist of a symbolic heap and a symbolic permission table, which have the same meaning as in concrete memories.
  Symbolic heaps, $\sheap : (\sexps \times \sexps) \pmap \Smvals$, model locations and offsets using logical expressions, similarly to \while and JavaScript. Symbolic memory values, $\Smval \in \Smvals$, are three-element lists, $[\sexp, k, n]$, denoting the $k^{\mathrm{th}}$ out of $n$ bytes of the C value represented by $\hat{e}$. Luckily, we always \emph{statically} know the size of a value, meaning that $k$ and $n$ will always be concrete. The signature of symbolic permission tables, expectedly, is $\sptable : \sexps \times \sexps \pmap \sexps$.

\begin{wrapfigure}{R}{0.5\textwidth}
\vspace*{-0.5cm}
\begin{minipage}{0.5\textwidth}
 {\footnotesize
\begin{mathpar} 
      \inferrule[\textsc{SLoad - Valid Access}]
   { 
    \smem = (\sheap, \sptable)
	\quad
	\pc \vdash \sexprl = \sexprl' \gand \sexp_o = \sexp_o'
	\\\\
	\pc \vdash \sexp_o'~\text{mod}~\calignment = 0
       \\\\
       (\pc \vdash \sptable(\sexprl', \sexp_o + i) \geq \mathtt{Readable})|_{i=0}^{\csize-1}
       \\\\
      (\sheap(\sexprl', \sexp_o + i) = [\sexp', i, \csize - 1])|_{i=0}^{\csize-1}
      \\\\
      \sexp = \cdecsymb(\sexp', \ctype)
   }{\action{\smem}{\cload}{\litlst{[\csize, \calignment, \ctype], \sexprl, \sexp_o}, \pc}{(\smem, \sexp, \true)}}
   %
 \end{mathpar}
 }
 \end{minipage}
  \vspace*{-0.5cm}
  \caption{Symbolic $\cload$~Action of C (excerpt)}\label{fig:cload}
 \vspace*{-0.3cm}
\end{wrapfigure}

In Figure~\ref{fig:cload}, we present one symbolic rule for the $\cload$ action, which 
retrieves a value from the memory. Note that, when \textit{loading} or \textit{storing} a value, a memory chunk has to be provided to indicate the size, alignment, and type of what should be read in the memory. For clarity, we present chunks as three-element lists, $\mchunk = [\csize, \calignment, \ctype]$. The $\cload$ function receives a memory chunk, the location, and the offset. First, it ensures that the value is correctly aligned and that it is allowed to be read. Next, it confirms that the read part of the memory represents the symbolic value $\sexp'$. Finally, it decodes $\sexp'$ using its type. The decoding, for example, understands if the result should be an integer or a floating-point, and of which precision.

\myparagraph{Implementation}
We import the concrete memory model of CompCert into \gillian by directly plugging in the appropriate OCaml module, extracted from the Coq development; the symbolic memory model we implement ourselves. We also implement directly in GIL the functions describing various internals of C, such as, for example, unary and binary operators, \textit{malloc} and \textit{free}. We note that, currently, \gillian-C assumes that, whenever memory is dynamically allocated, the size of the allocated chunk is known. This allows us to focus on making the analyses work and leave the complex layer of reasoning about maps with unknown domains for immediate future work.

\myparagraph{Trustworthiness}
We compile C to \gil via \csm (read: C-sharp-minor), one of the intermediate representations of CompCert. Therefore, similarly to \gillian-JS, the correctness of compilation for \gillian-C rests on the correctness of C-to-\csm and our C\#m-to-\gil compiler. However, we have a stronger correctness guarantee for C-to-\csm than for JS-2-JSIL, as this compilation step has been already verified in Coq as part of CompCert. Moreover, as \csm and \gillian-C share the same memory model by design, we could formalise \gil in Coq and re-use the proof techniques of CompCert to obtain a fully certified C-to-\gil compiler.

\subsection{Evaluation: Data-Structure Libraries}

We evaluate \gillian-JS and \gillian-C on data-structure libraries: binary search trees, key-value maps, priority queues, singly-and doubly-linked lists, and sorted lists. To demonstrate that \gillian can handle the complexity of the memory models of JavaScript and C, we write these libraries using the programming idioms specific to the two languages: for JavaScript, we use prototype inheritance and function closures; for C, we use pointers and structures. 

\begin{wraptable}{r}{0.53\textwidth}
{
\vspace*{-0.48cm}
\caption{Whole-program Symbolic Testing:\\Gillian-JS (left); Gillian-C (right)}
\label{res:wpst}
\small
\centering
\setlength\tabcolsep{3pt}
\vspace*{-0.2cm}
\begin{tabular}{lcc>{\bfseries}r}
\toprule
Name & \makecell{\#T} & \makecell{\gil\\cmds} & \makecell{Time} \\
\midrule
\texttt{BST}   &  6 & 137,379 & 1.36s \\
\texttt{KVMap} &  3 &  67,216 & 0.35s \\
\texttt{PriQ}  &  6 &  61,338 & 0.42s \\
\texttt{SLL}   &  6 &  28,389 & 0.21s \\
\texttt{DLL}   &  6 &  30,929 & 0.21s \\
\texttt{SL}    &  6 &  37,251 & 0.42s \\
\bottomrule
\end{tabular}
~
\begin{tabular}{lcc>{\bfseries}r}
\toprule
Name & \makecell{\#T} & \makecell{\gil\\cmds} & \makecell{Time} \\
\midrule
\texttt{BST}      &  5 & 15,256 & 0.33s \\
\texttt{KVMap}    &  1 & 1,165  & 0.13s \\
\texttt{PriQ}     &  4 & 1,008  & 0.13s \\
\texttt{SLL}      &  3 & 4,218  & 0.18s \\
\texttt{DLL}      &  5 & 7,000  & 0.22s \\
\texttt{SL}       &  3 & 2,688  & 0.11s \\
\bottomrule
\end{tabular}
}
\end{wraptable}

\myparagraph{Whole-Program Symbolic Testing}
We write symbolic tests with the goal of achieving full line coverage for each of the data structures. The results are given in Table~\ref{res:wpst} and they include, for each  data structure: the number of tests, the number of executed \gil commands, and the obtained times.

The results indicate that whole-program symbolic testing can scale to  larger codebases. The obtained times for \gillian-JS are comparable to those of JaVerT 2.0~\cite{javert2}, with the number of commands run being higher, as \gil is more low-level than JSIL. The analysis of \gillian-C demonstrates the difference in complexity between the JavaScript and the C semantics: the large number of executed commands for \gillian-JS stems from the repeated execution of numerous internal functions, of which C has far fewer. It also shows that the core of the obtained time does not lie with the number of executed commands, but rather with the time spent in the first-order solver, as highlighted by the BST example, which yields more complex entailments  than the other~examples.

\myparagraph{Verification} 
We specify and verify the data structures. For JS, we manually write the  required predicates, using the specification techniques of JaVerT~\cite{javert}. For C, we use the type information to automatically generate recursive predicates describing structures in memory, which we then extend with required information about the values. The results are shown in Table~\ref{res:verif}. For each data structure, we give: the number of verified functions, the total number of specifications for those functions, the number of executed \gil commands, and the verification time. 

\begin{wraptable}{r}{0.58\textwidth}
{
\caption{Verification: Gillian-JS (left); Gillian-C (right)}
\label{res:verif}
\small
\centering
\vspace*{-0.2cm}
\setlength\tabcolsep{3pt}
\begin{tabular}{lccc>{\bfseries}r}
\toprule
Name & \makecell{\#F} & \makecell{\#S} & \makecell{\gil\\cmds} & 
\makecell{Time} \\
\midrule
\texttt{BST}    & 5 &  5 &  9,620 & 2.26s \\
\texttt{KVMap}  & 4 &  9 & 11,284 & 0.75s \\
\texttt{PriQ}   & 6 &  9 & 12,823 & 1.38s \\
\texttt{SLL}    & 3 &  3 &  2,978 & 0.61s \\
\texttt{DLL}    & 3 &  3 &  4,220 & 0.82s \\
\texttt{SL}     & 2 &  2 &  3,152 & 0.42s \\
\bottomrule
\end{tabular}
\begin{tabular}{lccc>{\bfseries}r}
\toprule
Name & \makecell{\#F} & \makecell{\#S} & \makecell{\gil\\cmds} & 
\makecell{Time} \\
\midrule
\texttt{BST}      & 5 &  6 & 3,968 & 5.57s \\
\texttt{KVMap}    & 3 &  5 &   733 & 0.99s \\
\texttt{PriQ}     & 5 &  9 & 1,161 & 1.08s \\
\texttt{SLL}      & 6 &  6 & 1,201 & 0.17s \\
\texttt{DLL}      & 5 &  5 & 2,980 & 0.39s \\
\texttt{SL}       & 4 &  4 &   702 & 0.82s \\
\bottomrule
\end{tabular}
}
\end{wraptable}

We can observe that, when compared to whole-program symbolic testing, the number of executed commands is dramatically lower, but the obtained times are longer. 
This is in line with the facts that whole-program symbolic testing does not use procedure summaries during execution and that verification involves considerable predicate manipulation.
We also note that obtained times for \gillian-JS are marginally slower than those of JaVerT 2.0. This is expected, given the more general nature of \gillian.  
The verification times for \gillian-C are fast across the board, with the exception of binary search trees, which, as mentioned earlier, yield complex entailments that Z3 has to solve.

\begin{wraptable}{r}{0.5\textwidth}
{
\vspace*{-0.45cm}
\caption{{Bi-abduction}: Gillian-JS (left); Gillian-C (right)}
\label{res:act}
\small
\centering
\setlength\tabcolsep{3pt}
\vspace*{-0.2cm}
\hspace*{0.2cm}\begin{tabular}{lc>{\bfseries}r}
\toprule
Name & \makecell{S/E/B\\specs} & \makecell{Time}
 \\
\midrule
\texttt{BST}      & 26/0/5   & 3.98s \\
\texttt{KVMap}    &  9/12/12 & 1.57s \\
\texttt{PriQ}     & 15/1/12  & 3.23s \\
\texttt{SLL}      &  6/0/2   & 0.78s \\
\texttt{DLL}      &  9/0/3   & 1.04s \\
\texttt{SL}       &  8/0/2   & 1.09s \\
\bottomrule
\end{tabular}
\quad
\begin{tabular}{lc>{\bfseries}r}
\toprule
Name & \makecell{S/B\\specs} & \makecell{Time} 
 \\
\midrule
\texttt{BST}      & 37/8 & 0.37s \\
\texttt{KVMap}    & 9/0  & 0.13s \\
\texttt{PriQ}     & 19/3 & 0.21s \\
\texttt{SLL}      & 20/3 & 0.16s \\
\texttt{DLL}      & 37/7 & 0.24s \\
\texttt{SL}       & 12/0 & 0.15s \\
\bottomrule
\end{tabular}
}
\end{wraptable}

\myparagraph{{Bi-abduction}} We use bi-abduction to automatically create specifications for the data structures. We note that these specifications describe the behaviour of the up to a given bound. 
The results are presented in Table~\ref{res:wpst}. For both languages, we give the number of success and bug specs found (S/B), and the time required to find them. For JavaScript, we additionally include the number of error specs (as JavaScript programs can terminate with a user-thrown error). 
It is important to clarify that, for JavaScript, a bug spec means that the program terminated with a native error (e.g. a TypeError or ReferenceError); for C, a bug spec means that the program has thrown an exception (e.g.~a segmentation fault). 
To obtain bug specs for C, we introduce errors into the code; for JavaScript, the lack of typing information in the language is sufficient on its own.
Finally, we note that the obtained specifications may describe behaviour outside of the use cases of the function and could contain false positive bug reports that need to be treated afterwards either automatically or by the developer. 

As for verification, the obtained times for \gillian-JS are slightly slower than those of JaVerT 2.0 due to \gillian being a general framework. We note that they are still quite long, as we have not yet solved the bi-abduction issues of JaVerT 2.0 related to the internal branching of the JavaScript semantics, and view this to be beyond the scope of this paper. The bi-abduction of \gillian-C, on the other hand, yields promising results, as it is able to detect the introduced bugs in very quick times.


\section{Related Work}
\label{sec:relwork}

There is a wide range of works on symbolic analyses for C and JavaScript, such as~\cite{jsai,park2015scalable,jensen:sas:2009,verasco,botincan:entcs:2009}. 
As this paper focusses on the design of \gillian and its associated meta-theory, we centre our discussion on parametric frameworks for obtaining modular symbolic analyses in general, {covering the fields of} symbolic execution, abstract interpretation, and logic-based analysis and verification. 

\newcommand{\rosette}[0]{\textsc{Rosette}\xspace}
\newcommand{\chef}[0]{\textsc{Chef}\xspace}
\newcommand{\racket}[0]{\textsc{Racket}\xspace}
\newcommand{\cosette}[0]{\textsc{Cosette}\xspace}
\newcommand{\javertwo}[0]{\textsc{JaVerT 2.0}\xspace}
\myparagraph{Symbolic Execution}
Building and maintaining new symbolic execution engines for real-world programming languages is 
known to be a daunting task~\cite{survey:acm,survey:ecse}. 
Researchers in the field have, therefore, tried to automate this process, with tools such 
as  \rosette \cite{torlak:pldi:2014,torlak:onward:2013} and \chef~\cite{bucur:asplos:2014},
%
which propose two different mechanisms to automatically lift 
a user-provided vanilla concrete interpreter for a given target language to a fully-fledged symbolic execution engine for the same language. 
%
%

More concretely, \rosette is an extension of \racket~\cite{racket} which additionally provides a set of solver-aided facilities for creating 
symbolic values and expressing constraints on those values. With \rosette, the concrete interpreter 
of the target language is written directly in \racket and is then \emph{symbolically interpreted} using 
\rosette's core symbolic execution engine.  
In contrast, \chef takes a specially-packaged interpreter as input and executes the target language 
programs symbolically by symbolically executing the interpreter's \emph{binary}.

Importantly, neither \rosette nor \chef can scale to real-world programming languages. 
\rosette is aimed at domain-specific languages with restricted expressive power, while \chef is limited to languages with 
``moderately-sized'' interpreters. 
Interestingly, by implementing a JavaScript symbolic execution engine 
natively in OCaml~\cite{javert2}, 
instead of running a concrete interpreter on top of \rosette~\cite{cosette}, 
Fragoso Santos et al. gained a performance speed-up of two orders of magnitude.  

\myparagraph{Abstract Interpretation}
When it comes to general abstract interpreters, we identify two main strands of work: 
those based on small-step semantics and those based on big-step semantics. 

\vspace*{-0.2cm}
\paragraph{Small-Step Abstract Interpreters} 
Might~\cite{might:sas:2010} proposes a \emph{methodology} for automatically deriving a family 
of sound, computable abstract interpreters from a given concrete interpreter written in 
small-step style. This methodology assumes an initial Galois connection~\cite{cousot:popl:1977}
between the concrete and abstract domains in order to guarantee the \emph{optimality} of the derived abstract interpreters. 
Later, Might and others~\cite{varhorn:icfp:2010,vanhorn:jfuncprogram:2012} 
apply the proposed methodology to a family of abstract machines, obtaining  
the so called \emph{abstract abstract machines}, which include expressive programming 
language features, such as: first-class control, exception handling, and state. 
This technique is further refined in~\cite{sergey:pldi:2013}, where the authors 
show how to derive \emph{monadically}-parameterised abstract interpreters from  
concrete interpreters. The additional level of parameterisation is used to 
capture in a single unified formalism a number of different styles of program analyses, 
making it easy to instrument an analysis with parameterisable strategies for improving 
precision and performance. 
In~\cite{darais:ooplsa:2015}, the authors propose a library of
\emph{Galois tranformers} to streamline the construction of such strategies. 
The appeal of this approach is that each Galois transformer can be proved 
sound once and for all, making it much easier to prove the soundness of the 
obtained analysis.

We observe that, despite offering a general \emph{methodology} for designing 
abstract interpreters, none of the works mentioned above automates that methodology, 
in that one always has to \emph{manually} follow it to obtain an abstract 
interpreter for a given language. 
We further note that none of these techniques has been applied to real world 
programming languages, such as JavaScript and~C. 

\vspace*{-0.2cm}
\paragraph{Big-Step Abstract Interpreters} 
In~\cite{schmidt:sas:1995}, Schmidt presents a general approach for 
designing abstract interpreters based on co-inductively defined big-step 
semantics. 
Following similar ideas, \citet{bodin:popl:2019} establish a general 
framework, the \emph{skeletal semantics}, for developing 
concrete and abstract big-step semantics, connected with a general consistency 
result, leaving the user to prove a number of simple language-dependent lemmas.
This work, however, has only been applied to a simple While 
language with no heap, making its broader applicability 
difficult to assess. 

\newcommand{\corestar}{\textsc{CoreStar}\xspace}
\newcommand{\silver}{\textsc{Silver}\xspace}
\newcommand{\iris}{\textsc{Iris}\xspace}
\newcommand{\viper}{\textsc{Viper}\xspace}
\myparagraph{Logic-based Analysis and Verification}
$\mathbb{K}$~\cite{rosu:jlap:2010} is a language-independent verification infrastructure instantiated to several real-world languages, such as Java, JavaScript, and C~\cite{bogdanas:kjava,park:pldi:2015,hathorn:pldi:2015,stefanescu:oopsla:2016}, and evaluated on data-structure libraries similar to those of \gillian. As in \gillian, the user of $\mathbb{K}$ gets the verification guarantee for free, by construction. \gillian, however, has several advantages over $\mathbb{K}$: \gillian specifications are language-tailored; it supports compositional analyses; and it is faster. For example, the functions of the BST example are verified in $\mathbb{K}$ for C in 88.5 seconds and for JavaScript in 25.5 seconds; we verify the same functions in approx.~8 seconds for both languages.

\corestar~\cite{botincan:boogie:2011}, was envisioned as a general back-end for tools based on separation logic, where the user would encode the assertions of their language as abstract predicates and provide CoreStar with a separation algebra describing the entailments specific to those predicates. However, CoreStar does not appear to be fit for reasoning about  highly complex real-world dynamic languages, such as JavaScript. For instance, in JaVerT~\cite{javert}, the authors obtained prohibitive performance even for simple examples.

\viper~\cite{muller:dsse:2017,summers:vmcai:2016} is also a verification framework designed to 
serve as a back-end for tools for permission-based verification, such as separation-logic-based tools. 
However, \silver, the intermediate language of \viper, is not parameterisable. Hence, in order to use 
\viper, users have to encode their memory models in the memory model of \silver. 
We believe that this approach does not scale well when there is a big mismatch between the memory model 
to encode and that of the host language, in this case, \silver. 

\iris~\cite{jung:popl:2015,jung:jfunc:2018} is a Coq-based framework for reasoning about the safety of 
concurrent programs. It provides a general methodology for designing new 
sound concurrent program logics, as users can encode their own program logics 
into Iris and leverage Iris's soundness result to prove the soundness of their 
logics. Iris is, however, mainly aimed at the mechanisation of 
meta-theory, while the main goal of \gillian  is to streamline the development 
of analysis tools. 

JaVerT 2.0~\cite{javert2} is a tool for compositional analysis of JavaScript programs, Similarly to \gillian, it supports whole-program symbolic testing, full verification, and
bi-abduction. While the analysis of JaVerT 2.0 is structured modularly, all of its meta-theoretical results, as well as its implementation,  are specific to its intermediate language and rely on the specific concrete and symbolic memory models of JavaScript. Gillian takes the highly non-trivial step of generalising both the theory and the implementation of JaVerT 2.0 to a fully language-independent setting.

\section{Conclusions and Further Work}
\label{sec:concls}

We have presented \gillian, a language-independent framework for the development of compositional symbolic analysis tools, and demonstrated that it can be used to reason about real-world programming languages. Thanks to its parametric meta-theory and modular implementation, \gillian can readily be used by developers to create analysis tools for their language of choice, be it toy, domain-specific, or real-world. We believe that \gillian will be of interest to a broad range of users wishing to obtain correctness guarantees for their code.

The avenues for further work on \gillian are numerous. First of all, our immediate next steps are to streamline the bi-abduction of \gillian-JS and extend the reasoning about the symbolic memory model of \gillian-C to include arbitrary dynamic memory allocation, taking inspiration for the latter from the work of~\cite{Kirchner2015} done for Frama-C.

We are also investigating ways of extending \gillian with support for reasoning about complex language features, such as events and concurrency, as well as with additional forms of analysis, such as concolic execution. The modular design of \gillian lends itself well to these purposes.

Moreover, the infrastructure of \gillian can be extended to support the analysis of systems running code written in multiple programming languages that inter-operate with each other. A noteworthy goal in this vein of research would be a joint analysis of JavaScript and WebAssembly~\cite{wasm}, the emerging low-level language for the Web.

Finally, we plan to improve \gillian's error reporting mechanisms and develop interactive tools, such as a trace visualiser and a code-stepper, for the debugging of programs analysed by \gillian, in order to make \gillian more accessible to a wider audience of interested users.

\newpage
\bibliography{gillian} 

\newpage
\appendix 

\newpage
\section{Section 2: Parametric Execution}
\label{app:s2}

\smallskip
\begin{display}{The Syntax of \gil}\label{app:gil:syntax}
\hspace*{-0.15cm}
\begin{tabular}{l@{\quad\hspace*{0.2cm}}l}
   $\vl \in \vals \defeq \num \in \nums \mid i \in \ints \mid \str \in \strs \mid \bool \in \bools \mid \varsigma \in \locs \mid \type \in \types \mid \fid \in \fids \mid \lst{\vl}$
   & 
   $\e \in \exprs \defeq \vl \mid \x \in \xs \mid \unop{\e} \mid \binop{\e_1}{\e_2}$
   \\[3pt]
    $\cm \in \cmds{\actions}$ $\defeq$ $\x := \e \mid \ifgoto{\e}{i} \mid \x := \e(\e') 
                                     \mid \x := \act(\e) \mid$ &    $\proc \in \procs{\actions}$ $\defeq$ $\procedure{f}{\x}{\lst{\cm}}$
 \vspace*{3pt}
    \\
    \hspace*{1.28cm}$\x := \symb{j} \mid \x := \fresh{j} \mid \return{\e} \mid  \fail{e} \mid \vanish $
    & 
    $\prog \in \progs{\actions}$ $:$ $\fids \pmap \procs{\actions}$
 \end{tabular}
\end{display}

\begin{definition}[State Model]\label{app:sm}
A state model $\gstate \in \gstates$ is a triple $\tup{\sset{\gstate}, \gval, \actions}$, consisting of: 
\etag{1} a set of states on which \gil programs operate, $\sset{\gstate}$, \etag{2} a set of values stored in those states, $\gval$, and \etag{3} a set of 
actions that can be performed on those states, $\actions$. A state model defines the following functions for acting on states ($\equiv_{pp}$ denotes pretty-printing for readability): 

\begin{itemize}
   \item $\kwT{setVar} : \sset{\gstate} \tmap \xs \tmap \gval \tmap \sset{\gstate}$ 
            \hfill ($\kwT{setVar}(\st, \x, \vl) \equiv_{pp} \st.\kwT{setVar}(\x, \vl)$) 
   \item $\kwT{setStore} : \sset{\gstate} \tmap (\xs \pmap \gval) \tmap \sset{\gstate}$ 
            \hfill ($\kwT{setStore}(\st, \sto) \equiv_{pp} \st.\kwT{setStore}(\sto)$) 
   \item $\kwT{store} : \sset{\gstate} \tmap (\xs \pmap \gval)$ 
            \hfill ($\kwT{store}(\st) \equiv_{pp} \st.\kwT{store}$) 
   \item $\kwT{ee} : \sset{\gstate} \tmap \exprs \pmap \gval$ 
            \hfill ($\kwT{ee}(\st, \e) \equiv_{pp} \st.\kwT{ee}(\e)$) 
   \item $\kwT{ea} : \actions \tmap \sset{\gstate} \tmap \gval \pmap \power{\sset{\gstate} \times \gval}$ 
            \hfill ($(\st', \vl') \in \kwT{ea}(\act, \st, \vl) \equiv_{pp} \action{\st}{\act}{v}{(\st', v')}$) 
\end{itemize}
\end{definition}

\noindent A state model $\gstate =  \tup{\sset{\gstate}, \gval, \actions}$ is said to be \emph{proper} 
\emph{iff} it defines the following three distinguished actions: $\kwT{assume}$,
 $\kwT{symb}$, and $\kwT{fresh}$.

\medskip
\begin{display}{\gil Semantic Domains for $\gstate =  \tup{\sset{\gstate}, \gval, \actions}$}
\begin{tabular}{lr@{\ \ }c@{\ \ }l@{\quad}l}
   Call stacks: & $\cs \in \css{\gstate} $ & $\defeq$ & $ \vtup{\f} \mid \vtup{\f, \x, \sto, i} \cons \cs$ & where: $\f \in \fids$, $\x \in \xs$, $\sto : \xs \pmap \gval$, $i \in \ints$\\
   Configurations: & $\cf \in \cfs{\gstate}$    & $\defeq$ & $\tup{\prog, \st, \cs, i}$     & where: $\prog \in \progs{\actions}$, $\st \in \sset{\gstate}$, $\cs \in \css{\gstate}$, $i \in \ints$ \\
   Outcomes: & $\outcome \in \outcomes$  & $\defeq$ &  $\cont \mid \onormal{v} \mid \ofail{v}$ & where: $v \in \gval$
 \end{tabular}
\end{display}

\medskip
\noindent \gil semantic transitions: $\rightsquigarrow_\gstate~:~\sset{\gstate} \times \css\gstate \times \mathbb{N} \times \outcomes \pmap \power{\sset{\gstate} \times \css\gstate \times \mathbb{N} \times \outcomes}$.

\medskip
\begin{display}{Semantics of \gil: $\fullsemtrans{\st, \cs, i}{\st', \cs', j}{}{\prog}{\outcome'}{\outcome}$}
{\footnotesize
\begin{mathpar}
\inferrule[\textsc{Assignment}]
  {
    \cmd(\prog, \cs, i) = \x := \e
    \and
    \ee{\st}{\e} = v
  }{\semtrans{\st, \cs, i}{\st.\kwT{setVar}(\x, v), \cs, i{+}1}{}{\prog}{}}  	 
 \and
\inferrule[\textsc{Action}]
  {
    \cmd(\prog, \cs, i) = \x := \act(\e)
    \quad
    \ee{\st}{\e} = v
    \quad
     \action{\st}{\act}{v}{(\st', v')}
  }{\semtrans{\st, \cs, i}{\st'.\kwT{setVar}(\x, v'), \cs, i{+}1}{}{\prog}{}}  	 
\\
\inferrule[\textsc{IfGoto - True}]
  {
    \cmd(\prog, \cs, i) = \ifgoto{\e}{j}
    \\\\
    \ee{\st}{\e} = v
    \\\\
    \actionT{\st}{assume}{v}{\st'}
  }{\semtrans{\st, \cs, i}{\st', \cs, j}{}{\prog}{}}  	 
\and
\inferrule[\textsc{IfGoto - False}]
  {
    \cmd(\prog, \cs, i) = \ifgoto{\e}{j}
    \\\\
    \ee{\st}{\lnot\e} = v
    \\\\
    \actionT{\st}{assume}{v}{\st'}
  }{\semtrans{\st, \cs, i}{\st', \cs, i+1}{}{\prog}{}}  	 
\and
\vspace*{-0.1cm}
\inferrule[\textsc{Symb}]
  {
      \cmd(\prog, \cs, i) = \x := \symb{j}
      \\\\ 
      \actionT{\st}{fresh}{j}{(\st', v')}
  }{
    \semtrans{\st, \cs, i}{\st'.\kwT{setVar}(\x, v'), \cs, i{+}1}{}{\prog}{}
  }
 \and 
 \inferrule[\textsc{Fresh}]
  {
     \cmd(\prog, \cs, i) = \x := \fresh{j}
      \\\\ 
    \actionT{\st}{symb}{j}{(\st', v')}
  }{
    \semtrans{\st, \cs, i}{\st'.\kwT{setVar}(\x, v'), \cs, i{+}1}{}{\prog}{}
  }
  \and
\inferrule[\textsc{Call}]
  {
    \cmd(\prog, \cs, i) = \e(\e')
    \\\\
    \ee{\st}{\e} = \f
    \and
    \ee{\st}{\e'} = v
    \\\\
    \cs' = \vtup{\f, \x, \st.\kwT{store}, i+1} \cons \cs
  }{\semtrans{\st, \cs, i}{\st.\kwT{setStore}(\btup{\f.\kwT{arg} \mapsto v}), \cs', 0}{}{\prog}{}}  	
\and
\inferrule[\textsc{Return}]
  {
     \cmd(\prog, \cs, i) = \return{\e}
      \and
    \ee{\st}{\e} = v   
     \\\\
    \cs = \vtup{-, \x, \sto, j} \cons \cs'
    \and 
    \st' = \st.\kwT{setStore}(\sto)
  }{
    \semtrans{\st, \cs, i}{\st'.\kwT{setVar}(\x, v'), \cs', j}{}{\prog}{}
  }
  \qquad
 \inferrule[\textsc{Top Return}]
  {
     \cmd(\prog, \cs, i) = \return{\e}
      \\\\
    \ee{\st}{e} = v
  }{
    \semtrans{\st, \vtup{\fid}, i}{\st, \vtup{\fid}, i}{}{\prog}{\onormal{v}}
  } 
  \qquad
 \inferrule[\textsc{Fail}]
  {
     \cmd(\prog, \cs, i) = \fail{e}
      \\\\
    \ee{\st}{\e} = \vl
  }{
    \semtrans{\st, \cs, i}{\st, \cs, i}{}{\prog}{\ofail{\vl}}
  } 
\end{mathpar}}
\end{display}

\begin{definition}[Allocator Model]
\label{app:def:am}
An allocator model $\allocator \in \allocators$ is a pair $\tup{\sset{\allocator}, \gval}$ consisting of: 
\etag{1} a set $\sset{\allocator} \ni \arec$ of allocators; and 
\etag{2} a set $\gval$ of values to allocate.
It exposes the following function:
$$\allocf : \sset{\allocator} \tmap \mathbb{N} \tmap \power{\gval} \pmap \sset{\allocator} \times \gval \qquad\quad 
$$
which satisfies the well-formedness constraint: 
$$(\arec', \gvl) = \allocf(\arec, j, Y) \implies \gvl \in Y.$$
\end{definition}

\begin{tabular}{r@{~}c@{~}l}
Logical variables & : & $\lx \in \lxs$. \\
Logical expressions & : & $\sexp \in \sexps \ \defeq \ \vl \mid \lx \in \lxs \mid \unop{\sexp} \mid \binop{\sexp_1}{\sexp_2}$.
\end{tabular}

\medskip
\noindent Additionally, we write $\pi \in \Pi$ to denote logical expressions that that be statically typed as boolean (for example, $\true$, $\false$, $\lx \text{~and~} \hat{y}$, etc.).

\begin{definition}[Concrete Memory Model]\label{app:def:cmem}
A concrete memory model $\cmemory \in \cmemories$ 
is a pair $\tup{\sset{\cmemory}, \actions}$,
consisting of a set of concrete memories, $\sset{\cmemory} \ni \cmem$, and a set of actions $\actions$. 
A concrete memory model additionally defines a function $\cea$ for acting 
on memories: 
$$\cea : \actions \tmap \sset{\cmemory} \tmap \vals \pmap \sset{\cmemory} \times \vals
\qquad \qquad \qquad ({\cea}(\act, \cmem, \vl) \equiv_{pp} \cmem.\act(\vl))$$
\end{definition}

\begin{definition}[Symbolic Memory Model]\label{app:def:smem}
A symbolic memory model $\smemory \in \smemories$ 
is a pair $\tup{\sset{\smemory}, \actions}$,
consisting of a set of symbolic memories, $\sset{\smemory} \ni \smem$, and a set of actions $\actions$. 
A symbolic memory model additionally defines a function $\sea$ for acting 
on memories: 
 $$\begin{array}{c}\sea : \actions \tmap \sset{\smemory} \tmap \sexps \tmap \pcs \pmap\power{\sset{\smemory} \times \sexps \times \pcs} \\
  ((\smem', \sexp', \pc') \in {\sea}(\act, \smem, \sexp, \pc) \equiv_{pp} \action{\smem}{\act}{\sexp, \pc}{(\smem', \sexp', \pc')})\end{array}$$.
\end{definition}

\begin{tabular}{r@{~}c@{~}l@{\qquad}r@{~}c@{~}l}
Concrete stores & : & $\sto : \xs \pmap \vals$ &
Symbolic stores & : & $\ssto : \xs \pmap \sexps$ \\
Concrete allocators & : & $\arec \in \sset{\allocator}_{\vals}$ &
Symbolic allocators & : & $\sarec \in \sset{\allocator}_{\sexps}$ \\
Mandatory actions & : & $\actionsZ \defeq \set{\kwT{assume}, \kwT{fresh}, \kwT{symb}}$
\end{tabular}

\begin{definition}[Concrete State Constructor (\cstateConstr)]\label{app:def:concrete:lifting}
Given an allocator $\allocator = \tup{\sset{\allocator}, \vals}$, the 
concrete state constructor $\cstateConstr : \cmemories \tmap \gstates$ is defined as  
$\cstateConstr(\tup{\sset{\cmemory}, \actions}) \semeq \tup{\sset{\gstate}, \vals, \actions  \dunion \actionsZ}$,
where: 

{\small
\begin{tabular}{lll}
   \multicolumn{3}{l}{\mybullet $\sset{\gstate} = \sset{\cmemory} \times \pfunset{\xs}{\vals} \times \sset{\allocator}_{\vals}$} \\ 
  \mybullet $\kwT{setVar}(\tup{\cmem, \sto, \arec}, \x, \vl)$ 
       & $\semeq$ & 
       $\tup{\cmem, \sto[\x \mapsto \vl], \arec}$ \\ 
  \mybullet $\kwT{setStore}(\tup{\cmem, \_, \arec}, \sto)$
     & $\semeq$ & 
     $\tup{\cmem, \sto, \arec}$ \\ 
  \mybullet $\kwT{store}(\tup{\_, \sto, \_})$
    & $\semeq$ & 
    $\sto$ \\
  \mybullet $\kwT{ee}(\tup{\_, \sto, \_}, \e)$
    & $\semeq$ & 
    $\ceval{\e}{\sto}$ \\
 \mybullet $\kwT{ea}(\act, \tup{\cmem, \sto, \arec}, \vl)$
    & $\semeq$ & 
    $\lbrace (\tup{\cmem', \sto, \arec}, \vl') \mid (\cmem', \vl') = \cea(\act, \cmem, \vl) \rbrace$ \\
   \mybullet $\kwT{assume}(\st, \vl)$
    & $\semeq$ & 
    $\lbrace (\st, \vl) \mid \vl = \true \rbrace$ \\
    \mybullet $\kwT{fresh}(\tup{\cmem, \sto, \arec}, i)$
    & $\semeq$ &
     $\lbrace (\tup{\cmem, \sto, \arec'}, \varsigma) \mid \alloc{\arec}{i}{\arec', \varsigma}{\locs} \rbrace $ \\
    \mybullet $\kwT{symb}(\tup{\cmem, \sto, \arec}, i)$
    & $\semeq$ &
     $\lbrace (\tup{\smem, \ssto, \arec'}, \vl) \mid \alloc{\arec}{i}{\arec', \vl}{\vals} \rbrace $ \\
    
\end{tabular}}
\end{definition}

\begin{definition}[Symbolic State Constructor ($\sstateConstr$)]\label{app:def:symbolic:lifting}
Given an allocator $\hat\allocator = \tup{\sset{\hat\allocator}, \sexps}$, the 
symbolic state constructor $\sstateConstr : \smemories \tmap \gstates$ is defined as 
$\sstateConstr(\tup{\sset{\smemory}, \actions}) \semeq \tup{\gsstate, \sexps, \actions  \dunion \actionsZ}$,
where: 

{\small
\begin{tabular}{lll}
   \multicolumn{3}{l}{\mybullet $\gsstate = \sset{\smemory} \times \pfunset{\xs}{\sexps} \times \sset{\allocator}_{\sexps} \times \pcs$} \\ 
  \mybullet $\kwT{setVar}(\tup{\smem, \ssto, \sarec, \pc}, \x, \sexp)$ 
       & $\semeq$ & 
       $\tup{\smem, \ssto[\x \mapsto \sexp], \sarec, \pc}$ \\ 
  \mybullet $\kwT{setStore}(\tup{\smem, \_, \sarec, \pc}, \ssto)$
     & $\semeq$ & 
     $\tup{\smem, \ssto, \sarec, \pc}$ \\ 
  \mybullet $\kwT{store}(\tup{\_, \ssto, \_, \_})$
    & $\semeq$ & 
    $\ssto$ \\
  \mybullet $\kwT{ee}(\tup{\_, \ssto, \_, \_}, \e)$
    & $\semeq$ & 
    $\ssto(\e)$ \\
 \mybullet $\kwT{ea}(\act, \tup{\smem, \ssto, \sarec, \pc}, \sexp)$
    & $\semeq$ & 
    $\lbrace (\tup{\smem', \ssto, \sarec, \pc \, \wedge \pc'}, \sexp') \mid (\smem', \sexp', \pc') \in \sea(\act, \smem, \sexp) \rbrace$ \\
   \mybullet $\kwT{assume}(\tup{\smem, \ssto, \sarec, \pc}, \pc')$
    & $\semeq$ & 
    $\lbrace (\tup{\smem, \ssto, \sarec, \pc \gand \pc'}, \true) \mid \pc \gand \pc' \, \text{not \unsat} \rbrace$ \\
  \mybullet $\kwT{fresh}(\tup{\smem, \ssto, \sarec, \pc}, i)$
    & $\semeq$ &
    $\lbrace (\tup{\cmem, \sto, \sarec', \pc}, \varsigma) \mid \alloc{\sarec}{i}{\sarec', \varsigma}{\locs} \rbrace $ \\
  \mybullet $\kwT{symb}(\tup{\smem, \ssto, \sarec, \pc}, i)$
    & $\semeq$ &
    $\lbrace (\tup{\cmem, \sto, \sarec', \pc}, \lx) \mid \alloc{\sarec}{i}{\sarec', \lx}{\lxs} \rbrace $ \\
  
  %
\end{tabular}}
\end{definition}

\subsection{\while: Syntax, Actions, Compilation to \gil, Concrete/Symbolic memories}

\medskip
\begin{display}{\while: Syntax, Actions, Memories}
\begin{tabular}{r@{~}c@{~}l}
   $\cw \in \cws$ & $\defeq$ & $\x := \e \mid \wskip \mid \cw_1; \cw_2 \mid \wif{\e}{\cw_1}{\cw_2} \mid \wwhile{\e}{\cw} \mid \wreturn{\e} \mid \x := \fid(\e) \mid $  \\ 
   & & $\wassume{\e} \mid \wassert{\e} \mid \x := \wobj{\prop_i: \e_i\mid_{i=1}^n} \mid \wlookup{\x}{\e}{\prop} \mid \wmutate{\e}{\prop}{\e'} \mid \wdispose{\e}$ \\
   $\actionswhile$ & $\defeq$ & $\lbrace \lookupAction, \mutateAction, \disposeAction \rbrace$ \\
   $\cmem \in \wcmems$ & : & $\locs \times \strs \pmap \vals$ \\
   $\smem \in \wsmems$ & : & $\sexps \times \strs \pmap \sexps$
 \end{tabular}
\end{display}

\medskip
\begin{display}{\while-to-GIL Compiler: $\compWhile : \cws \tmap \ints \tmap \litlst{\cmds{\actionswhile}} \times \ints$}
{\footnotesize
\begin{mathpar}
\inferrule[\textsc{Assignment}]
  {}{
     {\begin{array}[t]{l}
      \compWhile(\x := \e, \wpc) \semeq \\
      {\begin{array}{l}
        \wpc:  \x := \e  \\ 
         \nextpc{\wpc + 1}
       \end{array}} 
      \end{array}}
  }
 \and 
 \inferrule[\textsc{Skip}]
  {}{
     {\begin{array}[t]{l}
      \compWhile(\wskip, \wpc) \semeq \\
      {\begin{array}{l}
        \wpc: \ifgoto{\true}{(\wpc + 1)} \\ 
         \nextpc{\wpc + 1}
       \end{array}} 
      \end{array}}
  }
 \and 
  \inferrule[\textsc{Return}]
  {}{
     {\begin{array}[t]{l}
      \compWhile(\wreturn{\e}, \wpc) \semeq \\
      {\begin{array}{l}
         \wpc: \return{\e} \\  
         \nextpc{\wpc+1}
       \end{array}} 
      \end{array}}
  }
 \and 
   \inferrule[\textsc{Call}]
  {}{
     {\begin{array}[t]{l}
      \compWhile(\x := \fid(\e), \wpc) \semeq \\
      {\begin{array}{l}
         \wpc: \x := \fid{\e} \\  
         \nextpc{\wpc+1}
       \end{array}} 
      \end{array}}
  }
 \and
 \inferrule[\textsc{Sequence}]
  {
     \compWhile(\cw_1, \wpc) = (\lst{\cm}_1, \wpc_1) \\\\
     \compWhile(\cw_2, \wpc_1) = (\lst{\cm}_2, \wpc_2)
  }{
     {\begin{array}[t]{l}
      \compWhile(\cw_1; \cw_2, \wpc) \semeq \\
      {\begin{array}{l}
         \lst{\cm}_1 \\ 
         \lst{\cm}_2 \\ 
         \nextpc{\wpc_2 + 1}
       \end{array}} 
      \end{array}}
  }
\and 
  \inferrule[\textsc{If}]
  {
     \compWhile(\cw_1, \wpc+1) = (\lst{\cm}_1, \wpc_1) \\\\
     \compWhile(\cw_2, \wpc_1+1) = (\lst{\cm}_2, \wpc_2)
  }{
     {\begin{array}[t]{l}
      \compWhile(\wif{\e}{\cw_1}{\cw_2}, \wpc) \semeq \\
      {\begin{array}{l}
         \wpc: \ifgoto{(\gnot \, \e)}{(\wpc_1+1)} \\ 
         \lst{\cm}_1 \\ 
         \wpc_1: \ifgoto{\true}{\wpc_2} \\ 
         \lst{\cm}_2 \\ 
         \nextpc{\wpc_2}
       \end{array}} 
      \end{array}}
  }
 \and 
  \inferrule[\textsc{While}]
  {
     \compWhile(\cw, \wpc+1) = (\lst{\cm}, \wpc') 
  }{
     {\begin{array}[t]{l}
      \compWhile(\wwhile{\e}{\cw}, \wpc) \semeq \\
      {\begin{array}{l}
         \wpc: \ifgoto{(\gnot \, \e)}{(\wpc' +1)} \\ 
         \lst{\cm} \\ 
         \wpc': \ifgoto{\true}{\wpc} \\ 
         \nextpc{\wpc'+1}
       \end{array}} 
      \end{array}}
  }
\and 
\inferrule[\textsc{Assume}]
  {}{
     {\begin{array}[t]{l}
      \compWhile(\wassume{\e}, \wpc) \semeq \\
      {\begin{array}{l}
         \wpc: \ifgoto{\e}{(\wpc + 2)} \\ 
          \wpc+1: \vanish \\
         \nextpc{\wpc + 2}
       \end{array}} 
      \end{array}}
  }
\and 
\inferrule[\textsc{Assert}]
  {}{
     {\begin{array}[t]{l}
      \compWhile(\wassert{\e}, \wpc) \semeq \\
      {\begin{array}{l}
         \wpc: \ifgoto{\e}{(\wpc + 2)} \\ 
          \wpc+1: \fail{\e} \\
         \nextpc{\wpc + 2}
       \end{array}} 
      \end{array}}
  }
\and 
\inferrule[\textsc{New}]
  {}{
     {\begin{array}[t]{l}
      \compWhile(\x := \wobj{\prop_i: \e_i\mid_{i=1}^n}, \wpc) \semeq \\
      {\begin{array}{l}
         \wpc: \x :=  \symb{} \\ 
          \wpc+i: \_ := \mutateAction(\litlst{\x, \prop_i, \e_i}) \mid_{i=1}^n \\
         \nextpc{\wpc + n + 1}
       \end{array}} 
      \end{array}}
  }
 \and 
 \inferrule[\textsc{Lookup}]
  {}{
     {\begin{array}[t]{l}
      \compWhile(\wlookup{\x}{\e}{\prop}, \wpc) \semeq \\
      {\begin{array}{l}
          \wpc: \x := \lookupAction(\x, \prop) \\
         \nextpc{\wpc + 1}
       \end{array}} 
      \end{array}}
  }
  \and 
  \inferrule[\textsc{Mutate}]
  {}{
     {\begin{array}[t]{l}
      \compWhile(\wmutate{\e_1}{\prop}{\e_2}, \wpc) \semeq \\
      {\begin{array}{l}
          \wpc: \_ := \mutateAction(\litlst{\e_1, \prop, \e_2}) \\
         \nextpc{\wpc + 1}
       \end{array}} 
      \end{array}}
  }
  \and 
    \inferrule[\textsc{Dispose}]
  {}{
     {\begin{array}[t]{l}
      \compWhile(\wdispose{\e}, \wpc) \semeq \\
      {\begin{array}{l}
          \wpc: \_ := \disposeAction(\e) \\
         \nextpc{\wpc + 1}
       \end{array}} 
      \end{array}}
  }
 \end{mathpar}}
\end{display}

\medskip
\begin{display}{\while: Actions in Concrete and Symbolic Memories}
\begin{minipage}{0.40\textwidth}
{\small
\begin{mathpar}
\inferrule[C-Lookup]
  {\cmem = \_ \dunion \wcell{\loc}{\prop}{\vl}}
  {\action{\cmem}{\lookupAction}{\litlst{\loc, \prop}}{(\cmem, \vl)}} 
 \\
\inferrule[C-Mutate-Present]
  {
    \cmem = \cmem' \dunion \wcell{\loc}{\prop}{\_}
    \quad
    \cmem'' = \cmem' \dunion \wcell{\loc}{\prop}{\vl}
  }{\action{\cmem}{\mutateAction}{\litlst{\loc, \prop, \vl}}{(\cmem'', \vl)}} 
 \\
\inferrule[C-Mutate-Absent]
  {
    (\loc, \prop) \notin \mathrm{dom}(\cmem)
    \and
    \cmem' = \cmem \dunion \wcell{\loc}{\prop}{\vl}
  }{\action{\cmem}{\mutateAction}{\litlst{\loc, \prop, \vl}}{(\cmem', \vl)}} 
  \\
\inferrule[C-Dispose]
  {
    \memproj{\cmem}{\loc} = (\_, \cmem')
  }{\action{\cmem}{\disposeAction}{\loc}{(\cmem', \true)}} 
\end{mathpar}}
\end{minipage}
\begin{minipage}{0.59\textwidth}
{\small
\begin{mathpar}
\inferrule[S-Lookup]
  {
     \pc \vdash \sexp = \sexp' 
     \and
     \smem = \_ \dunion \wcell{\sexp'}{\prop}{\sexp_v}
  }{\action{\smem}{\lookupAction}{\litlst{\sexp, \prop}, \pc}{\litset{(\smem, \sexp_v, \true)}}} 
 \\
\inferrule[S-Mutate-Present]
  {
  	\pc \vdash \sexp = \sexp'' 
	\quad
    \smem = \smem' \dunion \wcell{\sexp''}{\prop}{\_}
    \quad
    \smem'' = \smem' \dunion \wcell{\sexp''}{\prop}{\sexp'}
  }{\action{\smem}{\mutateAction}{\litlst{\sexp, \prop, \sexp'}, \pc}{\litset{(\smem'', \true, \true)}}} 
 \\
\inferrule[S-Mutate-Absent]
  {
  	\memproj{\smem}{\sexp, \prop, \pc} = \emptyset
	\quad
    \smem' = \smem \dunion \wcell{\sexp}{\prop}{\sexp'}
  }{\action{\smem}{\mutateAction}{\litlst{\sexp, \prop, \sexp'}, \pc}{\litset{(\smem', \true, \true)}}} 
  \\
\inferrule[S-Dispose]
  {
    \memproj{\smem}{\sexp, \pc} = (\_, \smem')
  }{\action{\smem}{\disposeAction}{\sexp, \pc}{\litset{(\smem', \true, \true)}}}
\end{mathpar}}
\end{minipage}
\end{display}

\newpage
\section{Section 3: Parametric Soundness}
\label{app:s3}

\subsection{Parametric Soundness}
\myparagraph{Restriction Operators} A \emph{restriction} operator $\frestriction : X \tmap X \pmap X$  on a set $X$, 
written $\rfun{x_1}{x_2}$ for $\frestriction\!(x_1, x_2)$, is a binary associative function 
satisfying the following properties: 
\begin{mathpar}
  \inferrule[Idempotence]{}{\rfun{x}{x} = x}
  \and
  \inferrule[Right Commutativity]{}{\rfun{(\rfun{x_1}{x_2})}{x_3} = \rfun{(\rfun{x_1}{x_3})}{x_2}}
  \and
  \inferrule[Weakening]{\rfun{x_1}{\rfun{x_2}{x_3}} = x}{\rfun{x_1}{x_2} = x_1 \, \wedge \, \rfun{x_1}{\x_3} = x_1}
\end{mathpar}

A restriction operator on states $\frestriction : \sset{\gstate} \tmap \sset{\gstate} \pmap \sset{\gstate}$ is said to be 
\emph{preserved} by a state model $\gstate = \tup{\sset{\gstate}, \gval, \actions}$ 
if all of the state-generating functions exposed by the state model are monotonic with respect to the pre-order induced by $\frestriction$; put formally: 
\begin{mathpar}
     {\small \inferrule[RMono-SetVar]{}{\st.\kwT{setVar}(\x, \vl) = \st' \implies \st' \rleq \st}
      \and
     \inferrule[RMono-SetStore]{}{\st.\kwT{setStore}(\sto) = \st' \implies \st' \rleq \st}
     \and
     \inferrule[RMono-Action]{}{\action{\st}{\act}{v}{(\st', -)} \implies \st' \rleq \st}}
   \end{mathpar}

We say that $\frestriction$ is a restriction operator on a state model
$\gstate = \tup{\sset{\gstate}, \gval, \actions}$, if $\frestriction$ is a restriction 
operator on the carrier set $\sset{\gstate}$ and $\frestriction$ is preserved by $\gstate$. 
Restriction operators are extended from states to configurations in the standard way: 
$\rfun{\tup{\prog, \st, \cs_1, i}}{\tup{\prog, \st', -, -}} \semeq \tup{\prog, \rfun{\st}{\st'}, \cs_1, i}$.

\myparagraph{Compatibility} A pre-order $(X, \leq)$ is \emph{compatible} with a restriction operator $\frestriction$ 
on $X$ \emph{iff} the following properties hold: 
\begin{mathpar}
\inferrule[$\frestriction$-$\leq$ Compatibility]{}{\rfun{x_1}{x_2} \, \leq \, x_1}
\and
\inferrule[$\leq$-$\frestriction$ Compatibility]{x_1 \leq x_2}{x_1 \rleq x_2}
\and
\inferrule[Strengthening]{x_1 \leq x_1' \quad x_2 \rleq x_2' }{\rfun{x_1}{x_2} \, \leq \, \rfun{x_1'}{x_2'}}
\end{mathpar}

\myparagraph{Soundness Relations} Given a relation $\rel \, \in X \times Y$ between two sets $X$ and $Y$, the pre-order 
on $X$ induced by $\rel$, written $\leq_{\rel}$, is defined as follows: 
$x_1 \leq_{\rel} x_2$ if and only if $\lbrace y \mid x_1 \, \rel \, y \rbrace \subseteq \lbrace y \mid x_2 \, \rel \, y \rbrace$. 

\begin{definition}[Soundness Relation - States]\label{app:def:soundness:relations}
Given two state models, $\hat\gstate = \tup{\sset{\hat\gstate}, \hat\gval, \actions}$ and $\gstate = \tup{\sset{\gstate}, \gval, \actions}$, 
a soundness relation $\soundRel$ for $\hat\gstate$ in terms of $\gstate$ is a triple
$\tup{\frestriction, \stRel, \vlRel}$, consisting of: 
\etag{1}~a restriction operator $\frestriction$ on $\hat\gstate$;
\etag{2}~a binary relation $\stRel \, \subseteq \sset{\hat\gstate} \times \sset{\gstate}$ between $\sset{\hat\gstate}$ and $\sset{\gstate}$; and 
\etag{3}~a ternary relation $\vlRel \subseteq \sset{\hat\gstate} \times \hat\gval \times \gval$ between  $\sset{\hat\gstate}$, $\hat\gval$, and $\gval$, 
such that $\frestriction$ is compatible with the pre-order induced by $\stRel$ (denoted by $\leq$) and the following constraints hold: 

\vspace*{-0.2cm}
{\small
\begin{mathpar}
     \inferrule[Store]{}{\stR{\sst}{\st} \implies \vlR{\sst}{\sst.\kwT{store}}{\st.\kwT{store}}}
     \and 
     \inferrule[EvalExpr]{}{\leqStR{\sst}{\sst'}  \gand \stR{\sst}{\st} 
            \implies \vlR{\rfun{\sst'}{\sst}}{\ee{\sst'}{\e}}{\ee{\st}{\e}}}
     \and
      \inferrule[SetVar]{}{
        {\begin{array}{l}
             \leqStR{\sst}{\sst'} \gand \stR{\sst}{\st}  \gand \vlR{\sst}{\hat\vl}{\vl} \\
              \quad \implies \stR{\rfun{\sst'.\kwT{setVar}(\x, \hat\vl)}{\sst}}{\st.\kwT{setVar}(\x, \vl)}
         \end{array}}}
     \and 
     \inferrule[SetStore]{}{
       {\begin{array}{l}
         \leqStR{\sst}{\sst'} \gand \stR{\sst}{\st}  \gand \vlR{\sst}{\hat\sto}{\sto} \\
            \quad  \implies  \stR{\rfun{\sst'.\kwT{setStore}(\hat\sto)}{\sst}}{\st.\kwT{setStore}(\sto)}
        \end{array}}}
    \and 
     \inferrule[Action]{}{\action{\sst'}{\act}{\hat\vl}{(\sst'', \hat\vl')} \gand \leqStR{\sst}{\rfun{\sst'}{\sst''}} \gand \ \stR{\sst}{\st} \gand  \vlR{\sst}{\hat\vl}{\vl} \\
            \qquad \implies \exists \, \st', \vl' \, . \, \action{\st}{\act}{\vl}{(\st', \vl')} \gand \stR{\rfun{\sst'}{\sst}}{\st'} \gand  \vlR{\rfun{\sst'}{\sst}}{\vl_1'}{\vl'}}
     \and
     \inferrule[Weakening]{}{
       {\begin{array}{l}
        \sst \rleq \sst' \gand \vlR{\sst}{\hat\vl}{\vl}  \\ 
        \qquad \implies \vlR{\sst'}{\hat\vl}{\vl}
        \end{array}}
     }
  \end{mathpar}}

\vspace*{-0.2cm}
\noindent where $\vlR{\sst}{\hat\sto}{\sto}$ is shorthand for: $\domain(\hat\sto) = \domain(\sto) = X$ and $\forall \x \in X \, . \, \vlR{\sst}{\hat\sto(\x)}{\sto(\x)}$.
\end{definition}

\medskip
\begin{display}{Soundness Relations - Call stacks and Configurations}
{
\small
 \begin{mathpar}
 \hspace*{-0.08cm}
 \inferrule[Top Frame]
   {}{\vlR{\sst}{\vtup{\f}}{\vtup{\f}}}
 \quad 
 \inferrule[Inner Frame]
      {
          \vlR{\sst}{\hat\sto}{\sto}
          \quad
           \vlR{\sst}{\hat\cs}{\cs}
      }{
          \vlR{\sst}{ \vtup{\f, \x, \hat\sto, i}  \cons \hat\cs}{\vtup{\f, \x, \sto, i}  \cons \cs}
      }\quad
   \inferrule[$\stRel$-Configuration]
      {
          \stR{\sst}{\st}
          \and
          \vlR{\sst}{\hat\cs}{\cs}
      }{
          \stR{\tup{\prog, \sst, \hat\cs, i}}{\tup{\prog, \st, \cs, i}}
      }
\quad
        \inferrule[$\leq$-Configuration]
      {
          \sst \leq \sst'
      }{
          \tup{\prog, \sst, \hat\cs, i} \leq \tup{\prog, \sst', \cs, i}
      }
      \end{mathpar}}
  \end{display}

\begin{theorem}[One-Step Soundness]
    Let $\soundRel = \tup{\frestriction, \stRel, \vlRel}$ be a soundness criterion
     for $\hat\gstate = \tup{\sset{\hat\gstate}, \hat\gval, \actions}$ in terms of $\gstate = \tup{\sset{\gstate}, \gval, \actions}$
     and $\leq$ the pre-order induced by $\stRel$. It holds that: 
     $$ 
     \hat\cf' \semarrow \hat\cf'' \gand \hat\cf \leq \rfun{\hat\cf'}{\hat\cf''} \gand \stR{\hat\cf}{\cf}
        \implies \exists \, \cf' \, . \, \cf \semarrow \cf' \gand \stR{\rfun{\hat\cf''}{\hat\cf}}{\cf'}
     $$
\end{theorem}
\begin{proofpf}
    \pf \pflongnumbers\\
    We proceed by case analysis on the rule that produced $\hat\cf' \semarrow \hat\cf''$. We only provide the proof for the variable assignment and action cases. The others cases are analogous.
  
    \step{assume}{
      \assume{
        \begin{pfenum}
          \item (H1) $\hat\cf' \semarrow \hat\cf''$
          \item (H2) $\hat\cf \leq \rfun{\hat\cf'}{\hat\cf''}$
          \item (H3) $\stR{\hat\cf}{\cf}$
        \end{pfenum}
      }
    }
    \step{1}{\case{Assignment}}
    \begin{proofpf}
      \step{1a}{$\hat\cf' = \tup{\sst', \hat\cs, i}$ [Destructing on H1 - assignment case]}
      \step{1b}{$\cmd(\prog, \hat\cs, i) = \x := \sexp$}
      \step{1c}{$\sst.\kwT{ee}(\sexp) \semarrow (\sst_1, \hat\vl)$}
      \step{1d}{$\sst_1.\kwT{setVar}(\x, \hat\vl_1) = \sst''$}
      \step{1e}{$\hat\cf'' = (\sst'', \hat\cs, i+1)$}
      \step{1f}{$\hat\cf = (\sst, \hat\cs, i)$ [From H2 + definition of $\leq_{cf}$]}
      \step{1g}{$\sst \leq \rfun{\sst'}{\sst''}$}
      \step{1h}{$\cf = (\st, \cs, i)$ [From H3 + \stepref{1g} + definition of $\rel_{\cf}$]}
      \step{1i}{$\stR{\hat\sst}{\st}$}
      \step{1j}{$\vlR{\sst}{\hat\cs}{\cs}$}
      \step{1k}{$\cmd(\prog, \cs, i) = \x := \e$ [From \stepref{1b} + \stepref{1k}]}
      \step{1l}{$\sst \leq \rfun{\sst'}{\sst_1}$ [From \stepref{1d} we know that $\sst'' \leq \sst_1$ and from \stepref{1g} we know that $\sst \leq \rfun{\sst'}{\sst''}$. From $\sst'' \rleq \sst_1$, it follows that $\rfun{\sst'}{\sst''} \leq \rfun{\sst'}{\sst_1}$, from which it follows, by transitivity that $\sst \leq \rfun{\sst'}{\sst_1}$.]}
      \step{1m}{$\exists \st_1, \vl \cdot$}
      \begin{pfenum*}
        \item $\st.\kwT{ee}(\e) \semarrow (\st_1, \vl)$ [From \stepref{1c} + \stepref{1i} + \stepref{1l} + Correctness relation] \label{1m1}
        \item $\stR{\rfun{\sst_1}{\sst}}{\st_1}$ \label{1m2}
        \item $\vlR{\rfun{\sst_1}{\sst}}{\hat\vl}{\vl}$ \label{1m3}
      \end{pfenum*}
      \step{1q}{$\exists \sst_2\cdot$ [From \stepref{1e} + SM(Trans. strenghtening)]}
      \begin{pfenum*}
        \item $(\rfun{\sst_1}{\sst}).\kwT{setVar}(\x, \hat\vl) = \sst_2$ \label{1q1}
        \item $\sst_2 \leq \sst''$ \label{1q2}
      \end{pfenum*}
      \step{1r}{$\exists \st':$ [From \stepref{1q}.1 + \stepref{1m}.2 + \stepref{1m}.3 + CR(\kwT{setVar})]}
      \begin{pfenum*}
        \item $\stR{\rfun{\sst_3}{\sst}}{\st''}$ \label{1r1}
        \item $\sst_1.\kwT{setVar}(\x, \vl) = \sst''$ \label{1r2}
      \end{pfenum*}
      \step{1s}{$\stR{\rfun{\sst''}{\sst}}{\st'}$ [From \stepref{1q}.2 + \stepref{1r}.1 + CR-Orders]}
      \step{1t}{$\stR{\cf}{(\st', \cs, i+1) = \cf'}$ [From \stepref{1h} + \stepref{1k} + \stepref{1m}.1 + \stepref{1r}.2]}
      \step{1u}{$\vlR{\rfun{\sst''}{\sst}}{\hat\cs}{\cs}$ [From \stepref{1j} + noting that $\rfun{\sst''}{\sst} \rleq \sst$]}
      \step{1v}{$\rfun{\hat\cf''}{\hat\cf} \rel \cf'$}
    \end{proofpf}
    \step{2}{\case{Action}}
    \begin{proofpf}
      \step{2a}{$\hat\cf' = \tup{\sst', \hat\cs, i}$ [Destructing on (H1) - action case]}
      \step{2b}{$\cmd(\prog, \hat\cs, i) = \x := \act(\sexp)$}
      \step{2c}{$\sst'.\kwT{ee}(\sexp) \semarrow (\sst_1, \hat\vl_1)$}
      \step{2d}{$\sst_1.\sea(\vl) \semarrow (\sst_2, \hat\vl_2)$}
      \step{2e}{$\sst_2.\kwT{setVar}(\x, \hat\vl_2) = \sst''$}
      \step{2f}{$\hat\cf'' = (\sst'', \hat\cs, i+1)$}
      \step{2g}{$\hat\cf = (\sst \hat\cs, i)$ [From H2 + definition of $\leq$]}
      \step{2h}{$\sst \leq \rfun{\sst'}{\sst''}$}
      \step{2i}{$\cf = (\st, \cs, i)$ [From H3 + \stepref{2f} + definition of $\rel{\cf}$]}
      \step{2j}{$\stR{\sst}{\st}$}
      \step{2k}{$\vlR{\sst}{\hat\cs}{\cs}$}
      \step{2l}{$\cmd(\prog, \cs, i) = \x := \act(\sexp)$ [From \stepref{2b} + \stepref{2k}]}
      \step{2m}{$\sst \leq \rfun{\sst'}{\sst_1}$[From \stepref{2d} we know that $\sst_2 \rleq \sst_1$ and from \stepref{2e} 
      we know that $\sst'' \rleq \sst_2$. If follows by transitivity that $\sst'' \rleq \sst_1$, from which it follows (together with $\sst'' \rleq \sst'$) that $\sst \leq \rfun{\sst'}{\sst_1}$.]}
      \step{2n}{$\exists \st_1, \vl_1 \cdot $ [From \stepref{2e} + \stepref{2k} + \stepref{2m} + CR($\kwT{ee}$)]}
      \begin{pfenum*}
        \item $\st.\kwT{ee}(\e) \semarrow (\st_1, \vl_1)$ \label{2n1}
        \item $\stR{\rfun{\sst_1}{\sst}}{\st_1}$ \label{2n2}
        \item $\vlR{\rfun{\sst_1}{\sst_2}}{\hat\vl_1}{\vl_1}$ \label{2n3}
      \end{pfenum*}
      \step{2o}{$\sst'' \rleq \sst_2$ [From \stepref{2d} + SM(monotonicity)]}
      \step{2o1}{$\sst \rleq \sst''$ [From \stepref{2h} + \stepref{2o}]}
      \step{2o2}{$\rfun{\sst_1}{\sst} \rleq \rfun{\sst_1}{\sst_2}$ [From \stepref{2o1}]}
      \step{2o3}{$\exists \st_2, \vl_2 \cdot$ [From \stepref{2d} + \stepref{2o3} + \stepref{2n}.2 + \stepref{2n}.3 + CR($\kwT{ea}$)]}
      \begin{pfenum*}
        \item $\st_2.\cea(\vl_2) \rel (\st_3, \vl_2)$ \label{10-1}
        \item $\rfun{\sst_2}{(\rfun{\sst'}{\sst})} \rel \st_2$ \label{10-2}
        \item $\rfun{\sst''}{(\rfun{\sst'}{\sst})} \vdash \vl_2 \rel_{v} \vl_2$ \label{10-3}
      \end{pfenum*}
      \step{11}{$\sst'' \rleq \sst'$ [From \stepref{2d} + SM(Monotonicity)]}
      \step{2p}{$\rfun{\sst''}{\sst} \rel \st_3$ [From \stepref{2n}.1 + \stepref{2n}.3 + \stepref{2o}, noting that: 
      $\rfun{\sst_3}{(\rfun{\sst'}{\sst})} = \rfun{(\rfun{\sst''}{\sst'})}{\sst} = \rfun{\sst''}{\sst}$ (since $\sst_2 \rleq \sst_1$)]}
      \step{2q}{$\exists \sst_4 \cdot$ [From \stepref{2e} + SM(Trans. strenghtening)]}
      \begin{pfenum*}
        \item $\rfun{\sst''}{\sst}.\kwT{setVar}(\x, \vl_1) = \sst_4$ \label{2q1}
        \item $\sst_4 \leq \sst''$ \label{2q2}
      \end{pfenum*}
      \step{2r}{$\exists \st' \cdot $ [From \stepref{2n} + \stepref{2o} + \stepref{2q}.1 + CR($\kwT{setVar}$)]}
      \begin{pfenum*}
        \item $\rfun{\sst''}{\sst} \stRel \st'$ \label{2r1}
        \item $\st_1.\kwT{setVar}(\x, \vl_1) = \st'$ \label{2r2}
      \end{pfenum*}
      \step{2s}{$\rfun{\sst''}{\sst}\stRel\st'$ [From \stepref{2q}.2 + \stepref{2r}.1 + CR - Orders]}
      \step{2t}{$\rfun{\sst''}{\sst} \vdash \cs \rel_{v} \cs$ [From \stepref{2k} + noting that $\rfun{\sst''}{\sst} \rleq \sst$]}
      \step{2u}{$\rfun{\hat\cf''}{\hat\cf} \rel \cf'$ [From \stepref{2f} + \stepref{2g} + \stepref{2s} + \stepref{2t}]}
    \end{proofpf}
\end{proofpf}

\begin{theorem}[Soundness - General]\label{app:theo:soundness}
Let $\soundRel = \tup{\frestriction, \stRel, \vlRel}$ be a soundness relation
 for $\hat\gstate = \tup{\sset{\hat\gstate}, \hat\gval, \actions}$ in terms of $\gstate = \tup{\sset{\gstate}, \gval, \actions}$
 and $\leq$ the pre-order induced by $\stRel$. It holds that: 
 $$ 
 \hat\cf' \ssemarrow \hat\cf'' \gand \hat\cf \leq \rfun{\hat\cf'}{\hat\cf''} \gand \stR{\hat\cf}{\cf}
    \implies \exists \, \cf' \, . \, \cf \ssemarrow \cf' \gand \stR{\rfun{\hat\cf''}{\hat\cf}}{\cf'}
 $$
\end{theorem}

\begin{proofpf}
	\pf \pflongnumbers \ We proceed by induction on $n$.
  \step{1}{Base Case: $n = 0$}
  \begin{proofpf}
    \step{1a}{
      \begin{pfenum}
        \item (H1) $\hat\cf' \semarrow^0 \hat\cf''$
        \item (H2) $\hat\cf \leq \rfun{\hat\cf'}{\hat\cf''}$
        \item (H3) $\stR{\hat\cf}{\cf}$
      \end{pfenum}
    }
		\step{1d}{$\hat\cf' = \hat\cf''$ [From H1]} 
		\step{1e}{$\cf \semarrow^0 \cf'$ [-]}
		\step{1f}{To prove: $\stR{\rfun{\hat\cf'}{\hat\cf}}{\cf'}$}
		\begin{proofpf}
			\step{1a1}{$\hat\cf \leq \rfun{\hat\cf'}{\hat\cf'}$ [H2 + \stepref{1d}]}
			\step{1a2}{$\hat\cf \leq \hat\cf'$ [\stepref{1a1}]}
			\step{1a3}{$\hat\cf \leq \hat\cf$}
			\step{1a4}{$\hat\cf \rleq \hat\cf$ [\stepref{1a3}]}
			\step{1a5}{$\hat\cf \leq \rfun{\hat\cf'}{\hat\cf''}$ [\stepref{1a2} + \stepref{1a4}]}
			\step{1a6}{$\stR{\rfun{\hat\cf'}{\hat\cf}}{\cf}$ [\stepref{1a5} + (H3)]}
		\end{proofpf}
	\end{proofpf}
	\step{2}{Inductive Step: $n = k + 1$}
  \begin{proofpf}
    \step{2ass}{
      \begin{pfenum}
        \item (H1) $\hat\cf' \semarrow^{k+1} \hat\cf''$
        \item (H2) $\hat\cf \leq \rfun{\hat\cf'}{\hat\cf''}$
        \item (H3) $\stR{\hat\cf}{\cf}$ [From H1]
      \end{pfenum}
    }
		\step{2a}{$\exists \hat\cf_1$:}
		\begin{pfenum*}
			\item $\hat\cf' \semarrow^k \hat\cf_1$ [To prove: $\rfun{\hat\cf_1}{\hat\cf} \leq \rfun{\hat\cf_1}{\hat\cf''}$] \label{2a1}
			\item $\hat\cf_1 \semarrow \hat\cf''$ \label{2a2}
		\end{pfenum*}
		\step{2b}{$\hat\cf'' \rleq \hat\cf_1$ [From \stepref{2a}.2]}
		\step{2c}{$\hat\cf' \leq \hat\cf'$}
		\step{2d}{$\rfun{\hat\cf'}{\hat\cf''} \leq \rfun{\hat\cf'}{\hat\cf_1}$ [From \stepref{2b} + \stepref{2c}]}
		\step{2e}{$\hat\cf \leq \rfun{\hat\cf'}{\hat\cf_1}$ [From H2 + \stepref{2d}]}
		\step{2f}{$\exists \cf'$: [Applying the IH on H3 + \stepref{2a}.1 + \stepref{2e}]} 
		\begin{pfenum*}
			\item $\cf \semarrow^k \cf'$ \label{2f1}
			\item $\stR{\rfun{\hat\cf_1}{\hat\cf}}{\cf''}$ \label{2f2}
		\end{pfenum*}
		\step{2g}{$\hat\cf_1 \leq \hat\cf_1$ [-]}
		\step{2h}{$\hat\cf \rleq \hat\cf''$ [From H2]}
		\step{2i}{$\rfun{\hat\cf_1}{\hat\cf} \leq \rfun{\hat\cf_1}{\hat\cf''}$ [From \stepref{2g} + \stepref{2h}]}
		\step{2j}{$\exists \cf$: [Applying Theorem \ref{theo:soundness} to \stepref{2a}.1 + \stepref{2f}.1 + \stepref{2i}]} 
		\begin{pfenum*}
			\item $\cf'' \semarrow \cf$ \label{2j1}
			\item $\stR{\rfun{\hat\cf''}{\hat\cf}}{\cf'}$ \label{2j2}
		\end{pfenum*}
		\step{2k}{$\cf \semarrow^{k+1} \cf'$ [From \stepref{2f}.1 + \stepref{2j}.1]}
	\end{proofpf}
\end{proofpf}

From there, by choosing $\hat\cf \equiv \rfun{\hat\cf'}{\hat\cf''}$, we obtain the desired soundness result.

\begin{corollary}[Soundness]\label{app:cor:soundness}
Let $\soundRel = \tup{\frestriction, \stRel, \vlRel}$ be a soundness relation
 for $\hat\gstate = \tup{\sset{\hat\gstate}, \hat\gval, \actions}$ in terms of $\gstate = \tup{\sset{\gstate}, \gval, \actions}$
 and $\leq$ the pre-order induced by $\stRel$. It holds that: 
 $$ 
 \hat\cf \ssemarrow \hat\cf' \gand \stR{(\rfun{\hat\cf}{\hat\cf'})}{\cf}
    \implies \exists \, \cf' \, . \, \cf \ssemarrow \cf' \gand \stR{\hat\cf'}{\cf'}
 $$
\end{corollary}

\subsection{Concrete-Symbolic Soundness}

\begin{definition}[Symbolic Memory Interpretation]\label{app:def:interpretation}
Given a symbolic memory model $\smemory = \tup{\sset{\smemory}, \actions, \sea}$ 
and a concrete memory model $\cmemory = \tup{\sset{\cmemory}, \actions, \cea}$, 
an interpretation of $\smemory$ with respect to $\cmemory$ is a function 
$\interp : \sset{\smemory} \tmap (\lxs \pmap \vals) \pmap \sset{\cmemory}$ 
such that:
\begin{equation}
\begin{array}{l}
\action{\smem}{\act}{\sexp, \pc}{(\smem', \sexp', \pc')} 
  \gand  \cmem = \interp(\smem, \lenv)
  \gand \seval{\pc \, \wedge \, \pc'}{\lenv} = \true \\
   \qquad \qquad \implies \exists \, \cmem' \, . \, 
     \cmem' = \interp(\smem', \lenv) 
     \gand 
     \caction{\cmem}{\act}{\seval{\sexp}{\lenv}}{(\cmem', \seval{\sexp'}{\lenv})}
\end{array}\label{app:eq:interp:restriction}
\end{equation}
\end{definition}

\myparagraph{Allocator Interpretation and Restriction}
\begin{definition}[Symbolic Allocator Interpretation]
Given a symbolic allocator model $\hat\allocator = \tup{\sset{\hat\allocator}, \hat\gval}$ 
and a concrete allocator model $\allocator = \tup{\sset{\allocator}, \gval}$, 
an interpretation of $\hat\allocator$ with respect to $\allocator$ is a function 
$\allocint : \sset{\hat\allocator} \tmap (\hat\gval \pmap \gval) \pmap \sset{\allocator}$ 
such that:
\[
\begin{array}{l}
\alloc{\hat\arec}{j}{\hat\arec', \hat\gvl}{Y}
  \gand \arec = \allocint(\hat\arec, \lenv) 
 \implies  
 	\exists \, \arec'. \, 
	 \arec' = \allocint(\hat\arec', \lenv) \gand
     \alloc{\arec}{j}{\arec', \lenv(\hat\gvl)}{\lenv(Y)}
\end{array}\]
\end{definition}

A restriction operator is said to be
\emph{preserved} by an allocator model $\allocator = \tup{\sset{\allocator}, \gval}$ 
if it satisfies the following two properties: 
\begin{mathpar}
     \inferrule[RMono-Alloc]{}{\alloc{\arec}{j}{\arec', \gvl}{Y} \implies \arec' \rleq \arec}
     \qquad
\inferrule[FutureToPastAlloc]{}{
     \alloc{\arec}{j}{\arec', \gvl}{Y} \gand  \arec'' \rleq \rfun{\arec}{\arec'}  \\\\
     \implies \alloc{\arec''}{j}{\rfun{\arec'}{\arec''}, \gvl}{Y}}
   \end{mathpar}
We say that $\frestriction$ is a restriction operator on an allocator model
$\allocator = \tup{\sset{\allocator}, \gval}$, if $\frestriction$ is a restriction 
operator on the carrier set $\sset{\allocator}$ and $\frestriction$ is preserved by $\allocator$. 

\myparagraph{Lifting Interpretations}
Given an interpretation $\interp : \sset{\smemory} \tmap (\lxs \pmap \vals) \pmap \sset{\cmemory}$  
of a symbolic memory model $\smemory = \tup{\sset{\smemory}, \actions, \sea}$  in terms of 
a concrete memory model $\cmemory = \tup{\sset{\cmemory}, \actions, \cea}$, 
the candidate soundness relation $\interpLift(\interp) = \tup{\frestriction, \stRel, \vlRel}$ for 
$\sstateConstr(\smemory)$ in terms of $\cstateConstr(\cmemory)$ is defined as~follows:  
$$
\begin{array}{lll}
\rfun{\tup{\smem, \ssto, \sarec, \pc}}{\tup{\_, \_, \sarec', \pc'}} & \semeq  & \tup{\smem, \ssto, \rfun{\sarec}{\sarec'}, \pc \gand \pc'}  \\ 
\stR{\sst}{\st} & \semeq & \exists \, \lenv \, . \, (\st, \lenv) \in \modls(\sst) \\
\vlR{\tup{\_, \_, \_, \pc}}{\sexp}{\vl} & \semeq & \exists \, \lenv \, . \, \seval{\pc}{\lenv} = \true \gand \seval{\sexp}{\lenv} = \vl
\end{array}
$$
where: 
$$
   \modls(\tup{\smem, \ssto, \pc, \sarec})  \semeq 
           \left\{ (\tup{\cmem, \sto, \arec}, \lenv) \mid
                   \seval{\pc}{\lenv} = \true \gand
                   \cmem = \interp(\smem, \lenv) \gand 
                   \sto = \seval{\ssto}{\lenv} \gand  
                   \arec = \allocint(\sarec, \seval{.}{\lenv})  
            \right\} 
$$

\begin{definition}[Env]
  The function $\Env : \pcs \tmap \lenvs$ is defined as follow :
  $$
    \Env(\pc) \semeq \left\{ \lenv \mid \seval{\pc}{\lenv} = \true \right\}
  $$

  If $\sst =  \tup{\smem, \ssto, \sarec, \pc}$, we also note
    $$\Env(\sst) = \Env(\pc)$$
\end{definition}

\begin{definition}[Restriction of Interpretations with Path Conditions]
  Given a path condition $\pc \in \pcs$, a concrete memory model $\cmemory \in \cmemories$, a symbolic memory model $\smemory \in \smemories$ and an intepretation $\interp :: \sset{\smemory} \tmap \lenvs \pmap \sset{\cmemory}$ of $\smemory$ with respect to $\cmemory$, we define
  $\interp_\pc :: \sset{\smemory} \tmap \Env(\pc) \pmap \sset{\cmemory}$, the restriction of $\interp$ to $\Env(\pc)$ :
  $$
    \interp_\pc = \interp\!\!\mid\!\Env(\pc)
  $$
\end{definition}

\begin{lemma}\label{app:lem:eq:rleq}
  Let $\cmemory \in \cmemories$ and $\smemory \in \smemories$ be a concrete and a symbolic memory model and $\interp :: \sset\smemory \tmap \lenvs \pmap \sset\cmemory$ an interpretation of $\smemory$ with respect to $\cmemory$, and let $\soundRel = \tup{\frestriction, \stRel, \vlRel}$ be the candidate soundness relation between $\cstateConstr(\cmemory)$ and $\sstateConstr(\smemory)$ induced by $\interp$. It holds that :
  $$
    \sst \rleq \sst' \iff \Env(\sst) \subseteq \Env(\sst') \gand \sst'.\sarec \rleq \sst.\sarec
  $$
\end{lemma}
\begin{proofpf}
  \pf \pflongnumbers\\
  First, we need to observe that $\sst \rleq \sst'$ is just a shorthand for $\rfun{\sst}{\sst'} = \sst$. We rewrite the equivalence we want to prove as
  $$
    \rfun{\sst}{\sst'} = \sst \iff \Env(\sst) \subseteq \Env(\sst') \gand \sst'.\sarec \rleq \sst.\sarec
  $$

  We analyse one direction of the equivalence at a time, first destructuring on $\sst$ and $\sst'$.
  \step{lets}{\pflet{Let $\sst = \tup{\smem, \ssto, \sarec, \pc}$ and $\sst' = \tup{\smem', \ssto', \sarec', \pc'}$}}
  \step{ltr}{\prove{$\rfun{\sst}{\sst'} = \sst \implies \Env(\sst) \subseteq \Env(\sst') \gand \sst'.\sarec \rleq \sst.\sarec$}}
  \begin{proofpf}
    \step{ltr_assume}{
      \assume{$\rfun{\sst}{\sst'} = \sst$}
    }
    \step{ltr_1}{$\rfun{\sst}{\sst'} = \tup{\smem, \ssto, \rfun{\sarec}{\sarec'}, \pc \land \pc'}$ [From \stepref{lets} and \stepref{ltr_assume}]}
    \step{ltr_2}{$\pc \implies \pc'$ [From \stepref{ltr_assume} and \stepref{ltr_1}]}
    \step{ltr_3}{$\rfun{\sarec}{\sarec'} = \sarec$ [From \stepref{ltr_assume} and \stepref{ltr_1}]}
    \step{ltr_4}{$\Env(\sst) \subseteq \Env(\sst')$ [From \stepref{ltr_2}]}
    \step{ltr_5}{$\sarec' \rleq \sarec$ [From \stepref{ltr_3}]}
  \end{proofpf}
  \step{rtl}{\prove{$\rfun{\sst}{\sst'} = \sst \impliedby \Env(\sst) \subseteq \Env(\sst') \gand \sst'.\sarec \rleq \sst.\sarec$}}
  \begin{proofpf}
    \step{rtl_assume}{
      \assume{
        \begin{pfenum}
          \item $\Env(\sst) \subseteq \Env(\sst')$
          \item $\sst'.\sarec \rleq \sst.\sarec$
        \end{pfenum}
      }
    }
    \step{rtl_1}{$\pc \implies \pc'$ [From \stepref{rtl_assume}.1]}
    \step{rtl_2}{$\rfun{\sst}{\sst'} = \sst$ [From \stepref{rtl_assume}.2 and \stepref{rtl_1}]}
  \end{proofpf}
\end{proofpf}

\begin{lemma}
  Let $\sst = \tup{\smem, \ssto, \sarec, \pc}$ and $\sst' = \tup{\smem', \ssto', \sarec', \pc'}$. Then it holds that
  $$
    \begin{array}{llll}
                      &      &       & \interp_\pc(\smem) = \interp_\pc(\smem') \\
    \sst \leq \sst'   & \iff & \gand & \interp_\pc(\ssto) = \interp_\pc(\ssto') \\
                      &      & \gand & \pc \implies \pc' \\
                      &      & \gand & \interp_\pc(\sarec') \rleq \interp_\pc(\sarec) \\
    \end{array}
  $$
  
\end{lemma}

\begin{lemma}
  Let $\sst = \tup{\smem, \ssto, \sarec, \pc}$ and $\sst' = \tup{\smem', \ssto', \sarec', \pc'}$. Then it holds that
  $$
    \sst \rleq \sst' \iff (\pc \implies \pc') \gand \interp_\pc(\sarec') \rleq \interp_\pc(\sarec)
  $$
  
\end{lemma}

\begin{lemma}[Lifted Restriction Operator]\label{app:lem:liftedro}
    Let $\interp$ be an interpretation of a symbolic memory model $\smemory$ in terms of a concrete memory model $\cmemory$, and $\tup{\frestriction, \stRel, \vlRel} = \interpLift(\interp)$. Then $\frestriction$ is a restriction operator on $\sstateConstr(\smemory)$.
\end{lemma}
\begin{proofpf}
  \pf \pflongnumbers\\
  Let $\cmemory \in \cmemories$, $\smemory \in \smemories$, $\interp$ an interpretation of $\smemory$ in term of $\cmemory$, $\tup{\frestriction, \_, \_} = \interpLift(\interp)$.
  We need to prove that $\frestriction$ is a restriction operator on the carrier set $\sset{\sstateConstr(\smemory)}$, and that it is preserved by $\sstateConstr(\smemory)$.
  \step{ro}{\prove{$\frestriction$ is a restriction operator}}
  \begin{proofpf}
    In order to prove that $\frestriction$ is a restriction operator, we need to prove its associativity, idempotence, right-commutativity and that it has the Weakening property.
    We only provide the proof for associativity, the other proofs are analogous.
    \prove{$\rfun{(\rfun{\sst}{\sst'})}{\sst'} = \rfun{\sst}{(\rfun{\sst'}{\sst''})}$}
    \step{ro_lets}{
      \pflet{
        \begin{pfenum}
          \item $\sst = \tup{\smem, \ssto, \sarec, \pc}$
          \item $\sst' = \tup{\smem', \ssto', \sarec', \pc'}$
          \item $\sst'' = \tup{\smem'', \ssto'', \sarec'', \pc''}$
        \end{pfenum}
      }
    }
    \step{ro_1}{
      $\rfun{\sst}{\sst'} = \tup{\smem, \ssto, \rfun{\sarec}{\sarec'}, \pc \land \pc'}$
      [From \stepref{ro_lets}.1 and \stepref{ro_lets}.2]
    }
    \step{ro_2}{
      $\rfun{(\rfun{\sst}{\sst'})}{\sst'} = \tup{\smem, \ssto, \rfun{(\rfun{\sarec}{\sarec'})}{\sarec''}, \pc \land \pc' \land \pc''}$ [From \stepref{ro_lets}.3 and\stepref{ro_1}]
    }
    \step{ro_3}{
      $\rfun{\sst'}{\sst''} = \tup{\smem, \ssto, \rfun{\sarec'}{\sarec''}, \pc' \land \pc''}$
      [From \stepref{ro_lets}.2 and \stepref{ro_lets}.3]
    }
    \step{ro_4}{
      $\rfun{\sst}{(\rfun{\sst'}{\sst''})} = \tup{\smem, \ssto, \rfun{\sarec}{(\rfun{\sarec'}{\sarec''})}, \pc' \land \pc''}$
      [From \stepref{ro_lets}.1 and \stepref{ro_3}]
    }
    \step{ro_5}{
      $\rfun{(\rfun{\sst}{\sst'})}{\sst'} = \rfun{\sst}{(\rfun{\sst'}{\sst''})}$
      [From \stepref{ro_2}, \stepref{ro_4} and associativity of allocator restriction operator]}
  \end{proofpf}
  \step{m}{\prove{$\frestriction$ is preserved by $\sstateConstr(\smemory)$}}
  \begin{proofpf}
    In order to prove that $\frestriction$ is preserved by $\sstateConstr(\smemory)$, we need to prove that all of the state-generating functions exposed by the state model are monotonic with respect to the pre-order $\rleq$ induced by $\frestriction$. We only provide the proof for the monotonicity with respect to action executions, the other cases are analogous.
    \case{Actions}
    \prove{$\action{\sst}{\act}{v}{(\sst', -)} \implies \sst' \rleq \sst$}
    \step{m_let}{
      \pflet{$\sst = \tup{\smem, \ssto, \sarec, \pc} \in \sstateConstr(\smemory)$}
    }
    \step{m_assume}{
      \assume{$\action{\sst}{\act}{\vl}{(\st', -)}$}
    }
    \step{m_1}{$\exists \smem', \pc' \cdot$
      \begin{pfenum*}
        \item $(\smem', \sexp', \pc') \in \sea(\act, \smem, \sexp)$
        \item $\sst' = \tup{\smem', \ssto, \sarec, \pc \gand \pc'}$
      \end{pfenum*}
      [From \stepref{m_assume} and \stepref{m_1}]
    }
    \step{m_2}{$\Env(\sst') \subseteq \Env(\sst)$ [From \stepref{m_1}.2]}
    \step{m_3}{$\sst.\sarec \leq \sst'.\sarec$ [From \stepref{m_1}.2]}
    \step{m_4}{$\sst' \rleq \sst$ [From \stepref{m_2}, \stepref{m_3} and Lemma \ref{app:lem:eq:rleq}]}
  \end{proofpf}
\end{proofpf}

\begin{theorem}[Soundness Relation - Lifting]
Let $\interp$ be an interpretation of a symbolic memory model  $\smemory$  in terms of 
a concrete memory model $\cmemory$; then, $\interpLift(\interp) = \tup{\frestriction, \stRel, \vlRel}$ 
is a soundness relation for $\sstateConstr(\smemory)$ in terms of $\cstateConstr(\cmemory)$.
\end{theorem}
\begin{proofpf}
  \pf \pflongnumbers\\
  In order to establish that $\interpLift(\interp) = \tup{\frestriction, \stRel, \vlRel}$ is a soudness relation for $\sstateConstr(\smemory)$ in terms of $\cstateConstr(\cmemory)$, we need to prove that
  \begin{pfenum*}
    \item $\frestriction$ is a restriction order on $\sstateConstr(\smemory)$
    \item $\stRel$ and $\vlRel$ are monotonic with respect to the functions exposed by the state model
    \item $\frestriction$ is compatible with $\leq$
  \end{pfenum*}
  The first property is the result of Lemma \ref{app:lem:liftedro}. Let us prove property 2 and 3.
  \step{m}{\prove{$\stRel$ and $\vlRel$ are monotonic with respect to the functions exposed by the state model}}
  \begin{proofpf}
    We need to prove that the property holds for $\kwT{store}$, $\kwT{ee}$, $\kwT{setVar}$, $\kwT{setStore}$ and actions. We only provide the proof for the action cases. The other cases are analogous.
    \case{Actions}
    \prove{$\action{\sst}{\act}{\hat\vl}{(\sst', \hat\vl')} \gand \leqStR{\sst''}{\rfun{\sst}{\sst'}} \gand \ \stR{\sst''}{\st} \gand  \vlR{\sst''}{\hat\vl}{\vl} \\
    \qquad \implies \exists \, \st', \vl' \, . \, \action{\st}{\act}{\vl}{(\st', \vl')} \gand \stR{\rfun{\sst'}{\sst''}}{\st'} \gand  \vlR{\rfun{\sst'}{\sst''}}{\hat\vl'}{\vl'}$}
    \step{m_assume}{
      \assume{
        \begin{pfenum}
          \item (H1) $\sst = \tup{\smem, \ssto, \sarec, \pc}$
          \item (H2) $\sst'' = \tup{\smem'', \ssto'', \sarec'', \pc''}$
          \item (H3) $\st = \tup{\cmem, \sto, \arec}$
          \item (H4) $\action{\sst}{\act}{\hat\vl}{(\sst', \hat\vl')}$
          \item (H5) $\leqStR{\sst''}{\rfun{\sst}{\sst'}}$
          \item (H6) $\sst \stRel \st$
          \item (H7) $\vlR{\sst}{\hat\vl}{\vl}$
        \end{pfenum}}
    }
    \step{m_1}{$\exists \smem', \hat\vl', \pc' \cdot$ 
    \begin{pfenum*}
      \item $\action{\smem}{\act}{\hat\vl, \pc}{(\smem', \hat\vl', \pc')}$
      \item $\action{\sst}{\act}{\hat\vl}{\tup{\smem, \ssto, \sarec, \pc \gand \pc'}, \hat\vl')}$
      \item $\sst' = \tup{\smem, \ssto, \sarec, \pc \gand \pc'}$
    \end{pfenum*}
    [From H1 and H4]}
    \step{m_2}{$\exists \lenv \cdot $
    \begin{pfenum*}
      \item $\seval{\pc''}{\lenv} = \true$
      \item $\seval{\hat\vl}{\lenv} = \vl$
    \end{pfenum*}
    [From H2 and H7]}
    \step{m_3}{$\exists \lenv' \cdot$ 
    \begin{pfenum*}
      \item $\lenv \leq \lenv'$
      \item $\seval{\pc''}{\lenv'} = \true$
      \item $\interp(\smem'',\lenv') = \cmem$
      \item $\interp{\ssto''}{\lenv'} = \sto$
      \item $\interp(\sarec'') \rleq \arec$
    \end{pfenum*}
    [From H6, \stepref{m_2}.1 ]}
    \step{m_4}{$\rfun{\sst}{\sst'} = \tup{\smem, \ssto, \sarec, \pc \gand \pc'}$ [From H1 and \stepref{m_1}.3]}
    \step{m_5}{
      \begin{pfenum*}
        \item $\pc'' \implies \pc \gand \pc'$
        \item $\sarec \rleq \sarec''$
        \item $\interp_{\pc''}(\smem) = \interp_{\pc''}(\smem'')$
        \item $\interp_{\pc''}(\ssto) = \interp_{\pc''}(\ssto'')$ 
      \end{pfenum*}
      [From H2, H5, \stepref{m_4} ]
    }
    \step{m_6}{$\seval{\hat\vl}{\lenv'} = \vl$[From \stepref{m_2}.2 and \stepref{m_3}.1]}
    \step{m_7}{$\seval{\pc \gand \pc'} = \true$ [From \stepref{m_3}.2 and \stepref{m_5}.1]}
    \step{m_8}{$\interp{\smem, \lenv'} = \cmem$ [From \stepref{m_3}.2, \stepref{m_3}.3 and \stepref{m_5}.3]}
    \step{m_9}{$\exists \lenv'', \cmem' \cdot$
    \begin{pfenum*}
      \item $\cmem' = \interp(\smem',\lenv'')$
      \item $\lenv' \leq \lenv''$
      \item $\action{\cmem}{\act}{\vl}{(\cmem', \seval{\hat\vl'}{\lenv''})}$
    \end{pfenum*}
    [From \stepref{m_1}.1, \stepref{m_6}, \stepref{m_7}, \stepref{m_8}]
    }
    \step{m_10}{\pflet{$\st' = \tup{\cmem', \sto, \arec}$}}
    \step{m_11}{$\action{\st}{\act}{\vl}{(\st', \seval{\hat\vl'}{\lenv''})}$[From H3, \stepref{m_9}.3 and \stepref{m_10}]}
    \step{m_12}{$\rfun{\sst'}{\sst''} = \tup{\smem', \ssto, \sarec'', \pc''}$ [From \stepref{m_1}.3 and H2]}
    \step{m_13}{\prove{$(\st', \lenv'') \in \modls(\rfun{\sst'}{\sst''})$}}
    \begin{proofpf}
      \step{m_13_1}{$\seval{\pc''} = \true$ [From \stepref{m_3}.2 and \stepref{m_9}.2]}
      \step{m_13_2}{$\smem' = \interp(\smem, \lenv'')$ [From \stepref{m_9}.1]}
      \step{m_13_3}{$\sto = \interp(\ssto, \lenv'')$ [From \stepref{m_3}.4 and \stepref{m_9}.2]}
      \step{m_13_4}{$\arec = \interp(\sarec, \lenv'')$ [From \stepref{m_3}.5 and \stepref{m_9}.2]}
      \step{m_13_5}{$(\sst', \lenv'') \in \modls(\rfun{\sst'}{\sst''})$ [From \stepref{m_13_1} to \stepref{m_13_4}]}
    \end{proofpf}
    \step{m_14}{$\seval{\pc''}{\lenv''} = \true$}
    \step{m_15}{
      $\vlR{\rfun{\sst'}{\sst''}}{\hat\vl'}{\seval{\hat\vl}{\lenv''}}$
      [From \stepref{m_12} and \stepref{m_14}]
    }
  \end{proofpf}
  \step{c}{\prove{$\frestriction$ is compatible with $\leq$}}
  \begin{proofpf}
    In order to prove that $\frestriction$ is compatible with $\leq$, we need to prove the $\frestriction-\leq$ compatibility, the $\leq-\frestriction$ compatibility and the strengthening property. We only provide the proof for the strengthening property. The other cases are analogous.
    \step{c_lets}{\pflet{
      \begin{pfenum}
        \item (H1) $\sst_1 = \tup{\smem_1, \ssto_1, \sarec_1, \pc_1}$
        \item (H2) $\sst_2 = \tup{\smem_2, \ssto_2, \sarec_2, \pc_2}$
        \item (H3) $\sst_1' = \tup{\smem_1', \ssto_1', \sarec_1', \pc_1'}$
        \item (H4) $\sst_2' = \tup{\smem_2', \ssto_2', \sarec_2', \pc_2'}$
      \end{pfenum}
      be 4 symbolic states}}
    \step{c_assume}{
      \assume{
        \begin{pfenum}
          \item (H5) $sst_1 \leq \sst_1'$
          \item (H6) $\sst_2 \rleq \sst_2'$
        \end{pfenum}
      }
      \prove{$\rfun{\sst_1}{\sst_2} \leq \rfun{\sst_1'}{\sst_2'}$}
    }
    \step{c_1}{$\modls{\tup{\smem_1, \ssto_1, \sarec_1, \pc_1}} \subseteq \modls{\tup{\smem_1', \ssto_1', \sarec_1', \pc_1'}}$ [From H1, H2 and H5]}
    \step{c_2}{$(\pc_2 \implies \pc_2') \gand \sarec_2' \rleq \sarec_2$ [From H6]}
    \step{c_3}{$\rfun{\sst_1}{\sst_2} = \tup{\smem_1, \ssto_1, \rfun{\sarec_1}{\sarec_2}, \pc_1 \gand \pc_2}$[From H1 and H2]}
    \step{c_4}{$\rfun{\sst_1'}{\sst_2'} = \tup{\smem_1', \ssto_1,' \rfun{\sarec_1'}{\sarec_2'}, \pc_1' \gand \pc_2'}$ [From H3 and H4]}
    \step{c_5}{\prove{$\modls{\tup{\smem_1, \ssto_1, \rfun{\sarec_1}{\sarec_2}, \pc_1 \gand \pc_2}} \subseteq \modls{\tup{\smem_1', \ssto_1', \rfun{\sarec_1'}{\sarec_2'}, \pc_1' \gand pc_2'}}$}}
    \begin{proofpf}
      \step{c_5_h}{\pflet{$(\tup{\cmem, \sto, \arec}, \lenv) \in \modls{\tup{\smem_1, \ssto_1, \rfun{\sarec_1}{\sarec_2}, \pc_1 \gand \pc_2}}$}}
      \step{c_5_1}{
        \begin{pfenum*}
          \item $\seval{\pc_1 \gand \pc_2}{\lenv} = \true$
          \item $\cmem = \interp{\smem, \lenv}$
          \item $\interp{\rfun{\sarec_1}{\sarec_2}} \rleq \arec$
          \item $\sto = \interp{\ssto_1}{\lenv}$
        \end{pfenum*}
        [From \stepref{c_5_h}]
      }
      \step{c_5_2}{$\modls{\tup{\smem_1, \ssto_1, \rfun{\sarec_1}{\sarec_2},\pc_1 \gand \pc_2}} \subseteq \modls{\tup{\smem_1, \ssto_1, \sarec_1, \pc_1}}$}
      \step{c_5_3}{$\modls{\tup{\smem_1, \ssto_1, \rfun{\sarec_1}{\sarec_2},\pc_1 \gand \pc_2}} \subseteq \modls{\tup{\smem_1', \ssto_1', \sarec_1', \pc_1'}}$
      [From \stepref{c_1} and \stepref{c_5_2}]}
      \step{c_5_4}{$(\tup{\cmem, \sto\ arec}) \in \modls{\tup{\smem_1', \ssto_1', \sarec_1', \pc_1'}}$
      [From \stepref{c_5_h} and \stepref{c_5_3}]}
      \step{c_5_5}{
        \begin{pfenum*}
          \item $\seval{\pc_1'}{\lenv} = \true$
          \item $\cmem = \interp{\smem', \lenv}$
          \item $\interp{\sarec_1'} \rleq \arec$
          \item $\sto = \interp{\ssto_1'}{\lenv}$
        \end{pfenum*}
        [From \stepref{c_5_4}]
      }
      \step{c_5_7}{
        $\rfun{\interp(\sarec_1,\lenv)}{\interp(\sarec_2,\lenv)} \rleq \arec$
        [From \stepref{c_5_1}.3 ]
      }
      \step{c_5_8}{
        $\interp(\sarec_2,\lenv) \rleq \arec$
        [From \stepref{c_5_7}]
      }
      \step{c_5_9}{
        $\interp(\sarec_2',\lenv) \rleq \interp(\sarec_2, \lenv)$
        [From \stepref{c_2}]
      }
      \step{c_5_10}{
        $\interp(\sarec_2', \lenv) \rleq \arec$
        [From \stepref{c_5_8} and \stepref{c_5_9}]
      }
      \step{c_5_11}{
        $\rfun{\interp(\sarec_1', \lenv)}{\interp(\sarec_2', \lenv)} \rleq \arec$
        [From \stepref{c_5_5}.4 and \stepref{c_5_10}]
      }
      \step{c_5_12}{
        $\interp(\rfun{\sarec_1'}{\sarec_2'}, \lenv) \rleq \arec$
        [From \stepref{c_5_11} ]
      }
      \step{c_5_13}{
        $\seval{\pc_2}{\lenv} = \true$
        [From \stepref{c_5_1}.1]
      }
      \step{c_5_14}{
        $\seval{\pc_2'}{\lenv} = \true$
        [From \stepref{c_2} and \stepref{c_5_13}]
      }
      \step{c_5_15}{
        $\seval{\pc_1' \gand \pc_2'}{\lenv} = \true$
        [From \stepref{c_5_1}.1 and \stepref{c_5_14}]
      }
      \step{c_5_16}{$(\tup{\cmem, \sto, \arec}, \lenv) \in \modls(\tup{\smem_1', \ssto_1', \rfun{\sarec_1'}{\sarec_2'}, \pc_1' \gand \pc_2'})$}
    \end{proofpf}
  \end{proofpf}
\end{proofpf}
\subsection{\while: Sound Symbolic Analysis}

The interpretation of \while symbolic memories in terms of \while concrete memories, $\winterp$, 
is inductively defined as follows: 
\begin{mathpar}
\inferrule[Empty]{}{\winterp(\emptymem, \lenv) \semeq \emptymem} 
\and 
\inferrule[Cell]
   {
     \loc = \seval{\sexp}{\lenv} 
     \and
     \vl = \seval{\sexp'}{\lenv}
   }{\winterp(\wcell{\sexp}{\prop}{\sexp'}, \lenv) \semeq \wcell{\loc}{\prop}{\vl}} 
 \and 
\inferrule[Union]
   {
     \cmem_1 = \winterp(\smem_1, \lenv) 
     \and
     \cmem_2 = \winterp(\smem_2, \lenv)
   }{\winterp(\smem_1 \dunion \smem_2, \lenv) \semeq  \cmem_1 \dunion \cmem_2}  
\end{mathpar}

\begin{lemma}[\while: Memory Interpretation]
  $\winterp$ is an interpretation of $\wsmems$ with respect to $\wcmems$. 
  \end{lemma}
  \begin{proofpf}
    \pf \pflongnumbers\\
    For every action $\act \in \actions$, we have to prove that:
    \begin{equation}
      \begin{array}{l}
      \action{\smem}{\act}{\sexp, \pc}{(\smem', \sexp', \pc')} 
        \gand  \cmem = \interp(\smem, \lenv)
        \gand \seval{\pc \, \wedge \, \pc'}{\lenv} = \true \\
        \qquad \qquad \implies \exists \, \cmem' \, . \, 
          \cmem' = \interp(\smem', \lenv) 
          \gand 
          \caction{\cmem}{\act}{\seval{\sexp}{\lenv}}{(\cmem', \seval{\sexp'}{\lenv})}
      \end{array}
      \end{equation}
    \\
    We proceed by case analysis on the rule that was used to derive $\action{\smem}{\act}{\sexp, \pc}{(\smem', \sexp', \pc')}$.
  
    \step{l}{\case{$\lookupAction$}}
    \begin{proofpf}
      \step{l_assume}{
        \assume{
          \begin{pfenum*}
            \item (H1) $\action{\smem}{\lookupAction}{\litlst{\sexp, \prop}, \pc}{(\smem', \sexp', \pc')}$
            \item (H2) $\seval{\sexp}{\lenv} = \vl$
            \item (H3) $\cmem = \winterp(\smem, \lenv)$
          \end{pfenum*}
        }
      }
      \step{l_1}{$\exists\, \sexp'' \cdot$
        \begin{pfenum*}
          \item $\pc \vdash \sexp'' = \sexp$
          \item $\smem = \_ \dunion (\sexp'', \prop) \mapsto \sexp'$
          \item $\pc' = \pc$
          \item $\smem' = \smem$
        \end{pfenum*}
        [From H1]
      }
      \step{l_2}{$\cmem = \_ \dunion \winterp((\sexp'', \prop) \mapsto \sexp', \lenv)$ [From H3 + \stepref{l_1}.2]}
      \step{l_3}{$\cmem = \_ \dunion (\seval{\sexp''}{\lenv}, \prop) \mapsto \seval{\sexp'}{\lenv}$ [From \stepref{l_2}]}
      \step{l_4}{$\seval{\sexp''}{\lenv} = \seval{\sexp}{\lenv}$ [From H4 + \stepref{l_1}]}
      \step{l_5}{$\cmem = \_ \dunion (\seval{\sexp}{\lenv}, \prop) \mapsto \seval{\sexp'}{\lenv}$ [From \stepref{l_3} and \stepref{l_4}]}
      \step{l_6}{$\action{\cmem}{\lookupAction}{\litlst{\seval{\sexp}{\lenv}, \prop}}{(\cmem, \seval{\sexp'}{\lenv})}$ [From \stepref{l_5}]}
    \end{proofpf}
    \step{m}{\case{$\mutateAction$}}
    \begin{proofpf}
      \step{m_assume}{
        \assume{
          \begin{pfenum*}
            \item (H1) $\action{\smem}{\mutateAction}{\litlst{\sexp, \prop, \sexp'}, \pc}{(\smem', \sexp'', \pc')}$
            \item (H2) $\seval{\sexp}{\lenv} = \vl$ and $\seval{\sexp'}{\lenv} = \vl'$
            \item (H3) $\cmem = \winterp(\smem, \lenv)$
          \end{pfenum*}
        }
      }
      \step{m_1}{
        $\exists \sexp'', \smem'' \cdot$
        \begin{pfenum*}
          \item $\smem = \smem'' \dunion (\sexp'', \prop) \mapsto \_$ \label{1-1}
          \item $\pc' \vdash \sexp = \sexp''$ \label{1-2}
          \item $\smem' = \smem'' \dunion (\sexp'', \prop) \mapsto \sexp'$ \label{1-3}
        \end{pfenum*}
        [From H1]
      }
      \step{m_2}{$\cmem = \winterp(\smem'', \lenv) \dunion (\seval{\sexp''}{\lenv}, \prop) \mapsto \_$ [From H3 + \stepref{m_1}.3]}
      \step{m_3}{$\seval{\sexp''}{\lenv} = \seval{\sexp}{\lenv} = \vl$ [From H2 + H4 + \stepref{m_1}.\ref{1-2}]}
      \step{m_4}{$\exists \loc \cdot \vl = \loc$ [We assume $\lenv$ is "well-formed"]}
      \step{m_5}{$\cmem = \winterp(\smem'', \lenv) \dunion (\loc, \prop) \mapsto \_$ [From \stepref{m_2} + \stepref{m_4}]}
      \step{m_6}{
        $\action{\cmem}{\mutateAction}{\litlst{\loc, \prop, \vl'}}{(\winterp(\smem'', \lenv) \dunion (\loc, \prop) \mapsto \vl', \vl')}$
        }
      \step{m_7}{
        \vspace{-0.8cm}
        $\begin{array}{lll}
          &\\ &\\
          \winterp(\smem', \lenv) & = \winterp(\smem'', \lenv) \dunion \winterp((\sexp'', \prop) \mapsto \sexp', \lenv) &\\
          &= \winterp(\smem'', \lenv) \dunion (\seval{\sexp''}{\lenv}, \prop) \mapsto \seval{\sexp'}{\lenv} & \\
          &= \winterp(\smem'', \lenv) \dunion (\loc, \prop) \mapsto \vl' &\textrm{[From H2 + \stepref{m_3} + \stepref{m_4}]}
            \end{array}$
      }
    \end{proofpf}
    \step{d}{\case{$\disposeAction$}}
    \begin{proofpf}
      \step{d_assume}{
        \assume{
          \begin{pfenum*}
            \item (H1) $\action{\smem}{\disposeAction}{\sexp, \pc}{(\smem', \true, \pc)}$
            \item (H2) $\seval{\sexp}{\lenv} = \vl$
            \item (H3) $\cmem = \winterp(\smem, \lenv)$
          \end{pfenum*}
        }
      }
      \step{1}{$\memproj{\smem}{\sexp, \pc} = (\_, \smem')$ [From H1]}
      \step{2}{$\winterp(\memproj{\smem}{\sexp, \pc}, \lenv) = (\_,\winterp(\smem', \lenv))$ [From \stepref{1}]}
      \step{3}{$\winterp(\memproj{\smem}{\sexp, \pc}, \lenv) = \memproj{\winterp(\smem, \lenv)}{\seval{\sexp}{\lenv}}$}
      \step{4}{$\memproj{\winterp(\smem, \lenv)}{\seval{\sexp}{\lenv}} = \memproj{\cmem}{\seval{\sexp}{\lenv}}$ [From H3]}
      \step{5}{$\memproj{\cmem}{\seval{\sexp}{\lenv}} = (\_, \winterp(\smem', \lenv))$ [From \stepref{2} - \stepref{4}]}
      \step{6}{$\action{\cmem}{\disposeAction}{\seval{\sexp}{\lenv}}{(\winterp(\smem', \lenv), \true)}$}
    \end{proofpf}
  \end{proofpf}

\begin{theorem}[\while: Soundness]\label{app:theo:while:soundness}
Given $\tup{\frestriction, \stRel, \vlRel} = \interpLift(\winterp)$, it holds that:
 $$ 
 \hat\cf \ssemarrow_{\wsub} \hat{\cf'} \gand  \stR{(\rfun{\hat\cf}{\hat{\cf'}})}{\cf}
    \implies \exists \, \cf'. \, \cf \semarrow_\wsub \cf' \gand \stR{\hat{\cf'}}{\cf_2'}
 $$
\end{theorem}

\newpage
\section{Section 4: Parametric Verification}
\label{app:s4}

\subsection{Parametric Assertion Language}

\smallskip
\begin{display}{\gil Parametric Assertions}
\begin{tabular}{rcl}
   $\Pass, \Qass \in \Passes{\corePreds}$ & $\defeq$ & $\pc \mid \cPred{\e} \mid \aPred{\pn}{\e} \mid \astar{\Pass}{\Qass}$
   \\
   $\pred \in \predicates$ & $\defeq$ & $\fullpredicate{\pn}{\x}{\Pass_0}{\Pass_n}$
   \\
   $\pval \in \pvals{\gval}$ & $\defeq$ & $\pn(\gvl)$, where $\gvl \in \gval$
\end{tabular}
\end{display}

\begin{definition}[Predicate State Constructor (\vpstateConstr)]\label{app:def:pred:lifting}
The predicate state constructor $\vpstateConstr : \gstates \tmap \gstates$ is defined as  
$\vpstateConstr(\tup{\sset{\gstate}, \gval, \actions}) \semeq \tup{\sset{\gstate'}, \gval, \actions \dunion \lbrace \kwT{setP}, \kwT{getP}\rbrace}$,
where: 

{\small
\begin{tabular}{lll}
   \mybullet $\sset{\gstate'}$ &$\semeq$ & $\sset{\gstate} \times \lst{\pvals{\gval}}$ \\ 
  \mybullet $\kwT{setVar}_{\predsub}(\tup{\st, \lst{\pval}}, \x, \gvl)$ 
       & $\semeq$ & 
       $\tup{\kwT{setVar}(\st,  \x, \gvl), \lst{\pval}}$ \\ 
  \mybullet $\kwT{setStore}_{\predsub}(\tup{\st, \lst{\pval}}, \sto)$
     & $\semeq$ & 
     $\tup{\kwT{setStore}(\st, \sto), \lst{\pval}}$ \\ 
  \mybullet $\kwT{store}_{\predsub}(\tup{\st, -})$
    & $\semeq$ & 
    $\kwT{store}(\st)$ \\
  \mybullet $\kwT{ee}_{\predsub}(\tup{\st, -}, \e)$
    & $\semeq$ & 
    $\kwT{ee}(\st, \e)$ \\
 \mybullet $\kwT{ea}_{\predsub}(\act, \tup{\st, \lst{\pval}}, \gvl)$
    & $\semeq$ & 
    $\lbrace (\tup{\st', \lst{\pval}}, \gvl') \mid (\st', \gvl') \in \kwT{ea}(\act, \st, \gvl) \rbrace$, \textbf{if} $\act \not\in \lbrace \kwT{getP}, \kwT{setP} \rbrace$ \\
 \mybullet $\kwT{ea}_{\predsub}(\kwT{setP}, \tup{\st, \lst{\pval}}, \litlst{\pn, \gvl})$
    & $\semeq$ & 
    $(\tup{\st, (\pn(\gvl)\cons\lst{\pval})}, -)$ \\
    \mybullet $\kwT{ea}_{\predsub}(\kwT{getP}, \tup{\st, \lst{\pval}}, \litlst{\pn, \gvl})$
    & $\semeq$ & 
    $(\tup{\st, \lst{\pval}_1 \cat \lst{\pval}_2}, -)$, 
    where $\lst{\pval} = \lst{\pval}_1 \cat \litlst{\, \pn(\gvl) \,} \cat \lst{\pval}_2$
\end{tabular}}
\end{definition}

\begin{definition}[Core Predicate Action Interpretation]
A core predicate action interpretation is a 4-tuple $\actInterp \in \actInterps = \tup{\corePreds, \actions, \fset, \fget}$
consisting of a set of core predicates $\corePreds$, a set of actions $\actions$, and two functions 
$\fset, \fget : \corePreds \tmap \actions$; we write $\fsetter{\delta}$ for $\fset(\delta)$ and $\fgetter{\delta}$ for $\fget(\delta)$. 
A core predicate action interpretation $\tup{\corePreds, \actions, \fset, \fget}$ is said to be \emph{well-formed} 
with respect to a state model 
 $\gstate = \tup{\sset{\gstate}, \gval, \actions}$ in and only if, 
 for all core predicates $\corePred \in \corePreds$, it holds that: 
\begin{equation}\label{app:eq:well:formedness}
   \action{\st}{\fgetter{\delta}}{\gvl}{\st'} \iff 
      \action{\st'}{\fsetter{\delta}}{\gvl}{\st}     
\end{equation}
\end{definition} 

Given a state model $\gstate = \tup{\sset{\gstate}, \gval, \actions}$ and a core action 
interpretation $\tup{\corePreds, \actions, \fset, \fget}$, the induced action interpretation of an 
assertion $\Pass \in \Passes{\corePreds}$ is a pair of functions consisting of the 
\emph{getter} and \emph{setter} of $\Pass$, respectively, $\fgetter{\Pass}$ and $\fsetter{\Pass}$. 
Formally, we define two induced functions: 
\begin{itemize}
   \item $\setInterpS{\corePreds}{\gstate} : \Passes{\corePreds} \tmap \sset{\gstate} \tmap (\lxs \dunion \xs \pmap \gval) \pmap \sset{\gstate}$
      \hfill
      ($\st' = \setInterpS{\corePreds}{\gstate}(\Pass, \st, \subst) \equiv_{pp} \actionSetterS{\st}{\Pass}{\subst}{\st'}$)
    \item $\getInterpS{\corePreds}{\gstate} : \Passes{\corePreds} \tmap \sset{\gstate} \tmap (\lxs \dunion \xs \pmap \gval) \pmap \sset{\gstate}$  
         \hfill
         ($\st' = \getInterpS{\corePreds}{\gstate}(\Pass, \st, \subst) \equiv_{pp}  \actionGetter{\st}{\Pass}{\subst}{\st'}$)
 \end{itemize}
mapping each assertion $\Pass \in \Passes{\corePreds}$ to its getter and setter, respectively, as follows:

\medskip
\begin{display}{Assertion Interpretation: $\actionSetterS{\st}{\Pass}{\subst}{\st'}$ and $\actionGetter{\st}{\Pass}{\subst}{\st'}$}
{\small
\begin{mathpar}
\inferrule[set - star]{
   \actionSetterS{\st}{\Pass}{\subst}{\st'}
   \\\\
   \actionSetterS{\st'}{\Qass}{\subst}{\st''}
}{
   \actionSetterS{\st}{(\astar{\Pass}{\Qass})}{\subst}{\st''}
}
\and 
\inferrule[get - star]{
   \actionGetter{\st}{\Pass}{\subst}{\st'}
   \\\\
   \actionGetter{\st'}{\Qass}{\subst}{\st''}
  }{
    \actionGetter{\st}{(\astar{\Pass}{\Qass})}{\subst}{\st''}
  } 
\and
\inferrule[set - boolean expr]
  {
    \ee{\st}{\subst(\e)} = \gvl
    \\\\
    \actionT{\st}{assume}{\gvl}{\st'}
  }{
    \actionSetterS{\st}{\pc}{\subst}{\st'}
  }
\and 
\inferrule[get - boolean expr]
  {
      \ee{\st}{\lnot \, \subst(\e)} = \gvl 
      \\\\
      \st.\kwT{assume}(\gvl) = \emptyset 
  }{
    \actionGetter{\st}{\pc}{\subst}{\st}
 } 
\and
\inferrule[set - pred]{
  \ee{\st}{\litlst{\pn, \subst(\e)}} = \gvl
  \\\\
  \action{\st}{setP}{\gvl}{\st'}
}{
   \actionSetterS{\st}{\aPred{\pn}{\e}}{\subst}{\st'}
}
\and 
\inferrule[get - pred]{
  \ee{\st}{\litlst{\pn, \subst(\e)}} = \gvl
  \\\\
  \action{\st}{getP}{\gvl}{\st'}
}{
    \actionGetter{\st}{\aPred{\pn}{\e}}{\subst}{\st'}
  } 
\and
\inferrule[set - core pred]{
   \ee{\st}{\subst(\e)} = \gvl
   \\\\
   \action{\st}{\fsetter{\delta}}{\gvl}{\st'} 
}{
   \actionSetterS{\st}{\cPred{\e}}{\subst}{\st'}
}
\and 
\inferrule[get - core pred]{
   \ee{\st}{\subst(\e)} = \gvl
   \\\\
   \action{\st}{\fgetter{\delta}}{\gvl}{\st'}
}{
   \actionGetter{\st}{\cPred{\e}}{\subst}{\st'}
}
\end{mathpar}}
\end{display}

\begin{lemma}[Assertion Interpretation]\label{app:lemma:asrt:interp}
Let $\tup{\corePreds, \actions, \fset, \fget}$ be a well-formed core predicate interpretation 
with respect to a predicate state model $\gstate = \tup{\sset{\gstate}, \gval, \actions}$; then, it holds that: 
$\actionSetterS{\st}{\Pass}{\subst}{\st'}$ if and only if 
$\actionGetter{\st'}{\Pass}{\subst}{\st}$. 
\end{lemma}

\begin{theorem}[Assertion Interpretation - Soundness]\label{app:teo:asrt:soundness}
Let $\soundRel = \tup{\frestriction, \stRel, \vlRel}$ be a soundness relation
 for $\hat\gstate = \tup{\sset{\hat\gstate}, \hat\gval, \actions}$ in terms of $\gstate = \tup{\sset{\gstate}, \gval, \actions}$
 and $\leq$ the pre-order induced by $\stRel$; and let $\tup{\corePreds, \actions, \fset, \fget}$ be a 
 well-formed core predicate action interpretation for $\hat\gstate$ and $\gstate$; 
 then, it holds that: 
 \begin{align}
\begin{array}{l}
 \actionSetterS{\sst}{\Pass}{\hat\subst}{\sst'} \gand  
   \sst'' \leq \rfun{\sst}{\sst'} \gand 
   \stR{\sst''}{\st} \gand
    \vlR{\sst''}{\hat\subst}{\subst}  \\ 
   \qquad \qquad \implies 
   \exists \, \st' . \ \
       \actionSetterS{\st}{\Pass}{\subst}{\st'} \gand 
       \stR{\rfun{\sst'}{\sst''}}{\st'} 
 \end{array}
  \\
 \begin{array}{l}
  \actionGetter{\sst}{\Pass}{\hat\subst}{\sst'} \gand 
     \sst'' \leq \rfun{\sst}{\sst'} \gand 
      \stR{\sst''}{\st} \gand 
      \vlR{\sst''}{\hat\subst}{\subst} \\
      \qquad \qquad \implies 
       \exists \, \st' . \ \ 
         \actionGetter{\st}{\Pass}{\subst}{\st'} \gand 
           \stR{\rfun{\sst'}{\sst''}}{\st'} 
 \end{array}
\end{align}
\end{theorem}

\begin{proofpf}
	\pf
	\pflongnumbers \\
	We are going to prove equations the two equations by induction on the structure of P, considering only the core predicate and separating conjunction cases for both proofs. The other cases are analogous.
	\step{set}{\prove{$\actionSetterS{\sst}{\Pass}{\hat\subst}{\sst'} \gand  
  \sst'' \leq \rfun{\sst}{\sst'} \gand 
  \stR{\sst''}{\st} \gand
   \vlR{\sst''}{\hat\subst}{\subst}  \\ 
  \qquad \qquad \implies 
  \exists \, \st' . \ \
      \actionSetterS{\st}{\Pass}{\subst}{\st'} \gand 
      \stR{\rfun{\sst'}{\sst''}}{\st'} $}}
	\begin{proofpf}
		\step{set_cpred}{\case{$\cPred\e$}}
		\begin{proofpf}
			\step{assume_set_cpred}{
				\assume{
					\begin{pfenum}
						\item (H1) $\actionSetterS{\sst}{\cPred\e}{\hat\subst}{\sst'}$
						\item (H2) $\sst'' \leq \rfun{\sst}{\sst'}$
						\item (H3) $\stR{\sst''}{\st}$
						\item (H4) $\vlR{\sst''}{\hat\subst}{\subst}$
					\end{pfenum}
				}
				\prove{
				 $\exists \, \st' . \ \
				 \actionSetterS{\st}{\cPred\e}{\subst}{\st'} \gand 
				 \stR{\rfun{\sst'}{\sst''}}{\st'}$
				}
			}
			\step{set_cpred_1}{
				$\exists \hat\vl.$
				\begin{pfenum*}
					\item $\ee{\sst}{\hat\subst(\e)} = \hat\vl$
					\item $\action{\sst}{\fsetter{\delta}}{\hat\vl}{\sst'}$
				\end{pfenum*}
				[From H1]
			}
			\step{set_cpred_2}{
				$\exists \vl.$
				\begin{pfenum*}
					\item $\ee{\st}{\subst(\e)} = \vl$
					\item $\vlR{\sst''}{\hat\vl}{\vl}$
				\end{pfenum*}
				[From H2, H3, H4 and \stepref{set_cpred_1}.1]
			}
			\step{set_cpred_3}{
				$\exists \hat\st'.$
				\begin{pfenum*}
					\item $\action{\st}{\fsetter{\delta}}{\vl}{\st'}$
					\item $\stR{\rfun{\sst'}{\sst''}}{\st'}$
				\end{pfenum*}
				[From H2, H3 \stepref{set_cpred_1}.2 and \stepref{set_cpred_2}.2]
			}
			\step{set_cpred_4}{
				$\actionSetterS{\st}{\cPred\e}{\subst}{\st'}$} [From \stepref{set_cpred_2}.1 and \stepref{set_cpred_3}.1]
		\end{proofpf}
		\step{set_star}{\case{$\astar{\Pass}{\Qass}$}}
		\begin{proofpf}
			\step{assume_set_star}{
				\assume{
					\begin{pfenum}
						\item (H1) $\actionSetterS{\sst}{\astar{\Pass}{\Qass}}{\hat\subst}{\sst'}$
						\item (H2) $\sst'' \leq \rfun{\sst}{\sst'}$
						\item (H3) $\stR{\sst''}{\st}$
						\item (H4) $\vlR{\sst''}{\hat\subst}{\subst}$
					\end{pfenum}
				}
				\prove{
				 $\exists \, \st' . \ \
				 \actionSetterS{\st}{\astar{\Pass}{\Qass}}{\subst}{\st'} \gand 
				 \stR{\rfun{\sst'}{\sst''}}{\st'}$
				}
			}
			\step{set_star_1}{
				$\exists \sst_1.$
				\begin{pfenum*}
					\item $\actionSetterS{\sst}{\Pass}{\hat\subst}{\sst_1}$
					\item $\actionSetterS{\sst_1}{\Qass}{\hat\subst}{\sst'}$
				\end{pfenum*}
			}
			\step{set_star_2}{
				$\sst' \rleq \sst_1$
				[From \stepref{set_star_1}.2]
			}
			\step{set_star_3}{
				$\st \leq \sst$
			}
			\step{set_star_4}{
				$\rfun{\sst}{\sst'} \leq \rfun{\sst}{\sst_1}$
				[From \stepref{set_star_2} and \stepref{set_star_3}]
			}
			\step{set_star_5}{
				$\sst'' \leq \rfun{\sst}{\sst_1}$
				[From (H2) and \stepref{set_star_4}]
			}
			\step{set_star_6}{
				$\exists \st_1.$
				\begin{pfenum*}
					\item $\actionSetterS{\st}{\Pass}{\subst}{\st_1}$
					\item $\stR{\rfun{\sst_1}{\sst''}}{\st_1}$
				\end{pfenum*}
				[From H3, H4, \stepref{set_star_1}.1 and IH]
			}
			\step{set_star_7}{
				$\sst_1 \leq \sst_1$
			}
			\step{set_star_8}{
				$\sst'' \rleq \sst'$ [From H2]
			}
			\step{set_star_9}{
				$\sst'' \rleq \sst'$ [From \stepref{set_star_7} and \stepref{set_star_8}]
			}
			\step{set_star_10}{
				$\sst'' \rleq \sst_1$ [From \stepref{set_star_5}]
			}
			\step{set_star_11}{
				$\sst'' \rleq \sst''$
			}
			\step{set_star_12}{
				$\sst'' \rleq \rfun{\sst_1}{\sst''}$ [From \stepref{set_star_10} and \stepref{set_star_11}]
			}
			\step{set_star_13}{
				$\vlR{\rfun{\sst_1}{\sst''}}{\hat\subst}{\subst}$ [From H4 and \stepref{set_star_12}]
			}
			\step{set_star_14}{
				$\exists \st'.$
				\begin{pfenum*}
					\item $\actionSetterS{\st_1}{\Qass}{\subst}{\st'}$
					\item $\stR{\rfun{\sst'}{(\rfun{\sst_1}{\sst''})}}{\st'}$
				\end{pfenum*}
				[From \stepref{set_star_1}.2 and \stepref{set_star_6}.2]
			}
			\step{set_star_15}{
				$\sst' \rleq \sst_1$ [From \stepref{set_star_1}.2]
			}
			\step{set_star_16}{
        \vspace{-0.4cm}
        $\begin{array}{lll}
          &&\\
					\rfun{\sst'}{\rfun{\sst_1}{\sst''}} & = & \rfun{(\rfun{\sst'}{\sst_1})}{\sst''}\\
					     																& = & \rfun{\sst'}{\sst''}
				\end{array}$
				[From \stepref{set_star_15}]
			}
			\step{set_star_17}{$\stR{\rfun{\sst'}{\sst''}}{\st'}$ [From \stepref{set_star_16} and \stepref{set_star_14}.2]}
			\step{set_star_18}{$\actionSetterS{\st}{\astar{\Pass}{\Qass}}{\subst}{\st'}$ [From \stepref{set_star_6}.1 and \stepref{set_star_14}.2]}
		\end{proofpf}
	\end{proofpf}
	\step{get}{\prove{$
  \actionGetter{\sst}{\Pass}{\hat\subst}{\sst'} \gand 
     \sst'' \leq \rfun{\sst}{\sst'} \gand 
      \stR{\sst''}{\st} \gand 
      \vlR{\sst''}{\hat\subst}{\subst} \\
      \qquad \qquad \implies 
       \exists \, \st' . \ \ 
         \actionGetter{\st}{\Pass}{\subst}{\st'} \gand 
           \stR{\rfun{\sst'}{\sst''}}{\st'} $}}
	\begin{proofpf}
		\step{get_cpred}{\case{$\cPred\e$}}
		\begin{proofpf}
			\step{assume_get_cpred}{
				\assume{
					\begin{pfenum}
						\item (H1) $\actionGetter{\sst}{\cPred\e}{\hat\subst}{\sst'}$
						\item (H2) $\sst'' \leq \rfun{\sst}{\sst'}$
						\item (H3) $\stR{\sst''}{\st}$
						\item (H4) $\vlR{\sst''}{\hat\subst}{\subst}$
					\end{pfenum}
				}
				\prove{
				 $\exists \, \st' . \ \
				 \actionGetter{\st}{\cPred\e}{\subst}{\st'} \gand 
				 \stR{\rfun{\sst'}{\sst''}}{\st'}$
				}
			}
			\step{get_cpred_1}{
				$\exists \hat\vl.$
				\begin{pfenum*}
					\item $\ee{\sst}{\hat\subst(\e)} = \hat\vl$
					\item $\action{\sst}{\fgetter{\delta}}{\hat\vl}{\sst'}$
				\end{pfenum*}
				[From H1]
			}
			\step{get_cpred_2}{
				$\exists \vl.$
				\begin{pfenum*}
					\item $\ee{\st}{\subst(\e)} = \vl$
					\item $\vlR{\sst''}{\hat\vl}{\vl}$
				\end{pfenum*}
				[From H2, H3, H4 and \stepref{get_cpred_1}.1]
			}
			\step{get_cpred_3}{
				$\exists \hat\st'.$
				\begin{pfenum*}
					\item $\action{\st}{\fgetter{\delta}}{\vl}{\st'}$
					\item $\stR{\rfun{\sst'}{\sst''}}{\st'}$
				\end{pfenum*}
				[From H2, H3 \stepref{get_cpred_1}.2 and \stepref{get_cpred_2}.2]
			}
			\step{get_cpred_4}{
				$\actionGetter{\st}{\cPred\e}{\subst}{\st'}$} [From \stepref{get_cpred_2}.1 and \stepref{get_cpred_3}.1]
		\end{proofpf}
		\step{get_star}{\case{$\astar{\Pass}{\Qass}$}}
		\begin{proofpf}
			\step{assume_get_star}{
				\assume{
					\begin{pfenum}
						\item (H1) $\actionGetter{\sst}{\astar{\Pass}{\Qass}}{\hat\subst}{\sst'}$
						\item (H2) $\sst'' \leq \rfun{\sst}{\sst'}$
						\item (H3) $\stR{\sst''}{\st}$
						\item (H4) $\vlR{\sst''}{\hat\subst}{\subst}$
					\end{pfenum}
				}
				\prove{
				 $\exists \, \st' . \ \
				 \actionSetterS{\st}{\astar{\Pass}{\Qass}}{\subst}{\st'} \gand 
				 \stR{\rfun{\sst'}{\sst''}}{\st'}$
				}
			}
			\step{get_star_1}{
				$\exists \sst_1.$
				\begin{pfenum*}
					\item $\actionGetter{\sst}{\Qass}{\hat\subst}{\sst_1}$
					\item $\actionGetter{\sst_1}{\Pass}{\hat\subst}{\sst'}$
				\end{pfenum*}
			}
			\step{get_star_2}{\prove{$\sst'' \leq \rfun{\sst}{\sst_1}$}}
			\begin{proofpf}
				\step{get_star_2_1}{$\sst'' \leq \sst$ [From H2]}
				\step{get_star_2_2}{$\sst' \rleq \sst_1$ [From \stepref{get_star_1}.2]}
				\step{get_star_2_3}{$\rfun{\sst''}{\sst'} \leq \rfun{\sst''}{\sst_1}$ [From \stepref{get_star_2_1} and \stepref{get_star_2_2}]}
				\step{get_star_2_4}{$\sst'' \leq \rfun{\sst''}{\sst_1}$ [From H2 and \stepref{get_star_2_3}]}
			\end{proofpf}
			\step{get_star_3}{
				$\exists \st_1.$
				\begin{pfenum*}
					\item $\actionGetter{\st}{\Qass}{\subst}{\st_1}$
					\item $\stR{\rfun{\sst_1}{\sst''}}{\st_1}$
				\end{pfenum*}
			}
			\step{get_star_4}{\prove{$\rfun{\sst_1}{\sst''} \leq \rfun{\sst_1}{\sst'}$}}
			\begin{proofpf}
				\step{get_star_4_1}{$\sst_1 \leq \sst_1$}
				\step{get_star_4_2}{$\sst'' \rleq \sst'$ [From H2]}
				\step{get_star_4_3}{$\rfun{\sst_1}{\sst''} \leq \rfun{\sst_1}{\sst'}$ [From \stepref{get_star_4_1} and \stepref{get_star_4_2}]}
			\end{proofpf}
			\step{get_star_5}{\prove{$\vlR{\rfun{\sst_1}{\sst''}}{\hat\subst}{\subst}$}}
			\begin{proofpf}
				\step{get_star_5_1}{$\sst' \rleq \sst_1$ [From \stepref{get_star_1}.2]}
				\step{get_star_5_2}{$\sst'' \rleq \sst'$ [From H2]}
				\step{get_star_5_3}{$\sst'' \rleq \sst_1$ [From \stepref{get_star_5_1} and \stepref{get_star_5_2}]}
				\step{get_star_5_4}{$\sst'' \rleq \rfun{\sst_1}{\sst''}$ [From \stepref{get_star_5_3}]}
				\step{get_star_5_5}{$\vlR{\rfun{\sst_1}{\sst''}}{\hat\subst}{\subst}$ [From H4 and \stepref{get_star_5_4}]}
			\end{proofpf}
			\step{get_star_6}{
				$\exists \st'.$
				\begin{pfenum*}
					\item $\actionGetter{\st_1}{\Pass}{\subst}{\st'}$
					\item $\stR{\rfun{\sst'}{(\rfun{\sst_1}{\sst''})}}{\st'}$
				\end{pfenum*}
				[From \stepref{get_star_1}.2, \stepref{get_star_3}.2, \stepref{get_star_4}, \stepref{get_star_5} and IH]
			}
			\step{get_star_7}{\prove{$\rfun{\sst'}{(\rfun{\sst_1}{\sst''})} = \rfun{\sst'}{\sst''}$}}
			\begin{proofpf}
				\step{get_star_7_1}{$\sst' \rleq \sst_1$ [From \stepref{get_star_1}.2]}
				\step{get_star_7_2}{$\rfun{\sst'}{(\rfun{\sst_1}{\sst''})} = \rfun{(\rfun{\sst'}{\sst_1})}{\sst''}$}
				\step{get_star_7_3}{$\rfun{\sst'}{(\rfun{\sst_1}{\sst''})} = \rfun{\sst'}{\sst''}$ [From \stepref{get_star_7_1} and \stepref{get_star_7_2}]}
			\end{proofpf}
			\step{get_star_8}{$\stR{\rfun{\sst'}{\sst''}}{\st'}$}
			\step{get_star_9}{$\actionGetter{\st}{\astar{\Pass}{\Qass}}{\subst}{\st'}$ [From \stepref{get_star_3}.1 and \stepref{get_star_6}.1]}
		\end{proofpf}
	\end{proofpf}
\end{proofpf}

\subsection{Parametric Verification Semantics}

\medskip
\begin{display}{Extended \gil Syntax}
\begin{tabular}{l@{\qquad\qquad\qquad\qquad}l}
  $\vcm \in \vcmds{\actions}$ $\defeq$ $\cm \in \cmds{\actions} \mid \vcall{\x}{\e_1}{\e_2}{j; \left(\lx_i: \e_i\right)\!\mid_{i=0}^n}\mid$ &   
                $\vproc \in \vprocs{\actions}$ $\defeq$ $\vprocedure{f}{\x}{\lst{\vcm}}$
  \\
    \hspace*{1.32cm}$\vfold{\pn}{\e}{j; \left(\lx_i: \e_i\right)\!\mid_{i=0}^n} \mid \vunfold{\pn}{\e}$
    & 
    $\vprog \in \vprogs{\actions}$ $:$ $\fids \pmap \vprocs{\actions}$
 \end{tabular}
\end{display}

\medskip
\begin{display}{Verification Semantics of \gil: $\vfullsemtrans{\st, \cs, i}{\st', \cs', j}{}{\vprog}{\outcome'}{\outcome}$}
{\footnotesize
\begin{mathpar}
\inferrule[\textsc{Non-logical Cmd}]
  {
     \cmd(\prog, \cs, i) \in \cmds{\actions}
      \\\\
      \semtrans{\st, \cs, i}{\st', \cs', j}{}{\downcast{\vprog}}{}  
  }{
    \vsemtrans{\st, \cs, i}{\st', \cs', j}{}{\vprog}{}
  }
\and 
  \inferrule[\textsc{UnFold}]
  {
     \cmd(\vprog, \cs, i) = \vunfold{\pn}{\e}
      \and
      \gvl = \ee{\st}{\e}
      \\\\
          \getppred{\vprog}{\pn}{} = \fullpredicate{\pn}{\x}{\Pass_0}{\Pass_n}
          \quad
          0 \leq j \leq n
      \\\\
        \action{\st}{getP}{\litlst{\pn, \gvl}}{\st'}
        \and
      {\actionSetter{\st'}{\Pass_j}{[x \mapsto \gvl]}{(\st'', -)}}
  }{
    \vsemtrans{\st, \cs, i}{\st'', \cs, i{+}1}{}{\vprog}{}
  }
\end{mathpar}}
\end{display}

\pagebreak
\begin{display}{Verification Semantics of \gil: $\vfullsemtrans{\st, \cs, i}{\st', \cs', j}{}{\vprog}{\outcome'}{\outcome}$ (continued)}
{\footnotesize
\begin{mathpar}
  \inferrule[\textsc{Fold}]
  {
     \cmd(\vprog, \cs, i) = \vfold{\pn}{\e_0}{j; \left(\lx_i: \e_i\right)\!\mid_{i=1}^n}
      \\\\
      \left(\gvl_i = \ee{\st}{\e_i}\right)\mid_{i=0}^n 
      \\\\
      \getppred{\vprog}{\pn}{} = \fullpredicate{\pn}{\x}{\Pass_0}{\Pass_n}
       \\\\
      \subst = [x \mapsto \gvl_0, \lx_1 \mapsto \gvl_1, ..., \lx_n \mapsto \gvl_n]
      \\\\
      \actionGetter{\st}{\Pass_j}{\subst}{\st'}
      \and 
      \action{\st'}{setP}{\litlst{\pn, \gvl}}{\st''}  
  }{
    \vsemtrans{\st, \cs, i}{\st'', \cs, i{+}1}{}{\vprog}{}
  }
  \and
\inferrule[\textsc{Spec Call}]
  {
     \cmd(\vprog, \cs, i) = \vcall{\x}{\e}{\e_0}{j; \left(\lx_i: \e_i\right)\!\mid_{i=1}^n}
      \\\\
      \f = \ee{\st}{\e}
      \and 
      \left(\gvl_i = \ee{\st}{\e_i}\right)\mid_{i=0}^n 
      \\\\
      \getspec{\vprog}{\f}{j} = \spec{\Pass}{\f}{\x_0}{\Qass}{\e'}
       \\\\
      \subst = [x_0 \mapsto \gvl_0, ..., \lx_n \mapsto \gvl_n]
      \\\\
      \ee{\st}{\subst(\e')} = \gvl
      \and
      \actionGetter{\st}{\Pass}{\subst}{\st'}
      \and 
      \actionSetterS{\st'}{\Qass}{\subst}{\st''}  
  }{
    \vsemtrans{\st, \cs, i}{\st''.\kwT{setVar}(\x, v), \cs, i{+}1}{}{\vprog}{}
  }
\end{mathpar}}
\end{display}

\begin{theorem}[Verification]\label{app:theo:verification:preds}
Let $\soundRel = \tup{\frestriction, \stRel, \vlRel}$ be a soundness relation
 for $\gsstate = \tup{\sset{\gsstate}, \gsval, \actions}$ in terms of $\gstate = \tup{\sset{\gstate}, \gval, \actions}$
 and $\leq$ the pre-order induced by $\stRel$. It holds that: 
 $$ 
 \hat\cf~\vssemarrow~\hat\cf' \gand \hat\cf = \rfun{\hat\cf}{\hat\cf'} \gand \stR{\hat\cf}{\cf}
    \implies \exists \, \cf' \, . \, \cf~\vssemarrow~\cf' \gand \stR{\hat\cf'}{\cf'}
 $$
\end{theorem}

\begin{proofpf}
	\pf
	\pflongnumbers \\
	In order to prove this theorem, we prove the following more general claim, from which the theorem immediately follows :
	$$ 
	 \hat\cf~\vssemarrow~\hat\cf' \gand \hat\cf'' \leq \rfun{\hat\cf}{\hat\cf'} \gand \stR{\hat\cf''}{\cf}
			\implies \exists \, \cf' \, . \, \cf~\vssemarrow~\cf' \gand \stR{\rfun{\hat\cf'}{\hat\cf''}}{\cf'}
	 $$
	We proceed by use analysis on the rule used to derive $\hat\cf~\vssemarrow~\hat\cf'$. We only cover the non-logic command, unfold and spec call cases, the other cases being analogous.
	\step{nlc}{\case{Non-logic commands}}
	\begin{proofpf}
		\step{assume_nlc}{
			\assume{
				\begin{pfenum}
					\item (H1) $\hat\cf~\vssemarrow~\hat\cf'$
					\item (H2) $\hat\cf'' \leq \rfun{\hat\cf}{\hat\cf'}$
					\item (H3) $\stR{\hat\cf''}{\cf}$
				\end{pfenum}
			}
			\prove{
				$\exists \, \cf' \, . \, \cf~\vssemarrow~\cf' \gand \stR{\hat\cf'}{\cf'}$
			}
		}
		\step{nlc_1}{$\hat\cf~\semarrow~\hat\cf'$ [From H1]}
		\step{nlc_2}{
			$\exists \cf'.$
			\begin{pfenum*}
				\item $\cf \semarrow \cf'$
				\item $\rfun{\hat\cf'}{\hat\cf''} \semarrow \cf'$
			\end{pfenum*}
			[From H2, H3, \stepref{nlc_1} and Theorem \ref{theo:verification}]
		}
		\step{nlc_3}{
			$\cf~\vssemarrow\cf'$ [From \stepref{nlc_2} and execution of non-logic command.]
		}
	\end{proofpf}
	\step{nlc}{\case{Unfold}}
	\begin{proofpf}
		\step{assume_u}{
			\assume{
				\begin{pfenum}
					\item (H1) $\hat\cf~\vssemarrow~\hat\cf'$
					\item (H2) $\hat\cf'' = \rfun{\hat\cf}{\hat\cf'}$
					\item (H3) $\stR{\hat\cf''}{\cf}$
				\end{pfenum}
			}
			\prove{
				$\exists \, \cf' \, . \, \cf~\vssemarrow~\cf' \gand \stR{\hat\cf'}{\cf'}$
			}
		}
		\step{u_1}{
			$\exists \, \vprog, \sst, \sst_1, \sst', \cs_1, i, j, \pn, \e, \x, {\Pass_i}\!\mid_{i=0}^n, \hat\gvl.$
			\begin{pfenum*}
				\item $\hat\cf = \tup{\sst, \cs_1, i}$
				\item $\cmd(\vprog, \cs_1, i) = \vunfold{\pn}{\e}$
				\item $\getppred{\vprog}{\pn}{} = \fullpredicate{\pn}{\x}{\Pass_0}{\Pass_n}$
				\item $\ee{\sst}{\e} = \hat\gvl$
				\item $\action{\sst}{getP}{\litlst{\pn, \hat\gvl}}{\sst_1}$
				\item $\actionSetter{\sst_1}{\Pass_j}{[x \mapsto \hat\gvl]}{(\sst', -)}$
				\item $\hat\cf' = \tup{\sst', \cs_1, i+1}$
			\end{pfenum*}
		}
		\step{u_2}{
			$\exists \, \sst''.$
			\begin{pfenum*}
				\item $\hat\cf'' = \tup{\sst'', \cs_1, i}$
				\item $\sst'' \leq \rfun{\sst}{\sst'}$
			\end{pfenum*}
			[From \stepref{u_1}.1 and H2]
		}
		\step{u_3}{
			$\exists \, \st, \cs_2.$
			\begin{pfenum*}
				\item $\hat\cf'' = \tup{\st, \cs_2, i}$
				\item $\str{\sst''}{\st}$
				\item $\vlR{\sst''}{\cs_1}{\cs_2}$
			\end{pfenum*}
			[From \stepref{u_2}.1 and H3]
		}
		\step{u_4}{$\cmd{\vprog,\cs_2, i} = \vunfold{\pn}{\e}$ [From \stepref{u_3}.3] and \stepref{u_1}.2}
		\step{u_5}{$\sst'' \leq \sst$ [From \stepref{u_2}.2]}
		\step{u_6}{
			$\exists \, \gvl.$
			\begin{pfenum*}
				\item $\ee{\st}{\e} = \gvl$
				\item $\vlR{\sst''}{\hat\gvl}{\gvl}$
			\end{pfenum*}
		}
		\step{u_7}{\prove{$\sst'' \leq \rfun{\sst}{\sst_1}$}}
		\begin{proofpf}
			\step{u_7_1}{$\sst \leq \sst$}
			\step{u_7_2}{$\sst' \rleq \sst_1$ [From \stepref{u_1}.6]}
			\step{u_7_3}{$\rfun{\sst}{\sst'} \leq \rfun{\sst}{\sst_1}$ [From \stepref{u_7_1} and \stepref{u_7_2}]}
			\step{u_7_4}{$\sst'' \leq \rfun{\sst}{\sst_1}$ [From \stepref{u_2}.2 and \stepref{u_7_3}]}
		\end{proofpf}
		\step{u_8}{
			$\exists \st_1.$
			\begin{pfenum*}
				\item $\action{\st}{getP}{\litlst{\pn, \gvl}}{\st_1}$
				\item $\stR{\rfun{\sst_1}{\sst''}}{\st_1}$
			\end{pfenum*}
			[From \stepref{u_1}.5, \stepref{u_3}.2, \stepref{u_6}.2, \stepref{u_7}]
		}
		\step{u_9}{\prove{$\rfun{\sst_1}{\sst''} \leq \rfun{\sst_1}{\sst'}$}}
		\begin{proofpf}
			\step{u_9_1}{$\sst_1 \leq \sst_1$}
			\step{u_9_2}{$\sst'' \rleq \sst'$ [From \stepref{u_2}.2]}
			\step{u_9_3}{$\rfun{\sst_1}{\sst''} \leq \rfun{\sst_1}{\sst'}$ [From \stepref{u_9_1} and \stepref{u_9_2}]}
		\end{proofpf}
		\step{u_10}{\prove{$\sst'' \rleq \rfun{\sst_1}{\sst''}$}}
		\begin{proofpf}
			\step{u_10_1}{$\sst'' \rleq \sst'$ [From \stepref{u_2}.2]}
			\step{u_10_2}{$\sst' \rleq \sst_2$ [From \stepref{u_1}.6]}
			\step{u_10_3}{$\sst'' \rleq \sst_1$ [From \stepref{u_10_1} and \stepref{u_10_2}]}
			\step{u_10_4}{
				\vspace{-0.8cm}
				$\begin{array}{lll}
					&\\ &\\
					\sst'' \rleq \rfun{\sst_1}{\sst''} & \\
					\rfun{\sst''}{(\rfun{\sst_1}{\sst''})} & =  \rfun{(\rfun{\sst''}{\sst_1})}{\sst''}\\
					=  \rfun{\sst''}{\sst''}     			     & = \sst'' 
				\end{array}$
				[From \stepref{u_10_3}]}
		\end{proofpf}
		\step{u_11}{$\vlR{\rfun{\sst_1}{\sst''}}{\hat\gvl}{\gvl}$ [From \stepref{u_6}.2 and \stepref{u_10}]}
		\step{u_12}{
			$\exists \st'.$
			\begin{pfenum*}
				\item $\actionSetter{\st_1}{\Pass_j}{[x \mapsto \gvl]}{(\st', -)}$
				\item $\stR{\rfun{\sst'}{(\rfun{\sst_1}{\sst''})}}{\st'}$
			\end{pfenum*}
			[From \stepref{u_1}.6, \stepref{u_8}.2, \stepref{u_9}, \stepref{u_11} and Theorem \ref{teo:asrt:soundness}]
		}
		\step{u_13}{\prove{$\rfun{\st'}{(\rfun{\sst_1}{\sst''})} = \rfun{\sst'}{\sst''}$}}
		\begin{proofpf}
			\step{u_13_1}{$\sst' \rleq \sst_1$ [From \stepref{u_1}.6]}
			\step{u_13_2}{$\rfun{\sst'}{(\rfun{\sst_1}{\sst''})} = \rfun{(\rfun{\sst'}{\sst_1})}{\sst''}$}
			\step{u_13_3}{$\rfun{(\rfun{\sst'}{\sst_1})}{\sst''} = \rfun{\sst'}{\sst''}$ [From \stepref{u_13_1}]}
			\step{u_13_4}{$\rfun{\sst'}{(\rfun{\sst_1}{\sst''})} = \rfun{\sst'}{\sst''}$ [From \stepref{u_13_2} and \stepref{u_13_3}]}
		\end{proofpf}
		\step{u_14}{$\stR{\rfun{\sst'}{\sst''}}{\st'}$ [From \stepref{u_12}.2 and \stepref{u_13}.3]}
		\step{u_15}{$\cf \vssemarrow \tup{\st', \cs_2, i+1} = \cf'$ [From \stepref{u_1}.3, \stepref{u_3}.1, \stepref{u_4}, \stepref{u_6}.1, \stepref{u_8}.1 and \stepref{u_12}.1]}
		\step{u_16}{\prove{$\sst'' \rleq \rfun{\sst'}{\sst''}$}}
		\begin{proofpf}
			\step{u_16_1}{$\sst'' \rleq \sst'$ [From \stepref{u_2}.2]}
			\step{u_16_2}{
				\vspace{-0.8cm}
				$\begin{array}{ll}
					&\\ &\\
					\sst'' \rleq \rfun{\sst'}{\sst''} & \\
					\rfun{\sst''}{(\rfun{\sst'}{\sst''})} & =  \rfun{(\rfun{\sst''}{\sst'})}{\sst''}\\
					=  \rfun{\sst''}{\sst''}     			     & = \sst'' 
				\end{array}$
				[From \stepref{u_16_1}]}
			\step{u_17}{$\vlR{\rfun{\sst'}{\sst''}}{\cs_1}{\cs_2}$ [From \stepref{u_3}.3 and \stepref{u_16_2}]}
			\step{u_18}{$\rfun{\hat\cf}{\hat\cf''} \semarrow \cf'$ [From \stepref{u_1}.7, \stepref{u_2}.1, \stepref{u_14}, \stepref{u_15} and \stepref{u_17}]}
		\end{proofpf}
	\end{proofpf}
	\step{ac}{\case{Spec Call}}
	\begin{proofpf}
		\step{ac_assume}{
			\assume{
				\begin{pfenum}
					\item (H1) $\hat\cf~\vssemarrow~\hat\cf'$
					\item (H2) $\hat\cf'' \leq \rfun{\hat\cf}{\hat\cf'}$
					\item (H3) $\stR{\hat\cf''}{\cf}$
				\end{pfenum}
			}
			\prove{$\exists \, \cf' \, . \, \cf~\vssemarrow~\cf' \gand \stR{\hat\cf'}{\cf'}$}
		}
		\step{ac_1}{
			$\exists \, \vprog, \sst, \sst_1, \sst_2, \sst', \cs_1, i, j, \e_R, \hat\gvl_R, \e_\f, \f, \x_0, \rep{\lx}{i}{1}{n}, \rep{\e}{i}{0}{n}, \rep{\hat\gvl}{i}{0}{n}, \Pass, \Qass, \hat\subst.$
			\begin{pfenum*}
				\item $\hat\cf = \tup{\sst, \cs_1, i}$
				\item $\cmd(\vprog, \cs_1, i) = \vcall{\x}{\e_f}{\e_0}{j; \left(\lx_i: \e_i\right)\!\mid_{i=1}^n} \gand \f = \ee{\sst}{\e_f}$
				\item $\getspec{\vprog}{\f}{j} = \spec{\Pass}{\f}{\x_0}{\Qass}{\e_R}$
				\item $\ee{\sst}{\e_0} = \hat\gvl_0$
				\item $\ee{\sst}{\e_i} = \hat\gvl_i\mid_{i=1}^n$
				\item $\ee{\sst}{\e_R} = \hat\gvl_R$
				\item $\hat\subst = [x_0 \mapsto \hat\gvl_0, ..., \lx_n \mapsto \hat\gvl_n]$
				\item $\actionGetter{\sst}{\Pass}{\hat\subst}{\sst_1}$
				\item $\actionSetterS{\sst_1}{\Qass}{\hat\subst}{\sst_2}$
				\item $\sst_2.\kwT{setVar}(\x_R, \hat\gvl_R) \vsemarrow \tup{\sst', \cs_1, i+1}$
				\item $\hat\cf' = \tup{\sst', \cs_1, i + 1}$
			\end{pfenum*}
		}
		\step{ac_2}{
			$\exists \, \sst''$
			\begin{pfenum*}
				\item $\hat\cf'' = \tup{\sst'', \cs_1, i}$
				\item $\sst'' \leq \rfun{\sst}{\sst'}$
			\end{pfenum*}
			[From \stepref{ac_1}.1, \stepref{ac_1}.11 and H2]			
		}
		\step{ac_3}{
			$\exists \, \st, \cs_2, $
			\begin{pfenum*}
				\item $\cf = \tup{\st, \cs_2, i}$
				\item $\stR{\sst''}{\st}$
				\item $\vlR{\sst''}{\cs_1}{\cs_2}$
			\end{pfenum*}
			[From \stepref{ac_2}.1 and H3]			
		}
		\step{ac_4}{$\cmd(\vprog, \cs_2, i) = \vcall{\x}{\e_f}{\e_0}{j; \left(\lx_i: \e_i\right)\!\mid_{i=1}^n} \gand \f = \ee{\st}{\e_f}$ [From \stepref{ac_1}.2 and \stepref{ac_3}.3] }
		\step{ac_5}{$\sst'' \leq \sst$ [From \stepref{ac_2}.2]}
		\step{ac_6}{
			$\exists \, \gvl_0.$
			\begin{pfenum*}
				\item $\ee{\st}{\e_0} = \gvl_0$
				\item $\vlR{\sst''}{\hat\gvl_0}{\gvl_0}$
			\end{pfenum*}
			[From \stepref{ac_1}.4, \stepref{ac_3}.2 and \stepref{ac_5}]
		}
		\step{ac_7}{
			$\exists \, \rep{\gvl}{i}{1}{n}. \, \forall 1 \leq i \leq n.$
			\begin{pfenum*}
				\item $\ee{\st}{\e_i} = \gvl_i$
				\item $\vlR{\sst''}{\hat\gvl_i}{\gvl_i}$
			\end{pfenum*}
			[From \stepref{ac_1}.5, \stepref{ac_3}.2 and \stepref{ac_5}]
		}
		\step{ac_8}{
			$\exists \, \gvl_R$
			\begin{pfenum*}
				\item $\ee{\st}{\e_R} = \gvl_R$
				\item $\vlR{\sst''}{\hat\gvl_R}{\gvl_R}$
			\end{pfenum*}
			[From \stepref{ac_1}.6, \stepref{ac_3}.2 and \stepref{ac_5}]
		}
		\step{ac_9}{\pflet{$\subst = [x_0 \mapsto \gvl_0, ..., \lx_n \mapsto \gvl_n]$}}
		\step{ac_10}{$\vlR{\sst''}{\hat\subst}{\subst}$ [From \stepref{ac_1}.7, \stepref{ac_6}.2, \stepref{ac_7}.2, \stepref{ac_8}.2, \stepref{ac_9}]}
		\step{ac_11}{\prove{$\sst'' \leq \rfun{\sst}{\sst_1}$}}
		\begin{proofpf}
			\step{ac_11_1}{$\sst \leq \sst$}
			\step{ac_11_2}{$\sst' \rleq \sst_1$ [From \stepref{ac_1}.9]}
			\step{ac_11_3}{$\rfun{\sst}{\sst'} \leq \rfun{\sst}{\sst_1}$ [From \stepref{ac_11_2} and \stepref{ac_11_3}]}
			\step{ac_11_4}{$\sst'' \leq \rfun{\sst}{\sst_1}$ [From \stepref{ac_2}.2 and \stepref{ac_11_3}]}
		\end{proofpf}
		\step{ac_12}{
			$\exists \, \st_1.$
			\begin{pfenum*}
				\item $\actionGetter{\st}{\Pass}{\subst}{\st_1}$		
				\item $\stR{\rfun{\sst_1}{\sst''}}{\st_1}$
			\end{pfenum*}
			[From \stepref{ac_1}.8, \stepref{ac_3}.2, \stepref{ac_10}, \stepref{ac_11} and Theorem \ref{teo:asrt:soundness}]
		}
		\step{ac_13}{\prove{$\rfun{\sst_1}{\sst''} \leq \rfun{\sst_1}{\sst_2}$}}
		\begin{proofpf}
			\step{ac_13_1}{$\sst_1 \leq \sst_1$}
			\step{ac_13_2}{$\sst'' \rleq \sst'$ [From \stepref{ac_2}.2]}
			\step{ac_13_3}{$\sst \rleq \sst_2$ [From \stepref{ac_1}.10]}
			\step{ac_13_4}{$\sst'' \rleq \sst_2$ [From \stepref{ac_13_2} and \stepref{ac_13_3}]}
			\step{ac_13_5}{$\rfun{\sst_1}{\sst''} \leq \rfun{\sst_1}{\sst_2}$ [From \stepref{ac_13_1} and \stepref{ac_13_4}]}
		\end{proofpf}
		\step{ac_14}{\prove{$\sst'' \rleq \rfun{\sst_1}{\sst''}$}}
		\begin{proofpf}
			\step{ac_14_1}{$\sst'' \rleq \sst'$ [From \stepref{ac_2}.2]}
			\step{ac_14_2}{$\sst' \rleq \sst_1$ [From \stepref{ac_1}.9, \stepref{ac_1}.10 and transitivity]}
			\step{ac_14_3}{$\sst'' \rleq \sst_1$ [From \stepref{ac_14_1} and \stepref{ac_14_2}]}
			\step{ac_14_4}{
				\vspace{-0.8cm}
				$\begin{array}{ll}
					&\\ &\\
					\sst'' \rleq \rfun{\sst_1}{\sst''} & \\
					\rfun{\sst''}{(\rfun{\sst_1}{\sst''})} & =  \rfun{(\rfun{\sst''}{\sst_1})}{\sst''}\\
					=  \rfun{\sst''}{\sst''}     			     & = \sst'' 
				\end{array}$
				[From \stepref{ac_14_3}]}
		\end{proofpf}
		\step{ac_15}{$\vlR{\rfun{\sst_1}{\sst''}}{\hat\subst}{\subst}$ [From \stepref{ac_10} and \stepref{ac_14}]}
		\step{ac_16}{
			$\exists \, \st_2$
			\begin{pfenum*}
				\item $\actionSetterS{\st_1}{\Qass}{\subst}{\st_2}$
				\item $\stR{\rfun{\sst_2}{\sst''}}{\st_2}$
			\end{pfenum*}
			[From \stepref{ac_1}.2, \stepref{ac_12}.2, \stepref{ac_13}, \stepref{ac_15} and Theorem \ref{teo:asrt:soundness}]
		}
		\step{ac_17}{\prove{$\rfun{\sst_2}{\sst''} \leq \rfun{\sst_2}{\sst'}$}}
		\begin{proofpf}
			\step{ac_17_1}{$\sst_2 \leq \sst_2$}
			\step{ac_17_2}{$\sst'' \rleq \sst'$ [From \stepref{ac_2}.2]}
			\step{ac_17_3}{$\rfun{\sst_2}{\sst''} \leq \rfun{\sst_2}{\sst'}$ [From \stepref{ac_17_1} and \stepref{ac_17_2}]}
		\end{proofpf}
		\step{ac_18}{\prove{$\sst'' \rleq \rfun{\sst_2}{\sst''}$}}
		\begin{proofpf}
			\step{ac_18_1}{$\sst'' \rleq \sst' \rleq \sst_2$ [From \stepref{ac_2}.2 and \stepref{ac_1}.10]}
			\step{ac_18_2}{
				\vspace{-0.4cm}
				$\begin{array}{ll}
					& \\
					\rfun{\sst''}{(\rfun{\sst_2}{\sst''})} & = \rfun{(\rfun{\sst''}{\sst_2})}{\sst''} \\
					= \rfun{\sst''}{\sst''} & = \sst''
				\end{array}$
				[From \stepref{ac_18_1}]
			}
		\end{proofpf}
		\step{ac_19}{$\vlR{\rfun{\sst_2}{\sst''}}{\hat\gvl_R}{\gvl_R}$ [From \stepref{ac_8}.2 and \stepref{ac_18}]}
		\step{ac_20}{
			$\exists \, \st'.$
			\begin{pfenum*}
				\item $\st_2.\kwT{setVar}(\x_R, \gvl_R) \semarrow \st'$
				\item $\stR{\rfun{\sst'}{(\rfun{\sst_2}{\sst''})}}{\st'}$
			\end{pfenum*}
		}
		\step{ac_21}{\prove{$\rfun{\sst'}{(\rfun{\sst_2}{\sst''})} = \rfun{\sst'}{\sst''}$}}
		\begin{proofpf}
			\step{ac_21_1}{
				\vspace{-0.4cm}
				$\begin{array}{lll}
					& &\\
					\rfun{\sst'}{(\rfun{\sst_2}{\sst''})} & = \rfun{(\rfun{\sst'}{\sst_2})}{\sst''} & \text{[Given that } \sst' \rleq \sst_2\text{]}\\
					 & = \rfun{\sst'}{\sst''}
				\end{array}$}
		\end{proofpf}
		\step{ac_22}{$\stR{\rfun{\sst'}{\sst''}}{\st'}$}
		\step{ac_23}{\prove{$\sst'' \rleq \rfun{\sst'}{\sst''}$}}
		\begin{proofpf}
			\step{ac_23_1}{$\sst'' \rleq \sst'$ [From \stepref{ac_2}.2]}
			\step{ac_23_2}{
				\vspace{-0.4cm}
				$\begin{array}{ll}
					& \\
					\rfun{\sst''}{(\rfun{\sst'}{\sst''})} & = \rfun{(\rfun{\sst''}{\sst'})}{\sst''}\\
					 & = \rfun{\sst''}{\sst''} = \sst''
				\end{array} $}
		\end{proofpf}
		\step{ac_24}{$\vlR{\rfun{\sst'}{\sst''}}{\cs_1}{\cs2}$ [From \stepref{ac_3}.3 and \stepref{ac_22}]}
		\step{ac_25}{$\cf = \tup{\st, \cs_2, i} \semarrow \tup{\st', \cs_2, i+1} = \cf'$}
		\step{ac_26}{$\stR{\rfun{\hat{\cf}}{\hat\cf'}}{\cf'}$}
	\end{proofpf}
\end{proofpf}

\subsection{\while: Verification}
\label{app:subsec:while:verification}

\begin{itemize}
\item $\corePreds_{\wsub} =  \lbrace \wpcell \rbrace$
\item $\actionswhile = \lbrace \lookupAction, \mutateAction, \disposeAction, \wscell, \wgcell \rbrace$
\end{itemize}

\medskip
\begin{display}{\while: Concrete and Symbolic Memories (Continued)}
\begin{minipage}{0.40\textwidth}
{\small
\begin{mathpar}
\inferrule[C-SetCell]
  {
     (\loc, \prop) \not\in \domain(\cmem)
     \quad 
     \cmem' = \cmem \dunion \wcell{\loc}{\prop}{\vl}
  }{\action{\cmem}{\wscell}{\litlst{\loc, \prop, \vl}}{(\cmem', \true)}} 
 \\
\inferrule[C-GetCell]
  {
    \cmem = \cmem' \dunion \wcell{\loc}{\prop}{\vl}
  }{\action{\cmem}{\wgcell}{\litlst{\loc, \prop, \vl}}{(\cmem', \true)}} 
\end{mathpar}}
\end{minipage}
\begin{minipage}{0.59\textwidth}
{\small
\begin{mathpar}
\inferrule[S-SetCell]
  {
   \memproj{\smem}{\sexp_l, \prop, \pc} = \emptyset
     \and 
      \smem' = \smem \dunion \wcell{\sexp_l}{\prop}{\sexp_v}
  }{\action{\smem}{\wscell}{\litlst{\sexp_l, \prop, \sexp_v}, \pc}{\litset{(\smem', \true, \true)}}} 
 \\
\inferrule[S-GetCell]
  {
       \pc \vdash (\sexp_l = \sexp_l' \gand \sexp_v = \sexp_v')
       \and
  	  \smem = \smem' \dunion \wcell{\sexp_l'}{\prop}{\sexp_v'}
  }{\action{\smem}{\wgcell}{\litlst{\sexp_l, \prop, \sexp_v}, \pc}{\litset{(\smem', \true, \true)}}} 
\end{mathpar}}
\end{minipage}
\end{display}

\begin{lemma}[\while: Concrete Core Predicate Interpretation]\label{lem:while:pred:concrete}
	The core predicate action interpretation $\tup{\corePreds_{\wsub}, \actionswhile, [\wpcell \mapsto \wscell], [\wpcell \mapsto \wgcell]}$ is well-formed 
	with respect to $\cstateConstr(\wcmems)$.
\end{lemma}
\begin{proofpf}
	\pf \pflongnumbers \\
	We need to prove that the following equivalent holds for any concrete memories $\cmem, \cmem' \in \wcmems$, and any triple $\litlst{\loc, \prop, \vl}$ :
	$$
		\action{\st}{\wgcell}{\litlst{\loc, \prop, \vl}}{\st'} \iff 
		\action{\st'}{\wscell}{\litlst{\loc, \prop, \vl}}{\st}
	$$
	We start by proving the left to right implication, before proving the right to left implication.
	\step{ltr}{\prove{$\action{\st}{\wgcell}{\litlst{\loc, \prop, \vl}}{\st'} \Rightarrow 
	\action{\st'}{\wscell}{\litlst{\loc, \prop, \vl}}{\st}$}}
	\begin{proofpf}
		\step{ltr_assume}{
			\assume{(H1) $\action{\st}{\wgcell}{\litlst{\loc, \prop, \vl}}{\st'}$}
			\prove{$\action{\st'}{\wscell}{\litlst{\loc, \prop, \vl}}{\st}$}
			}
		\step{ltr_1}{$\exists \, \cmem'. \cmem = \cmem' \dunion (\loc, \prop) \mapsto \vl$ [From (H1)]}
		\step{ltr_2}{
			$\exists \, \cmem'', \rep{\prop}{i}{1}{n}.$
			\begin{pfenum*}
				\item $\memproj{\cmem'}{\loc} = \cmem'' \dunion \_$
				\item $\cmem'' = \dunion_{i=1}^n (\loc, \prop_i) \mapsto \_$
				\item $(\prop_i \neq \prop) \mid_{i=1}^n$
			\end{pfenum*}
			[From \stepref{ltr_1}]
		}
		\step{ltr_3}{$\action{\st'}{\wscell}{\litlst{\loc, \prop, \vl}}{\st}$ [From \stepref{ltr_2}]}
	\end{proofpf}

	\step{rtl}{\prove{$\action{\st}{\wgcell}{\litlst{\loc, \prop, \vl}}{\st'} \Leftarrow 
	\action{\st'}{\wscell}{\litlst{\loc, \prop, \vl}}{\st}$}}
	\begin{proofpf}
		\step{ltr_assume}{
			\assume{(H1) $\action{\st}{\wscell}{\litlst{\loc, \prop, \vl}}{\st'}$}
			\prove{$\action{\st'}{\wgcell}{\litlst{\loc, \prop, \vl}}{\st}$}
			}
		\step{ltr_1}{
			$\exists \, \cmem'', \rep{\prop}{i}{1}{n}.$
			\begin{pfenum*}
				\item $\memproj{\cmem'}{\loc} = \cmem'' \dunion \_$
				\item $\cmem'' = \dunion_{i=1}^n (\loc, \prop_i) \mapsto \_$
				\item $\prop \in \left\{ \prop_1, ..., \prop_n \right\}$
				\item $\cmem' = \cmem \dunion (\loc, \prop) \mapsto \vl$
			\end{pfenum*}
			[From \stepref{ltr_1}]
		}
		\step{ltr_2}{$\action{\st'}{\wgcell}{\litlst{\loc, \prop, \vl}}{\st}$ [From \stepref{ltr_1}]}
		
	\end{proofpf}
\end{proofpf}

\begin{lemma}[\while: Symbolic Core Predicate Interpretation]
	The core predicate action interpretation $\tup{\corePreds_{\wsub}, \actionswhile, [\wpcell \mapsto \wscell], [\wpcell \mapsto \wgcell]}$ is well-formed 
	with respect to $\sstateConstr(\wsmems)$.
\end{lemma}
\begin{proofpf}
	\pf \ 
	The proof is analogous to the proof of Lemma \ref{lem:while:pred:concrete}.
\end{proofpf}

\begin{lemma}[Extended \while: Memory Interpretation]
	$\winterp$ is an intepretation of While symbolic memories $\wsmems$ with respect to the concrete memories $\wcmems$, for \while extended with specifications.
\end{lemma}
\begin{proofpf}
	\pf \pflongnumbers\\
	For every action $\act \in \actionswhile$, we have to prove that:

	$\forall \smem, \smem' \in \wsmems;\, \cmem \in \wcmems;\, \sexp, \sexp' \in \sexps;\, \lenv \in \lenvs;\, \pc, \pc' \in \pcs$. 
	\begin{equation}
		\begin{array}{l}
		\action{\smem}{\act}{\sexp, \pc}{(\smem', \sexp', \pc')} 
			\gand  \cmem = \winterp(\smem, \lenv)
			\gand \seval{\pc \, \wedge \, \pc'}{\lenv} = \true \\
			 \qquad \qquad \implies \exists \, \cmem' \, . \, 
				 \cmem' = \winterp(\smem', \lenv) 
				 \gand 
				 \caction{\cmem}{\act}{\seval{\sexp}{\lenv}}{(\cmem', \seval{\sexp'}{\lenv})}
		\end{array}
	\end{equation}
	We proceed by case analysis on the rule that was used to derive $\action{\smem}{\act}{\sexp, \pc}{(\smem', \sexp', \pc')}$.

	For the $\lookupAction$, $\mutateAction$ and $\disposeAction$ actions we refer to the Lemma \ref{lemma:while:soundness}.

	Only the $\wscell$ and $\wgcell$ cases remain. We prove the $\wscell$ case, the $\wgcell$ case being analogous. 
	\step{main_case}{\case{$\wscell$}}
	\begin{proofpf}
		\step{assume}{
			\assume{
				\begin{pfenum}
					\item (H1) $\action{\smem}{\wscell}{\litlst{\sexp_l, \prop, \sexp_v}, \pc}{(\smem', \sexp', \pc')}$
					\item (H2) \begin{pfenum}
						\item $\seval{\sexp_l}{\lenv} = \loc$
						\item $\seval{\sexp_v}{\lenv} = \vl$
					\end{pfenum}
					\item (H3) $\cmem = \winterp(\smem,\lenv)$
					\item (H4) $\seval{\pc \, \wedge \, \pc'}{\lenv} = \true$
				\end{pfenum}
			}
			\prove{ $\exists \, \cmem' \, . \, 
			\cmem' = \winterp(\smem', \lenv) 
			\gand 
			\caction{\cmem}{\act}{\seval{\sexp}{\lenv}}{(\cmem', \seval{\sexp'}{\lenv})}$ }
		}
		\step{1}{
			$\exists \, \smem_1, \smem_2, \rep{\prop}{i}{1}{n}$
			\begin{pfenum*}
				\item $\memproj{\smem}{\sexp_l, \pc} = \smem_1 \dunion \_$
				\item $\smem_1 = \dunion_{i=1}^n (\sexp_i, \prop_i) \mapsto \_$
				\item $(\prop \neq \prop_i)\!\mid_{i = 1}^n$
				\item $\smem' = \smem \dunion (\sexp_l, \prop) \mapsto \sexp_v$
				\item $\sexp' = \true$
				\item $\pc' = \true$
			\end{pfenum*}
			[From (H1)]
		}
		\step{2}{$\memproj{\winterp(\smem, \lenv)}{\seval{\sexp_l}{\lenv}} = \winterp(\smem, \lenv) \dunion \_$ [From \stepref{1}.1 and (H4)]}
		\step{3}{$\winterp(\smem_1, \lenv) = \dunion_{i=1}^n (\seval{\sexp_l}{\lenv}, \prop_i) \mapsto \_$ [From \stepref{1}.2]}
		\step{4}{$\memproj{\cmem}{\loc} = \winterp(\smem_1, \lenv) \dunion \_$ [From (H2.a), (H3) and \stepref{2}]}
		\step{5}{$\winterp(\smem_1, \lenv) = \dunion_{i=1}^n (\loc, \prop_i) \mapsto \_$ [From (H2.a) \stepref{3}]}
		\step{6}{$\action{\cmem}{\wscell}{\litlst{\loc, \prop, \vl}}{(\cmem \dunion (\loc, \prop) \mapsto \vl , \true)}$ [From \stepref{1}.3, \stepref{4} and \stepref{5}]}
		\step{7}{
			\vspace{-1.2cm}
			$\begin{array}{lll}
				  &     & \\
				  &     & \\
				  &     & \\
				\winterp(\smem', \lenv) & = \winterp(\smem \dunion (\sexp_l, \prop) \mapsto \sexp_v, \lenv) & \text{[From \stepref{1}.4]}\\
				  & = \winterp(\smem, \lenv) \dunion \winterp ((\sexp_l, \prop) \mapsto \sexp_v, \lenv) & \\
				  & = \smem \dunion (\seval{\sexp_l}{\lenv}, \prop) \mapsto \seval{\sexp_v}{\lenv} & \text{[From (H3)]}\\
				  & = \smem \dunion (\loc, \prop) \mapsto \vl & \text{[From (H2)]}\\
			\end{array}$
		}

	\end{proofpf}
\end{proofpf}

\newpage
\section{Section 5: Parametric Bi-Abductive Analysis}
\label{app:s5}

\subsection{Parametric Bi-abductive Analysis}

\begin{display}{Action Fixes}
\quad\begin{minipage}{0.95\textwidth}
 $\kwT{ea} : \actions \tmap \sset{\gstate} \tmap \gval \pmap \power{\sset{\gstate} \times \gval} \dunion \power{\Passes{\corePreds}}$ 
            \hfill ($(\st', \vl') \in \kwT{ea}(\act, \st, \vl) \equiv_{pp} \succAction{\st}{\act}{\gvl}{\st', v'}$) \\ 
            $\phantom{xx}$ \hfill  ($\Pass \in \kwT{ea}(\act, \st, \vl) \equiv_{pp} \failAction{\st}{\act}{v}{\Pass}$)
\end{minipage}
\end{display}

\begin{definition}[Well-formedness of Fixes] If the original 
state is extended with a generated fix, it must be possible to execute the action successfully. 
\begin{equation}
\label{app:eq:bi-abd}
\failAction{\st}{\act}{\gvl}{\Pass} \gand \actionSetterS{\st}{\Pass}{\zsubst}{\st''}
  \implies \exists \, \st', \gvl'.  \succAction{\st''}{\act}{\gvl}{\st', \gvl'}
\end{equation}
\end{definition}

\myparagraph{Bi-Abductive Analysis}
 
\begin{definition}[BiState Constructor (\bistateConstr)]\label{app:def:bi:lifting}
The bi-abductive state constructor $\bistateConstr : \gstates \tmap \gstates$ is defined as  
$\bistateConstr(\tup{\sset{\gstate}, \gval, \actions}) \semeq \tup{\sset{\gstate'}, \gval, \actions}$,
where: 

\medskip
{\small
\begin{tabular}{lll}
   \mybullet $\sset{\gstate'}$ &$\semeq$ & $\sset{\gstate} \times \Passes{\corePreds}$ \\ 
  \mybullet $\kwT{setVar}_{\bisub}(\tup{\st, \Pass}, \x, \gvl)$ 
       & $\semeq$ & 
       $\tup{\kwT{setVar}(\st,  \x, \gvl), \Pass}$ \\ 
  \mybullet $\kwT{setStore}_{\bisub}(\tup{\st, \Pass}, \sto)$
     & $\semeq$ & 
     $\tup{\kwT{setStore}(\st, \sto), \Pass}$ \\ 
  \mybullet $\kwT{store}_{\bisub}(\tup{\st, -, -})$
    & $\semeq$ & 
    $\kwT{store}(\st)$ \\
  \mybullet $\kwT{ee}_{\bisub}(\tup{\st, -, -}, \e)$
    & $\semeq$ & 
    $\kwT{ee}(\st, \e)$ \\
 \mybullet $\kwT{ea}_{\bisub}(\act, \tup{\st, \Pass}, \gvl)$
    & $\semeq$ & 
    $\left\{ (\tup{\st', \Pass}, \gvl') \mid \succAction{\st}{\act}{\vl}{\st', \gvl'} \right\}$ if $\act \neq \kwT{assume}$\\ 
 \mybullet $\kwT{ea}_{\bisub}(\kwT{assume}, \tup{\st, \Pass}, \gvl)$
    & $\semeq$ & 
    $\left\{ (\tup{\st', \Pass * \pc}, \gvl') \mid \upcast{\vl} = \pc \gand \succActionT{\st}{assume}{\vl}{\st', \gvl'} \right\}$\\ 
 \mybullet $\kwT{ea}_{\bisub}(\act, \tup{\st, \Pass}, \gvl)$
    & $\semeq$ & 
         $\left\{{\begin{array}{l}
               \tup{\st'', \Pass * \Qass}, \gvl') \\
               \qquad \mid
               \failAction{\st}{\act}{\vl}{\Qass} \gand 
               \actionSetterS{\st}{\Qass}{\zsubst}{\st'} \gand
               \succAction{\st'}{\act}{\vl}{\st'', \gvl'}
          \end{array}}\right\}$ 
\end{tabular}}
\end{definition}

\begin{lemma}[One-Step]
$$
\begin{array}{l}
  \vsemtrans{\tup{\st, \Pass}, \cs, i}{\tup{\st', \Qass}, \cs', j}{\bistateConstr(\gstate)}{\vprog}{\outcome}
  \\
  \qquad \implies 
  \exists \, \Pass'. \, \Qass \vdash \astar{\Pass}{\Pass'} \gand
  \actionSetterS{\st}{\Pass'}{\litlst{}}{\st''} \gand  
      \st'' \leq \rfun{\st}{\st'}
  \vsemtrans{\st'', \cs, i}{\st', \cs', j}{\gstate}{\vprog}{\outcome}
\end{array}
$$
\end{lemma}
\begin{proofpf}
  \pf \pflongnumbers\\
  We proceed by case analysis on the rule used to produce $\vsemtrans{\tup{\st, \Pass}, \cs, i}{\tup{\st', \Qass}, \cs', j}{\bistateConstr(\gstate)}{\vprog}{\outcome}$.

  We only provide the proof for the execute action case. The other cases are analogous.

  There are two rules concerning the execute action case :
  \begin{pfenum*}
    \item The case where the action does not generate an error
    \item The case where the action generates an error.
  \end{pfenum*}
  We are providing the proof for both cases.
  \step{ne}{\case{[No Error]}}
  \begin{proofpf}
    \step{ne_a}{\assume{$\vsemtrans{\tup{\st, \Pass}, \cs, i}{\tup{\st', \Qass}, \cs', j}{\bistateConstr(\gstate)}{\vprog}{\outcome}$}}
    \step{ne_1}{
      \begin{pfenum*}
        \item $\cmd(\vprog, \cs, i) = \x := \act(\exp)$
        \item $\ee{\st}{\exp} = \vl$
        \item $\succAction{\st}{\act}{\vl}{\st'', \gvl'}$
        \item $\st' = \st''.\kwT{setVar}(x, \vl')$
      \end{pfenum*}
      [From \stepref{ne_a}]
    }
    \step{ne_2}{
      $\vsemtrans{\st, \cs, i}{\st', \cs, i+1}{\gstate}{\vprog}{\outcome}$
      [From \stepref{ne_1}]
    }
    \step{ne_3}{
      $\Pass \vdash \astar{\Pass}{\emp}$
    }
    \step{ne_4}{
      $\actionSetterS{\st}{\emp}{\litlst{}}{\st}$
    }
  \end{proofpf}
  \step{e}{\case{[Action Error]}}
  \begin{proofpf}
    \step{e_a}{\assume{$\vsemtrans{\tup{\st, \Pass}, \cs, i}{\tup{\st', \Qass}, \cs', j}{\bistateConstr(\gstate)}{\vprog}{\outcome}$}}
    \step{e_1}{
      \begin{pfenum*}
        \item $\cmd(\vprog, \cs, i) = \x := \act(\exp)$
        \item $\ee{\st}{\exp} = \vl$
        \item $\failAction{\st}{\act}{\vl}{\Qass}$
        \item $\st' = \st''.\kwT{setVar}(x, \vl')$
        \item $\actionSetterS{\st}{\Pass'}{\litlst{}}{\st''}$
        \item $\succAction{\st''}{\act}{\vl}{\st_1, \vl_1}$
        \item $\st' = \st_1.\kwT{setVar}(x, \vl_1)$
        \item $\Qass = \astar{\Pass}{\Pass'}$
      \end{pfenum*}
      [From \stepref{e_a}]
    }
    \step{e_2}{
      $\ee{\st''}{\exp} = \vl$
      [From \stepref{e_1}.2 and \stepref{e_1}.4]
    }
    \step{e_3}{
      $\vsemtrans{\st'', \cs, i}{\st', \cs, i+1}{\gstate}{\vprog}{\outcome}$
      [From \stepref{e_1}.1, \stepref{e_2} and \stepref{e_1}.3]
    }
  \end{proofpf}
\end{proofpf}

\begin{theorem}[Multiple-Step]\label{multi-step-bi}
$$
  \begin{array}{l}
    \vsemtrans{\tup{\st, \Pass}, \cs, i}{\tup{\st', \Qass}, \cs', j}[*]{\bistateConstr(\gstate)}{\vprog}{\outcome}
    \\
    \qquad \implies 
    \exists \, \Pass'. \, \Qass \vdash \astar{\Pass}{\Pass'} \gand
    \actionSetterS{\st}{\Pass'}{\litlst{}}{\st''} \gand  
        \st'' \leq \rfun{\st}{\st'}
    \vsemtrans{\st'', \cs, i}{\st', \cs', j}[*]{\gstate}{\vprog}{\outcome}
  \end{array}
$$
\end{theorem}
\begin{proofpf}
  \pf \pflongnumbers\\
  We proceed by induction on the length of the derivation $\vsemtrans{\tup{\st, \Pass}, \cs, i}{\tup{\st', \Qass}, \cs', j}[*]{\bistateConstr(\gstate)}{\vprog}{\outcome}$.

  \step{b}{\case{$n = 0$}}
  \begin{proofpf}
    \step{b_a}{\assume{$\vsemtrans{\tup{\st, \Pass}, \cs, i}{\tup{\st', \Qass}, \cs', j}[0]{\bistateConstr(\gstate)}{\vprog}{\outcome}$}}
    \step{b_1}{
      \begin{pfenum}
        \item $\st' = \st$
        \item $\Qass = \Pass$
        \item $\cs' = \cs$
        \item $j = i$
      \end{pfenum}
    }
    \step{b_2}{
      $\vsemtrans{\st, \cs, i}{\st, \cs, i}[0]{\gstate}{\vprog}{\outcome}$
    }
    \step{b_3}{$\Pass \vdash \astar{\Pass}{\emp}$}
    \step{b_4}{$\actionSetterS{\st}{\emp}{\litlst{}}{\st}$}
  \end{proofpf}
  \step{b}{\case{$n = k+1$}}
  \begin{proofpf}
    \step{b_a}{\assume{
      $\vsemtrans{\tup{\st, \Pass}, \cs, i}{\tup{\st', \Qass}, \cs', j}[k+1]{\bistateConstr(\gstate)}{\vprog}{\outcome}$
    }}
    \step{b_1}{
      $\exists\, \st_1, \Qass_1, \cs_1, j_1.$
      \begin{pfenum*}
        \item  $\vsemtrans{\tup{\st, \Pass}, \cs, i}{\tup{\st_1, \Qass_1}, \cs_1, j_1}[k]{\bistateConstr(\gstate)}{\vprog}{\outcome}$
        \item  $\vsemtrans{\tup{\st_1, \Qass_1}, \cs_1, j_1}{\tup{\st', \Qass}, \cs', j'}{\bistateConstr(\gstate)}{\vprog}{\outcome}$
      \end{pfenum*}
      [From \stepref{b_1}]
    }
    \step{b_2}{
      $\exists\, \st_2, \Pass_2$
      \begin{pfenum*}
        \item $\Qass_1 \vdash \astar{\Pass}{\Pass_2}$
        \item $\actionSetterS{\st}{\Pass_2}{\litlst{}}{\st_2}$
        \item  $\vsemtrans{\st_2, \cs, ji}[K]{\st_1, \cs_1, j_1}{\gstate}{\vprog}{\outcome}$
      \end{pfenum*}
      [From IH and \stepref{b_1}.1]
    }
    \step{b_3}{
      $\exists\, \Pass_3, \st_3$
      \begin{pfenum*}
        \item $\Qass \vdash \astar{\Qass_1}{\Pass_3}$
        \item $\actionSetterS{\st_1}{\Pass_3}{\litlst{}}{\st_3}$
        \item  $\vsemtrans{\st_3, \cs, i}{\st', \cs', j}{\gstate}{\vprog}{\outcome}$
      \end{pfenum*}
      [From previous Lemma]
    }
    \step{b_4}{
      $\exists\, \st_4$
      \begin{pfenum*}
        \item $\actionSetterS{\st_2}{\Pass_3}{\litlst{}}{\st_4}$
        \item $\vsemtrans{\st_4, \cs, i}{\st_3, \cs_1, j_1}[k]{\gstate}{\vprog}{\outcome}$
      \end{pfenum*}
      [From Frame rule, \stepref{b_3}.2 and \stepref{b_2}.3]
    }
    \step{b_5}{
      $\vsemtrans{\st_4, \cs, i}{\st', \cs', j}[k+1]{\gstate}{\vprog}{\outcome}$
      [From \stepref{b_3}.3 and \stepref{b_4}.2]
    }
    \step{b_6}{
      $\actionSetterS{\st}{\star{\Pass_2}{\Pass_3}}{\litlst{}}{\st_5}$
      [From \stepref{b_2}.2 and \stepref{b_4}.1]
    }
    \step{b_7}{
      $\Qass \vdash \astar{\astar{\Pass}{\Pass_2}{\Pass_3}}$
      [From \stepref{b_2}.1 and \stepref{b_3}.1]
    }
  \end{proofpf}
\end{proofpf}

\begin{theorem}[Bi-Abduction]
$$
\begin{array}{l}
\vsemtrans{\tup{\st, \emp}, \cs, i}{\tup{\st', \Pass}, \cs', j}[*]{\bistateConstr(\gstate)}{\vprog}{\outcome}
  \gand 
  \actionSetterS{\st}{\Pass}{\zsubst}{\st''} \\
  \qquad \implies 
  \vsemtrans{\st'', \cs, i}{\st', \cs', j}{\gstate}{\vprog}{\outcome}
\end{array}
$$
\end{theorem}
\begin{proofpf}
  \pf: \ Immediate corollary from Theorem \ref{multi-step-bi}
\end{proofpf}

\end{document}